\documentclass{aa}  

\usepackage{graphicx}
\usepackage{txfonts}
\usepackage{natbib}
\bibpunct{(}{)}{;}{a}{}{,} 

\usepackage{longtable}

\usepackage{lscape}
\usepackage{footnote}
\makesavenoteenv{tabular}
\usepackage{twoopt}
\usepackage{morefloats}

\usepackage{marvosym}
\usepackage{color,hyperref}
\definecolor{darkblue}{rgb}{0.0,0.0,0.3}
\hypersetup{colorlinks,breaklinks,
            linkcolor=blue,urlcolor=blue,
            anchorcolor=blue,citecolor=blue}

\begin{document} 

\newcommand{\new}{\textbf}
\newcommand{\kms}{km s$^{-1}$}

   \title{Physical and chemical differentiation of the luminous star-forming region W49A}

   \subtitle{Results from the JCMT Spectral Legacy Survey}
   
   \author{Z. Nagy\inst{1,2,3,4},
        F. F. S. van der Tak\inst{4,3},
        G. A. Fuller\inst{5},
        \and
		R. Plume\inst{6}
        }
          
\authorrunning{Z. Nagy et al.}     
\titlerunning{Physical and chemical differentiation of W49A}     

\institute
{Department of Physics and Astronomy, University of Toledo, 2801 West Bancroft Street, Toledo, OH 43606, USA \\
\email{zsofia.nagy.astro@gmail.com} 
\and
I. Physikalisches Institut, Universit\"at zu K\"oln, Z\"ulpicher Str. 77, 50937 K\"oln, Germany
\and
Kapteyn Astronomical Institute, University of Groningen, P.O. Box 800, 9700AV, Groningen, The Netherlands
\and
SRON Netherlands Institute for Space Research, P.O. Box 800, 9700AV, Groningen, The Netherlands
\and
Jodrell Bank Centre for Astrophysics, School of Physics and Astronomy, University of Manchester, Manchester, M13 9PL, UK
\and
Department of Physics and Astronomy, University of Calgary, Calgary, T2N 1N4, AB, Canada
}


 
\abstract
{The massive and luminous star-forming region W49A is a well known Galactic candidate to probe the physical conditions and chemistry similar to those expected in external starburst galaxies. 
}
{We aim to probe the physical and chemical structure of W49A on a spatial scale of $\sim$0.8 pc based on the JCMT Spectral Legacy Survey, which covers the frequency range between 330 and 373 GHz.}
{
The wide $2\times2$ arcminutes field and the high spectral resolution of the HARP instrument on JCMT provides information on the spatial structure and kinematics of the cloud traced by the observed molecular lines.
For species where multiple transitions are available, we estimate excitation temperatures and column densities using a population diagram method that takes beam dilution and optical depth corrections into account.
}
{We detected 255 transitions corresponding to 60 species in the 330-373 GHz range at the center position of W49A. Excitation conditions can be probed for 16 molecules, including the complex organic molecules CH$_3$CCH, CH$_3$CN,  and CH$_3$OH. 
The chemical composition suggests the importance of shock-, PDR-, and hot core chemistry.
Many molecular lines show a significant spatial extent across the maps including CO and its isotopologues, high density tracers (e.g. HCN, HNC, CS, HCO$^+$), and tracers of UV-irradiation (e.g. CN and C$_2$H).
The spatially extended species reveal a complex velocity-structure of W49A with possible infall and outflow motions. Large variations are seen between the sub-regions with mostly blue-shifted emission toward the Eastern tail, mostly red-shifted emission toward the Northern clump, and emission peaking around the expected source velocity toward the South-west clump.
}
{
A comparison of column density ratios of characteristic species observed toward W49A to Galactic PDRs suggests that while the chemistry toward the W49A center is driven by a combination of UV-irradiation and shocks, UV-irradiation dominates for the Northern Clump, Eastern tail, and South-west clump regions.
A comparison to a starburst galaxy and an AGN suggests similar C$_2$H, CN, and H$_2$CO abundances (with respect to the dense gas tracer $^{34}$CS) between the $\sim$0.8 pc scale probed for W49A and the $>$1 kpc regions in external galaxies with global star-formation.  
} 
\keywords{stars: formation -- ISM: molecules -- ISM: individual objects: W49A}
\maketitle

\section{Introduction} 

Star-formation in galaxies occurs on different scales from isolated cores / globules with sizes of $<$1 pc through giant molecular clouds of a few 10 pc up to the scales of starburst galaxies representing star-formation on scales of a few 100 pc and with luminosities of $L_{\rm{IR}} > 10^{11}$ $L_\odot$ for luminous infrared galaxies (LIRGs) and $L_{\rm{IR}} > 10^{12}$ $L_\odot$ for ultraluminous infrared galaxies (ULIRGs). These regions also represent a large spread in star-formation rates with the 3 $M_\odot$/yr in the Milky Way to $\sim$10$^3$ $M_\odot$/yr in starburst galaxies
(e.g. \citealp{solomonvandenbout2005}).
The different types of star-forming regions in galaxies may trace different physical processes that control the star-formation in the different environments.

The Galactic starburst-analogue W49A with its more than $\sim$100 pc size \citep{simon2001} represents an intermediate case between the star-formation seen in Galactic star-forming regions and in starburst galaxies.
With its intense star-formation compared to other Galactic sources, studying its physical and chemical properties can give insights into the mechanism behind the starburst phenomenon. In our previous work based on part of the data presented in this paper we focused on the physical properties of W49A \citep{nagy2012}. In this work we focus more on the chemistry of this Galactic starburst analogue.
W49A is one of the most massive ($M\sim10^6$ $M_\odot$, \citealp{sievers1991}) and luminous ($>$10$^7$ $L_\odot$, \citealp{wardthompson1990}) star-forming regions in the Galaxy. 
Even though its luminosity is lower than that of starburst galaxies, W49A is more comparable to ULIRGs in terms of its luminosity per unit mass of gas. In ULIRGs values of $\sim$100 $L_\odot$ / $M_\odot$ are measured  (e.g. \citealp{solomon1997}) while the value for W49A is $>$10 $L_\odot$ / $M_\odot$. 
The luminosity of W49A is a result of an embedded stellar cluster containing the equivalent of about 100 O7 stars \citep{contiblum2002}, corresponding to four stellar clusters in a 5$'\times$5$'$ (16$\times$16 pc) region around the center based on $J$, $H$, and $K_S$ images \citep{alveshomeier2003}. \citet{wu2014} discovered a very massive star corresponding to the central cluster of W49A, and estimated its mass to be in the range between 90 and 250 $M_\odot$.
Apart from the stellar population observed at near-infrared wavelengths, the ongoing star formation which appears to be forming a comparable mass of stars can be studied in the radio continuum (e.g. \citealp{depree1997,depree2004,depree2005}) and at mm- and sub-mm (e.g. \citealp{wilner2001}, \citealp{galvanmadrid2013}) wavelengths as hot cores and Ultra-Compact (UC) H{\sc{ii}} regions.

The distance of W49A was recently estimated to be 11.11$^{+0.79}_{-0.69}$ kpc \citep{zhang2013}, which is consistent with the 11.4 kpc value derived by \citet{gwinn1992}.
Due to the large distance of W49A its small-scale structure including hot cores, outflows, and UC H{\sc{ii}} regions require interferometric observations to be fully resolved (e.g. \citealp{wilner2001}, \citealp{depree2005}, \citealp{galvanmadrid2013}). However, previous studies using single-dish telescopes have already provided information on the chemical complexity of W49A, and the physical and chemical properties of the region.

HCO$^+$ 1-0 lines with red-shifted absorption and blue-shifted emission were observed by \citet{welch1987} and interpreted as evidence of global collapse toward the central $\sim$2 pc region of W49A.
An alternative explanation for the large number of O-stars contributing to the luminosity of W49A was proposed by \citet{serabyn1993}, who suggested a cloud-cloud collision based on multiple transitions of CS and C$^{34}$S (from $J$=3-2 to $J$=10-9). However, based on more recent observations, this explanation is less probable than the global cloud collapse scenario. Large-scale maps of W49A (e.g. \citealp{galvanmadrid2013}) show a hierarchical network of filaments, converging from larger scales to the center of W49A (W49N).

Signatures of the high star-formation activity were discovered using Spitzer mid-IR images and position-velocity diagrams based on $^{13}$CO $J=2-1$ and C$^{18}$O $J=2-1$ maps by \citet{peng2010}, who have identified two expanding shells toward W49A.

\citet{vastel2001} have studied the physical conditions in the PDR component of W49A using the FIR lines of [O{\sc{i}}] and [C{\sc{ii}}] observed with the Long Wavelength Spectrometer of the Infrared Space Observatory as well as rotational lines of CO (CO 1-0, C$^{18}$O 2-1) observed with the 15-m SEST
telescope, and derived a radiation field of $G_0=3\times10^5$ (in units of 1.3$\times$10$^{-4}$ erg s$^{-1}$ cm$^{-2}$ sr$^{-1}$) and an average gas density of 10$^4$ cm$^{-3}$. 

A recent study based on data from the JCMT Spectral Legacy Survey (SLS, \citealp{plume2007}) focused on an extended warm and dense gas component toward W49A seen in H$_2$CO \citep{nagy2012}, and characterized the physical properties and excitation of the region using CH$_3$OH, SO$_2$, H$_2$CO, and HCN transitions. Another study that makes use of the SLS line survey data focuses on dense gas tracers \citep{roberts2011}, such as HCN, HNC, DCN, HCO$^+$, and their isotopologues, and by a comparison to line ratios measured toward starburst galaxies and Active Galactic Nuclei (AGN), as well as to chemical models, finds that W49A is a template for starburst galaxies rather than for AGN.

In this paper, we present results on the chemical inventory of W49A based on the SLS, carried out with the James Clerk Maxwell Telescope (JCMT) at a resolution of $\sim$15$''$. This line survey provides the largest frequency coverage data toward W49A up to date, and as such, it provides a useful starting point for future higher resolution studies with instruments such as ALMA.

\section{Observations and data reduction}

The SLS observations have been carried out using the 16-receptor (spatial pixel) Heterodyne Array Receiver Programme B (HARP-B, 325-375 GHz) and the Auto-Correlation Spectral Imaging System (ACSIS) correlator \citep{buckle2009} at the James Clerk Maxwell Telescope\footnote{The James Clerk Maxwell Telescope is operated by the Joint Astronomy Centre on behalf of the Science and Technology Facilities
Council of the United Kingdom, the Netherlands Organisation for Scientific Research, and the National Research Council of Canada.} (JCMT) on Mauna Kea, Hawai'i. The observations were carried out in jiggle position switch mode, sampled every 7.5$''$ for a 2$\times$2 arcminutes field centered on  RA(J2000) = 19$^{\rm{h}}$10$^{\rm{m}}$13$.^{\rm{s}}$4; Dec(J2000) = $09^\circ06'14''$.
The spectra were calibrated using an off-position corresponding to an offset (+840$''$, +840$''$) compared to the central position. Observations toward the reference position have been carried out to confirm that the off position is reliable and is without emission at the observed frequencies.
The pointing was checked every hour and is estimated to be accurate to 1.5$''$.
The angular resolution of the JCMT is $\sim$15$''$ at 345 GHz, equivalent to $\sim$0.8 pc at the distance of W49. The spectral resolution at the observed frequencies is $\sim$0.8 km s$^{-1}$ and the beam efficiency is 0.63 \citep{buckle2009}. The original line survey was carried out in the 330-360 GHz frequency range and was later extended to 373 GHz. This paper summarizes the results from the whole 330-373 GHz line survey.

The data reduction was done using a combination of tasks from the Starlink package and the ORAC Data Reduction pipeline (ORAC-DR). 
We reduced the observed time-series cubes using the ORAC-DR pipeline, which creates three-dimensional cubes after checking for consistency between the calculated $T_{\rm{sys}}$ and observed rms noise in the data; checking for variations in the rms noise measured by each receptor across the map, removes baselines from every spectrum and co-adds spectra that corresponds to the same position and frequency.
The results of the pipeline were checked and were corrected for remaining bad data, such as spectra with high rms noise level, bad baselines, and spectral ranges affected by spikes, using a combination of Starlink tasks.

\begin{figure}[!h]
\centering
\includegraphics[width=8 cm,trim=0cm -0.1cm 0cm 0cm,clip=true]{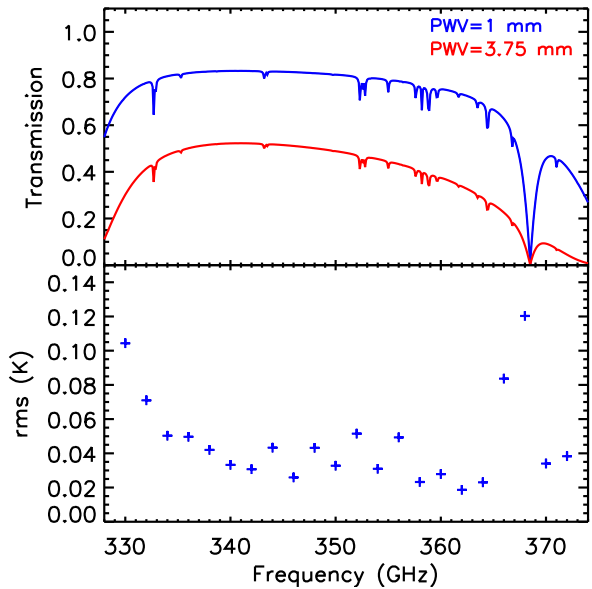}
\caption{The typical rms noise levels in $T_{\rm{A}}^*$ units measured in the SLS survey as a function of the frequency (bottom panel) and the corresponding atmospheric transmission for typical levels of water vapor at Mauna Kea (top panel).}
\label{rms_w49survey}
\end{figure}

Figure \ref{rms_w49survey} shows the typical rms noise levels as a function of frequency in the 330-373 GHz frequency range measured at every 2 GHz. Typical noise levels in $T_{\rm{A}}^*$ units are in the range between 0.02 and 0.12 K. The highest noise level measured at $\sim$368 GHz is due to the poor atmospheric transmission as shown in the top panel of Fig. \ref{rms_w49survey}. Atmospheric transmissions are shown for typical precipitable water vapor (PWV) levels of 1 mm and 3.75 mm on Mauna Kea.

\section{Results}

\subsection{Line identification}

We have identified the lines detected toward the center of W49A (RA(J2000) = 19$^{\rm{h}}$10$^{\rm{m}}$13$.^{\rm{s}}$4; Dec(J2000) = 09$^\circ$06$'$14$''$) in the frequency range between 330 and 373 GHz.
Analysed positions other then the center include the eastern tail (RA(J2000) = 19$^{\rm{h}}$10$^{\rm{m}}$16$.^{\rm{s}}$6; Dec(J2000) = 09$^\circ$05$'$48$''$), northern clump (RA(J2000) = 19$^{\rm{h}}$10$^{\rm{m}}$13$.^{\rm{s}}$6; Dec(J2000) = 09$^\circ$06$'$48$''$) and southwest clump (RA(J2000) = 19$^{\rm{h}}$10$^{\rm{m}}$10$.^{\rm{s}}$6; Dec(J2000) = 09$^\circ$05$'$18$''$). These positions were selected based on different kinematical signatures traced by dense gas tracers such as HCO$^+$, HCN, and HNC \citep{roberts2011} and are shown for example on Fig. \ref{w49_maps_shock}.
Appendix \ref{line_ident_details} shows details on the detected species.
We added a comment when the lines were detected in the high-mass protostar AFGL 2591, another source in the JCMT SLS, based on \citet{vanderwiel2011}. The spectroscopic data are based on the Cologne Database for Molecular Spectroscopy (CDMS, \citealp{muller2005})\footnote{\texttt{http://www.astro.uni-koeln.de/cdms/catalog}} and the Jet Propulsion Laboratory (JPL, \citealp{pickett1998})\footnote{\texttt{http://spec.jpl.nasa.gov}} molecular databases.
The identification is based on an initial search range of transitions with an upper level energy of 400 K, and was extended in particular cases such as vibrationally excited lines of HCN and HNC. In the case of weak ($\lesssim$3$\sigma$ detections) lines, an additional check of the spatial distribution and a cross-check of the detection in data corresponding to different rest frequencies was applied to avoid possible artifacts e.g. spurs or baseline errors. For the identification we considered velocities in a range between $\pm$5 \kms around the expected source velocity of 10 \kms and extended it for the most asymmetric lines.

\begin{figure}[!h]
\centering
\includegraphics[width=6.5cm,angle=-90,trim=0cm 0cm 0cm 0cm,clip=true]{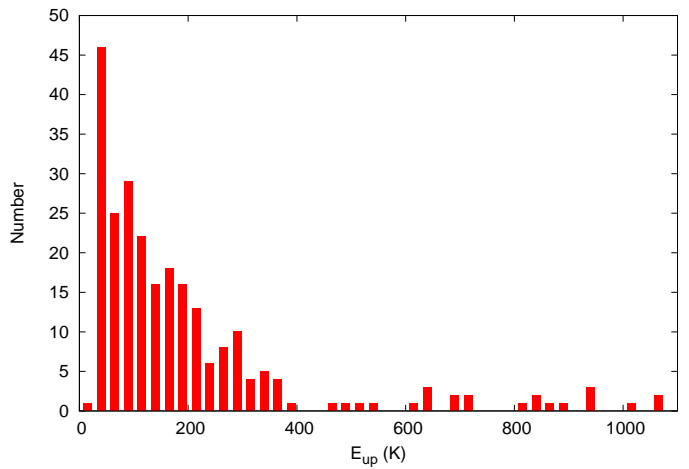}
\caption{
The number of transitions detected versus the upper state energy of the transition in 25 K wide bins.}
\label{w49_eup}
\end{figure}

Toward the centre position we detected 255 lines in the 330-373 GHz frequency range, that belong to 60 molecular species summarized in Table \ref{table:spatial_extent}. Most of the detected lines are related to sulphur-bearing species, such as SO$_2$ and its isotopologues. Various species are detected that indicate the possible importance of shock- and hot core chemistry and UV irradiation (see Sect. \ref{discussion}). The detected transitions cover a large energy range up to $\sim$1067 K corresponding to vibrationally excited HCN (Fuller et al, in prep.). However, most transitions have upper level energies below 400 K (Fig. \ref{w49_eup}). Excitation conditions can be probed for 16 species (Table \ref{lines_summary1}) based on multiple detected transitions with a sufficient energy range.
Among the 255 detected lines, there are 3 unidentified lines (U-lines in Appendix \ref{appendix_line_parameters}). 
With the large number of spectral features misassignment to species is possible. Also some of the line profiles are non-Gaussian, and the FWHM (full width at half maximum) width of the lines vary over a relatively large range ($\sim$7-19 \kms toward the central position). This may suggest that some of the identified lines are unresolved blends and have contribution to the line emission from other molecules / transitions. 
However the excitation diagram analysis provides a cross-check on the line identifications (and indeed highlighted some initial mis-assignments) and we believe that misindentified lines are not a signifiant issue.

To measure the spatial extent of the detected species, we apply a 2-dimensional Gaussian fit to the integrated line intensities, using the BEAMFIT task in Starlink. The values shown in Table \ref{table:spatial_extent} correspond to the FWHM values of the fit. For molecules with multiple detected transitions we include their largest spatial extent. 
For molecules with measured source sizes of $\sim$23$\times$23$''$, the lines were only detected toward a few positions around the source center. In the following sections we quote sizes of $\sim$17$\times$17$''$ for these species, to correct for the $\sim$15$''$ beam size of JCMT around the observed frequencies.
The center positions of the Gaussian fits corresponding to the different species are within a few arcsecond distance around the W49A center.
In addition to CO and its isotopologues, high density tracers (e.g. HCN, HNC, CS, HCO$^+$) and tracers of UV-irradiation (e.g. CN and C$_2$H) show the largest spatial extent (Table \ref{table:spatial_extent}).
The measured position angles cover a large range indicating a complex source structure.
Section \ref{section_species} includes a summary of the detected molecules. We discuss the excitation conditions and column densities of the species with multiple detected transitions in Section \ref{excitation}.

\begin{savenotes}
\begin{table*}
\begin{minipage}[h!]{\linewidth}
\centering
\caption{The measured spatial extent of the species detected toward the center of W49A. The position angle is measured from North through East.}             
\label{table:spatial_extent}
\resizebox{\textwidth}{!}{%
\renewcommand{\thefootnote}{\alph{footnote}}  
\begin{tabular}{llrllr|llrllr}
\hline     
\hline
\\[-0.2cm]

\footnotetext[1]{The error corresponding to the fitted major and minor axis values is a few arcseconds.}

\footnotetext[2]{Position angles are only derived for species with significant spatial extent and elongation in order to minimise the uncertainty.}

Species& Number of&     $E_{\rm{up}}$&    \multicolumn{3}{c|}{Spatial extent\footnotemark[1]}& Species& Number of&     $E_{\rm{up}}$&  \multicolumn{3}{c}{Spatial extent\footnotemark[1]}\\
       & transitions&   (K)&    Major& Minor& Pos. angle\footnotemark[2]& & transitions&   (K)&    Major& Minor& Pos. angle\\ 
       &       &                &        axis ($''$)& axis ($''$)&  (deg)& &       &                &        axis ($''$)& axis ($''$)&  (deg)\\ 
\hline\hline 

CO&                     1&  33&           98&   60&     90&
$^{13}$CO&              1&  32&           72&   54&     87\\[0.05cm]

C$^{17}$O&              1&  32&           39&   28&     69&
C$_2$H&                 3&  42&           61&   46&     79\\[0.05cm]

CN&                     3&  33&           54&   39&     77&
HCN&                    1&  43&           37&   27&     76\\[0.05cm]

CS&                     1&  66&           39&   24&     70&
DCN&                    1&  52&           36&   23&     146\\[0.05cm]

CH$_3$OH&              26&  $17-372$&     30&   23&     64&
H$_2$CO&                7&  $52-241$&     29&   23&     64\\[0.05cm] 

C$^{34}$S&              1&  65&           28&   23&     69&
H$_2$CS&                6&  $91-209$&     30&   23&     142\\[0.05cm]

NS&                     1&  70&           29&   23&     9&  
NO&                     3&  $36-209$&     29&   23&     62\\[0.05cm]

$^{33}$SO&              3&  $78-87$&      33&   23&     112&
HNC&                    1&    43&         31&   23&     67\\[0.05cm]

HCO$^+$&                1&    43&         32&   26&     179&
H&                      2&      &         35&   23&     77\\[0.05cm]


SO$_2$&                          41&  31$-$612&     23&  23&  & 
N$_2$H$^+$&                       1&   45&           23&  23&  \\[0.05cm]

$^{33}$SO$_2$&                   18&  $35-338$&     23&  23&  &
$^{13}$CS&                        1&   80&           23&  23&  \\[0.05cm]

$^{34}$SO$_2$&                   33&  $35-547$&     23&  23&  &
H$_3$O$^+$&                       1&  140&          23&  23&  \\[0.05cm] 

SO$^{18}$O&                       4&  $117-326$&    23&  23&  &
SiO&                              1&  75&           23&  23&  \\[0.05cm]

SO$^{17}$O&                       1&  $58-180$&     23&  23&  &
SO$_2$, v$_2$=1&                  9&  $805-998$&    23&  23&  \\[0.05cm]

$^{13}$CH$_3$OH&                  1&  45&           23&  23&   &
SO&                               6&  $26-143$&     23&  23&  \\[0.05cm]

HCS$^+$&                          1&  74&           23&   23&  & 
$^{34}$SO&                        5&  $25-85$&      23&   23&  \\[0.05cm]

H$^{13}$CN&                       1&  41&           23&   23&  &
OCS&                              3&  $237-271$&    25&   23&  \\[0.05cm]

SO$^+$&                           1&  70&           23&   23&  &  
HNCO&                             7&  $127-204$&    23&   23&  \\[0.05cm]

O$^{13}$C$^{34}$S&                1&  247&          23&   23&  &     
CH$_3$CN&                         5&  $151-215$&    23&   23&  \\[0.05cm]

HC$^{15}$N&                       1&  41&           23&   23&  &
H$^{13}$CO$^+$&                   1&  42&           23&   23&  \\[0.05cm]

HN$^{13}$C&                       1&  42&           23&   23&  &
CH$_3$CCH&                        4&  $172-254$&    23&   23&  \\[0.05cm]

HCN, v$_2$=1&                     1&  1067&         23&   23&  &
HC$_3$N&                          3&  $307-376$&    23&   23&  \\[0.05cm]

HC$_3$N, $v_7=1$&                 4&  629--663&     23&   23&  & 
HCO&                              1&  74&           23&   23&  \\[0.05cm]

CH$_3$OCH$_3$&                    2& $48-49$&       23&   23&  &  
S$^{18}$O&                        3& $91-99$&       23&   23&  \\[0.05cm]

CO$^+$&                           1&    34&         23&   23&  &
H$_2$S&                           2& $135-263$&     23&   23&  \\[0.05cm]

S$^{17}$O&                        2& $76-103$&      23&   23&  &
H$_2^{13}$CO&                     2& $61-65$&       23&   23&  \\[0.05cm]

CH$_3$CHO&                        1& $155$&         23&   23&  & 
HNC, v$_2$=1&                     2& $709-710$&     23&   23&  \\[0.05cm]

H$^{15}$NC&                       1& 43&            23&   23&  &
HC$^{17}$O$^+$&                   1& 42&            23&   23&  \\[0.05cm]

H$_2$CN&                          3& $53-100$&      23&   23&  &
HO$^{13}$C$^+$&                   1&    41&         23&   23&  \\[0.05cm]

HC$^{18}$O$^+$&                   1&    41&         23&   23&  & 
H$_2$C$^{34}$S&                   1&   118&         23&   23&  \\[0.05cm]

CH$_3$COCH$_3$&                   1&   282&         23&   23&  &
H$_2^{33}$S&                      1&   154&         23&   23&  \\[0.05cm] 
 
H$_2^{34}$S&                      1&   154&         23&   23&  &
                                   &      &           &     &  \\          
\hline   
               
\end{tabular}
}
\end{minipage}
\end{table*}
\end{savenotes}

\subsection{The detected species}
\label{section_species}

The detected species summarized in Table \ref{table:spatial_extent} can be related to various processes including shocks, UV-irradiation by the embedded OB stars, and to hot core chemistry that includes complex organic molecules that are released from the grains into the gas-phase at temperatures of 100-300 K and are shielded from the dissociative UV-radiation.  

\subsubsection{Shock tracers}
\label{shock_tracers}

\begin{figure*}[ht]
\centering

\includegraphics[width=15cm,trim=0cm 0cm 0cm 0cm,clip=true]{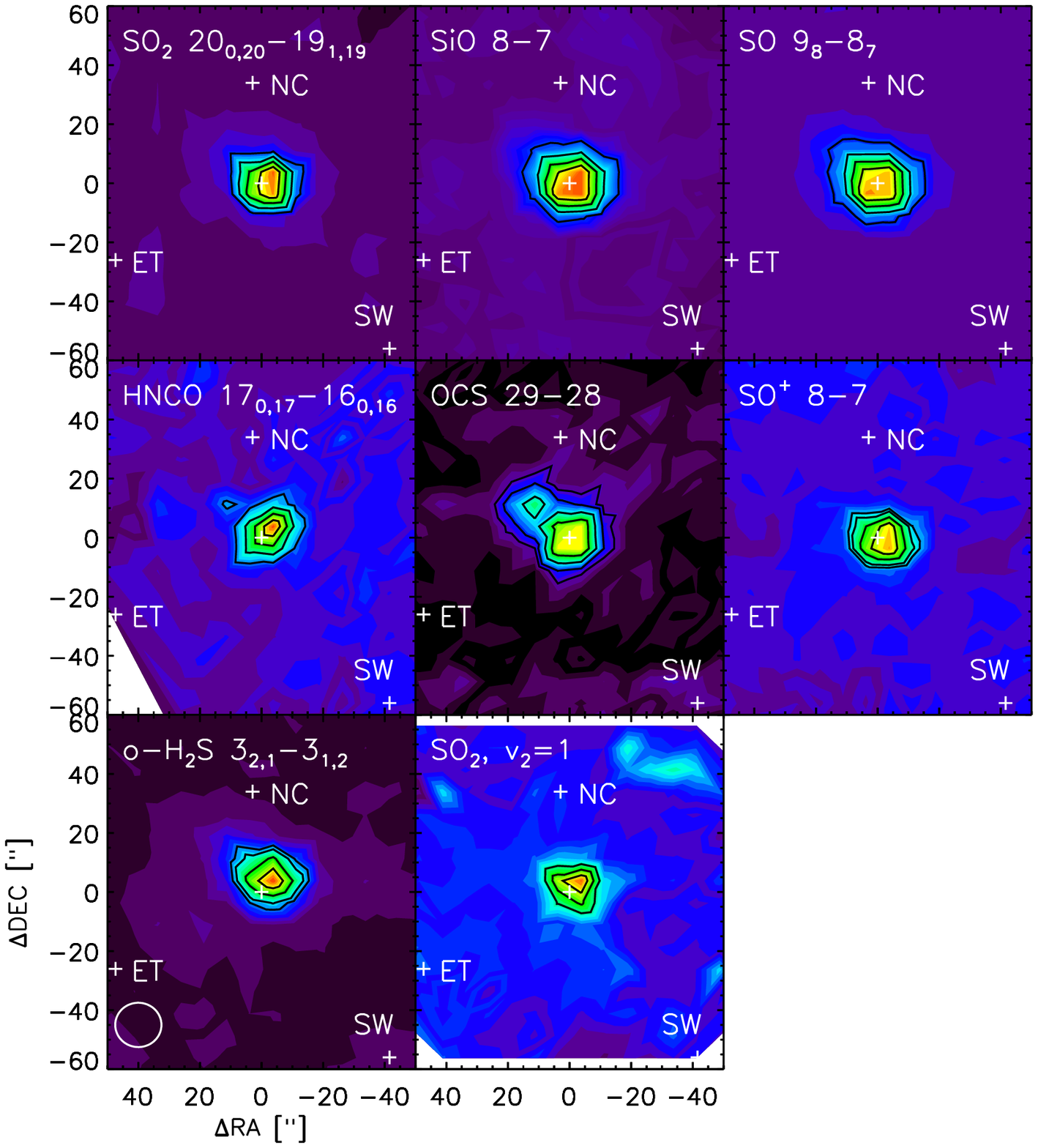}
\caption{The spatial distribution of species that can be related to shock chemistry. The crosses correspond to the subregions discussed in this paper: source center, eastern tail, northern clump, and south-west clump. The contours correspond to the 20\%, 40\%, 60\%, and 80\% of the maximum line intensity. The SO$_2$, v$_2$=1 integrated intensity was calculated as the total integrated intensity of the three blended SO$_2$, v$_2$=1 lines around 366.1 GHz.}
\label{w49_maps_shock}

\includegraphics[width=15cm,trim=0.5cm 0cm 0cm 0cm,clip=true]{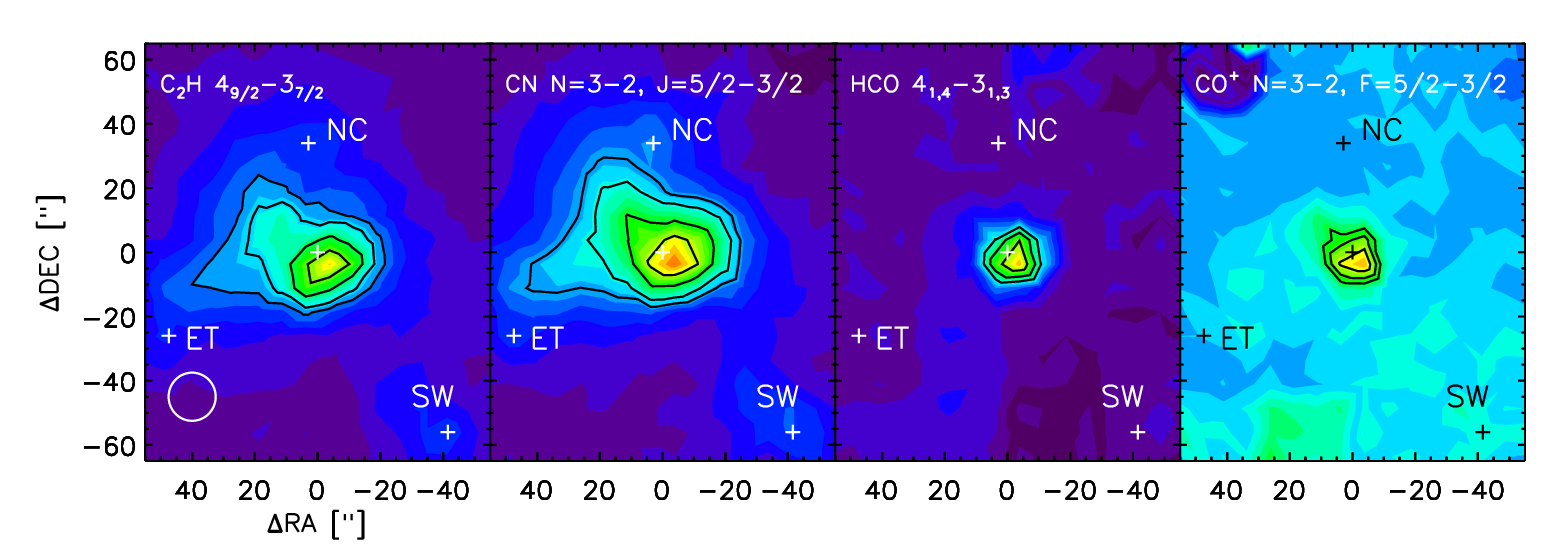}
\caption{The spatial distribution of species that can be related to PDR chemistry.
The crosses correspond to the subregions discussed in this paper: source center, eastern tail, northern clump, and south-west clump. For C$_2$H and CN the contours correspond to the 30\%, 40\%, 60\%, and 80\% of the maximum line intensity. For CO$^+$ and HCO the contours correspond to the 45\%, 60\%, and 80\% of the maximum line intensity. The different percentages used for the contour levels for the CO$^+$ and HCO integrated intensities is due to the lower signal-to-noise of these species compared to C$_2$H and CN.}
\label{w49_maps_2}
\end{figure*}

A large number of the detected molecules can be related to shock chemistry (Fig. \ref{w49_maps_shock}). All of these molecules are observed toward a $\lesssim20'' \times 20''$ field around the center, but not detected toward the other subregions analysed in this paper.
\begin{itemize}

\item SiO was detected in hot and shocked regions, such as molecular outflows (e.g. \citealp{Nisini2007}) and supernova remnants (e.g. \citealp{{Ziurys1989}}). The observations suggest that silicon is released from grain mantles in these regions. A process that can release silicon into the gas phase is the sputtering of (charged) grains by heavy neutral particles in C-shocks (\citealp{Schilke1997}, \citealp{Gusdorf2008}). 
SiO is detected toward W49A with a spatial extent of $\sim$17$''\times17''$ (corrected for the 15$''$ beam size of JCMT) covering the positions corresponding to the central stellar cluster in its $J=8-7$ transition. Its $J=2-1$ transition has previously been detected toward W49A by \citet{lucasliszt2000}.

\item Most of the detected species are sulphur-bearing molecules, such as SO and SO$_2$ and their isotopologues $^{34}$SO$_2$, $^{33}$SO$_2$, SO$^{17}$O, SO$^{18}$O, $^{34}$SO, $^{33}$SO, S$^{17}$O, and S$^{18}$O.
Sulphur is frozen onto grain mantles and can be sputtered in shocks, leading to the formation of the species mentioned above (e.g. \citealp{wakelam2004}).
We also detected 8 transitions of vibrationally excited SO$_2$. These molecules are only detected toward the center of W49A and spatially confined to a region of $\sim17'' \times 17''$, similar to SiO.

\item Other detected sulphur-bearing species include H$_2$S, OCS (and its isotopologue O$^{13}$C$^{34}$S). H$_2$S and OCS show spatial extents similar to that of SO$_2$, SO, and their isotopologues; and to SiO.

\item We also detected SO$^+$, which has been reported to be a diagnostic of dissociative shock chemistry \citep{turner1992}, but has also been detected toward various PDRs (see Sect. \ref{pdr_tracers}).

\item We detected 6 transitions of HNCO, which has also been reported to trace shocks toward the Galactic Center and in molecular outflows, such as L1157 \citep{rodriguezfernandez2010}.

\end{itemize}

\clearpage

\subsubsection{Tracers of UV-irradiation}
\label{pdr_tracers}

The embedded clusters of O and B stars in the W49 region (e.g. \citealp{alveshomeier2003}) are expected to create PDRs. Several species that can be related to PDR chemistry are detected in the SLS frequency range (Fig. \ref{w49_maps_2}), some of them with a significant spatial extent.

\begin{itemize}

\item We detected CN with a large extent ($\sim$54$''\times39''$), including all subregions. This suggests the possible importance of PDR chemistry in W49A and may be tested using the ratio of CN/HCN, which is a well-known tracer of PDRs (e.g. \citealp{fuente1996}). The spatial distribution of HCN and its isotopomer HNC based on SLS data were analysed in \citet{roberts2011}.

\item C$_2$H (ethynyl) is a commonly observed molecule in the interstellar medium, including in PDRs. In star-forming regions it was proposed to be related to the earliest stages of massive star-formation (Beuther et al., 2008). We detected three transitions toward W49A, with a very large spatial extent ($\sim$61$''\times46''$).

\item HCO (formyl radical) is also detected in a $\sim$17$''\times17''$ region around the center. HCO was found to be a tracer of illuminated cloud interfaces as it was detected in several PDRs including the Orion Bar, NGC 2023, NGC 7023, and S140 \citep{schilke2001} and the Horsehead PDR \citep{gerin2009}.

\item The SO$^+$ and CO$^+$ reactive ions have been detected in various PDRs, such as the Orion Bar \citep{fuente2003}, and in other regions of high FUV- and X-ray irradiation. We detected both CO$^+$ and SO$^+$ toward the center of W49A with a spatial extent of $\sim$17$''\times17''$.

\end{itemize}

\subsubsection{Complex organic molecules}
\label{complex_mol}

We have detected a number of complex organic molecules (e.g. \citealp{herbstvandishoeck2009}), that are most likely related to hot cores.

\begin{itemize}

\item CH$_3$OH is detected over a $30''\times20''$ region around the center.
Its spatial distribution is shown in \citet{nagy2012}. Its $^{13}$CH$_3$OH isotopologue is also detected.

\item We detected 5 transitions of CH$_3$CN (methyl cyanide), that has been detected in hot cores, including Orion KL. The $J=12-11$ transition of CH$_3$CN was previously detected toward W49 using the BIMA array and is likely connected to hot cores and also follows the compact dust concentrations \citep{wilner2001}. 

\item We detected four transitions of HC$_3$N (cyanoacetylene), that has been detected toward various Galactic star forming regions as well as in a number of dust-rich galaxies \citep{lindberg2011}. As it is destroyed by UV radiation and in reactions with the C$^+$ ion, HC$_3$N is expected to trace warm, dense, and shielded regions. This may explain the fact that HC$_3$N is seen to be confined toward the central $\sim$17$''\times17''$ region. 
Some of the vibrational transitions of HC$_3$N ($v_7$=1) were also detected toward the W49A center. Vibrationally excited HC$_3$N has also been detected toward several hot cores \citep{wyrowski1999}, as well as  the galaxy NGC4418 \citep{costagliolaaalto2010}.  

\item One transition of CH$_3$CHO (acetaldehyde) are detected toward the W49A center. A previous survey by \citet{ikeda2001} detected this molecule toward various hot molecular clouds.

\item Four transitions of CH$_3$CCH (propyne) are detected in our line survey at a few central pixels with a spatial extent of $\sim$17$''\times17''$. This molecule has been detected in a number of starburst galaxies, such as M82 \citep{aladro2011a}. 

\end{itemize}

\begin{figure}[ht]
\centering
\includegraphics[width=9.3cm,trim=0.3cm 0cm 0cm 0cm,clip=true]{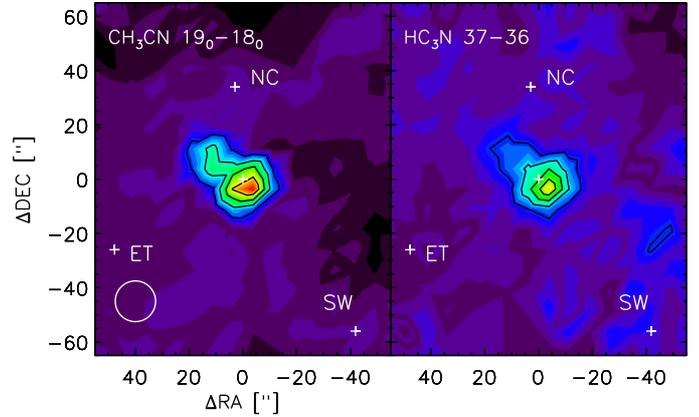}
\caption{The spatial distribution of the detected most spatially extended complex organic molecules.
The crosses correspond to the subregions discussed in this paper: source center, eastern tail, northern clump, and south-west clump. The contours correspond to the 20\%, 40\%, 60\%, and 80\% of the maximum line intensity.
}
\label{w49_maps_3}
\end{figure}

\subsubsection{Other detected species}
\label{other_species}

\begin{figure*}[ht]
\centering

\includegraphics[width=13.5cm,trim=0.3cm 0cm 0cm 0cm,clip=true]{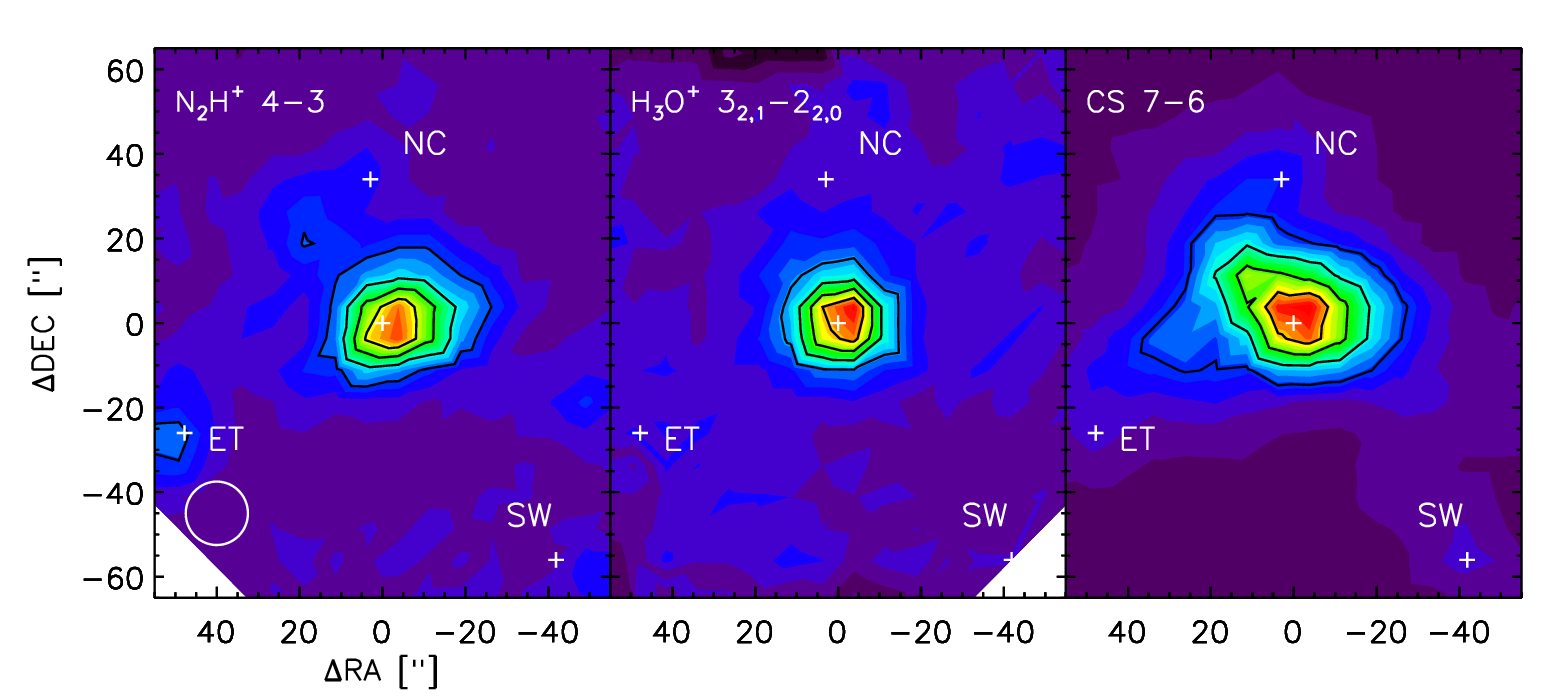}
\caption{The spatial distribution of N$_2$H$^+$ and H$_3$O$^+$ ions and CS. 
The crosses correspond to the subregions discussed in this paper: source center, eastern tail (ET), northern clump (NC), and south-west clump (SW). The contours correspond to the 20\%, 40\%, 60\%, and 80\% of the maximum line intensity.
}

\label{w49_maps_4}

\includegraphics[width=13.5cm,trim=0.3cm 0cm 0cm 0cm,clip=true]{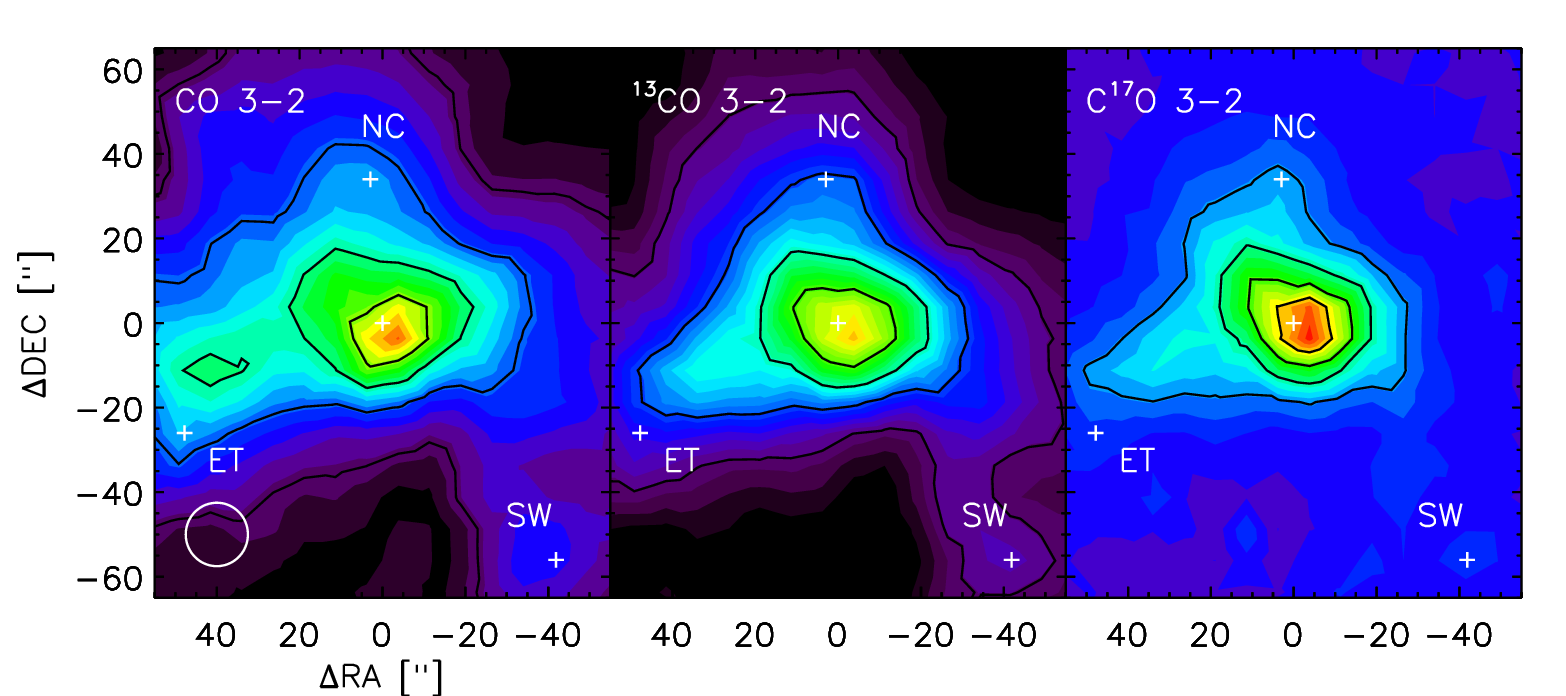}
\caption{The spatial distribution of CO, $^{13}$CO, and C$^{17}$O line emission. 
The crosses correspond to the subregions discussed in this paper: source center, eastern tail (ET), northern clump (NC), and south-west clump (SW). The contours correspond to the 20\%, 40\%, 60\%, and 80\% of the maximum line intensity.
}
\label{co_distribution}
\end{figure*}

\begin{itemize}

\item CO, $^{13}$CO, and C$^{17}$O are covered by the SLS survey, and are detected at each position of the SLS field (Figure \ref{co_distribution}), allowing us to probe the gas kinematics and column density.

\item The only deuterated species we detected toward W49A is DCN \citep{roberts2011}. 
An upper limit for the covered but not detected DCO$^+$ $J$=5-4 transition was also reported by \citet{roberts2011}. A low abundance of deuterated ions is expected as deuterated molecular ions are destroyed rapidly at gas temperatures above 30 K (e.g. \citealp{robertsmillar2007}).

\item H$_2$CO (formaldehyde) is a commonly used temperature and density tracer in Galactic star-forming regions (e.g. \citealp{mangumwootten1993}) and in external galaxies (e.g. \citealp{mangum2008}) as well. Previous results from this line survey focused on this molecule, which revealed a large $\sim$3$\times3$ pc region around the W49A center with kinetic temperatures of $>$100 K and densities of $>$10$^5$ cm$^{-3}$ \citep{nagy2012}.  
Its isotopologue H$_2^{13}$CO is also detected toward the center.

\item Similar to H$_2$CO, H$_2$CS (thio-formaldehyde) is a near-prolate rotor, which allows estimation of both kinetic temperature and density. Though its spatial extent ($30''\times17''$) given by a 2D Gaussian fit toward the center is similar to that of H$_2$CO, H$_2$CS was not detected toward the other main regions (Eastern tail, Northern clump, and Southwest clump). 
Its isotopologue H$_2$C$^{34}$S is also detected.

\item We detected a few ions, including HCO$^+$ and its isotopologues HC$^{18}$O$^+$, H$^{13}$CO$^+$, and HC$^{17}$O$^+$. As a dense gas tracer, HCO$^+$ data in the SLS survey were analysed by \citet{roberts2011}. Other detected ions not mentioned in the sections above include H$_3$O$^+$ and N$_2$H$^+$. N$_2$H$^+$ shows a larger spatial extent than H$_3$O$^+$ covering the Eastern tail and Northern clump regions. H$_3$O$^+$ is an important chemical ingredient of molecular clouds as it can be related to the formation of water in the gas phase and is formed via OH$^+$ and H$_2$O$^+$ ions, that have recently been found to be important tracers of the cosmic-ray ionization rate in the diffuse interstellar medium (e.g. \citealp{2010A&A...521L..10N}, \citealp{2010A&A...518L.110G}, \citealp{hollenbach2012}). Strong emission lines of OH$^+$, H$_2$O$^+$, and H$_3$O$^+$ ions are also found in AGNs (e.g. \citealp{2010A&A...518L..42V}).
Detections of H$_3$O$^+$ transitions toward starburst galaxies and AGNs have also been reported with the JCMT (e.g. \citealp{aalto2011} and \citealp{vandertak2008}).

\item CS is a well-known tracer of dense gas that has been analysed both in Galactic star-forming regions (e.g. \citealp{helmichvandishoeck1997}) and in external galaxies, including starburst galaxies (e.g. \citealp{aladro2011b}). Multiple rotational transitions of CS have previously been studied in W49A by \citet{serabyn1993}, leading to the idea that the large star-formation activity in W49A has been triggered by a cloud-cloud collision.
CS $J$=7-6 has been detected toward W49A with a spatial extent of $\sim40''\times24''$. Its isotopologues $^{13}$CS and C$^{34}$S have also been detected in the SLS survey.

\item We detected two hydrogen recombination lines: H26$\alpha$ at 353.6 GHz and H33$\beta$ at 335.2 GHz. Their measured widths of $\sim$37 \kms and $\sim$31 \kms are larger than the widths found for molecular lines detected toward the W49A center in the line survey and are similar to those of other H recombination lines detected toward W49A (e.g. \citealp{galvanmadrid2013}). The velocity of the H26$\alpha$ line is consistent with that of the H41$\alpha$ line detected by \citet{galvanmadrid2013}.
Other hydrogen recombination lines were previously detected in W49A by \citet{gordonwalmsley1990} and \citet{galvanmadrid2013}.

\end{itemize}

\subsection{Line profiles and velocity structure}

\begin{figure*}[ht]
\centering
\includegraphics[width=16cm,trim=0.0cm 0cm 0cm 0cm,clip=true]{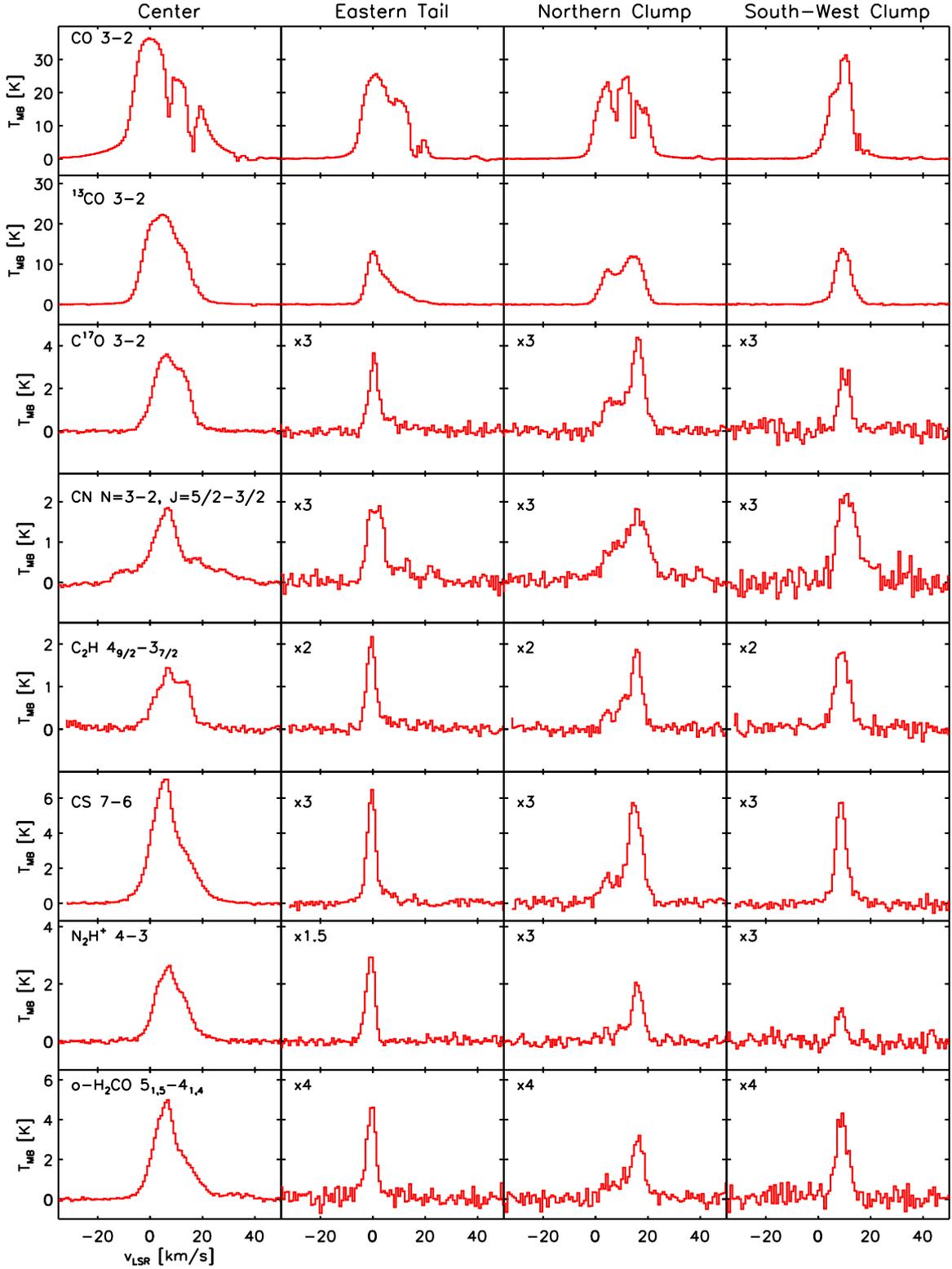}
\caption{Line profiles of molecules observed toward the various subregions of W49A.}
\label{line_profiles_w49a}
\end{figure*}

Figure \ref{line_profiles_w49a} shows typical line profiles observed toward the various subregions of W49A for the most spatially extended molecular lines, with the exception of HCN, HNC, and HCO$^+$ that were analysed in \citet{roberts2011}. Most of the detected lines are double-peaked or asymmetric. While lines observed toward the center and the South-West clump peak around the source velocity of $\sim$7-8 km s$^{-1}$ (e.g. \citealp{roberts2011} and references therein), line emission is seen mainly blue-shifted toward the Eastern tail, and red-shifted toward the Northern clump, similar to what is seen for HNC, HCN, and HCO$^+$, and their isotopologues \citep{roberts2011}. 
Some of the asymmetric, double-peaked line profiles have a brighter blue-shifted than a red-shifted peak, which can be interpreted as gas infalling toward the W49A center.
Such line profiles that are consistent with infall signatures have been observed toward the source center in HCO$^+$ 1-0 (\citealp{welch1987}, \citealp{serabyn1993}) and in the SLS survey traced by HCN 4-3 and 3-2, HNC 4-3, and HCO$^+$ 4-3 \citep{roberts2011}. Though most observed lines show an excess emission in the blue-shifted wing of the line profile toward the source center, the line profiles which are most consistent with infall signatures are seen in HCN 4-3 and 3-2, HNC 4-3, and HCO$^+$ 4-3.
\begin{figure*}[ht]
\centering
\includegraphics[width=15cm,trim=0cm 0cm 0cm 0cm,clip=true]{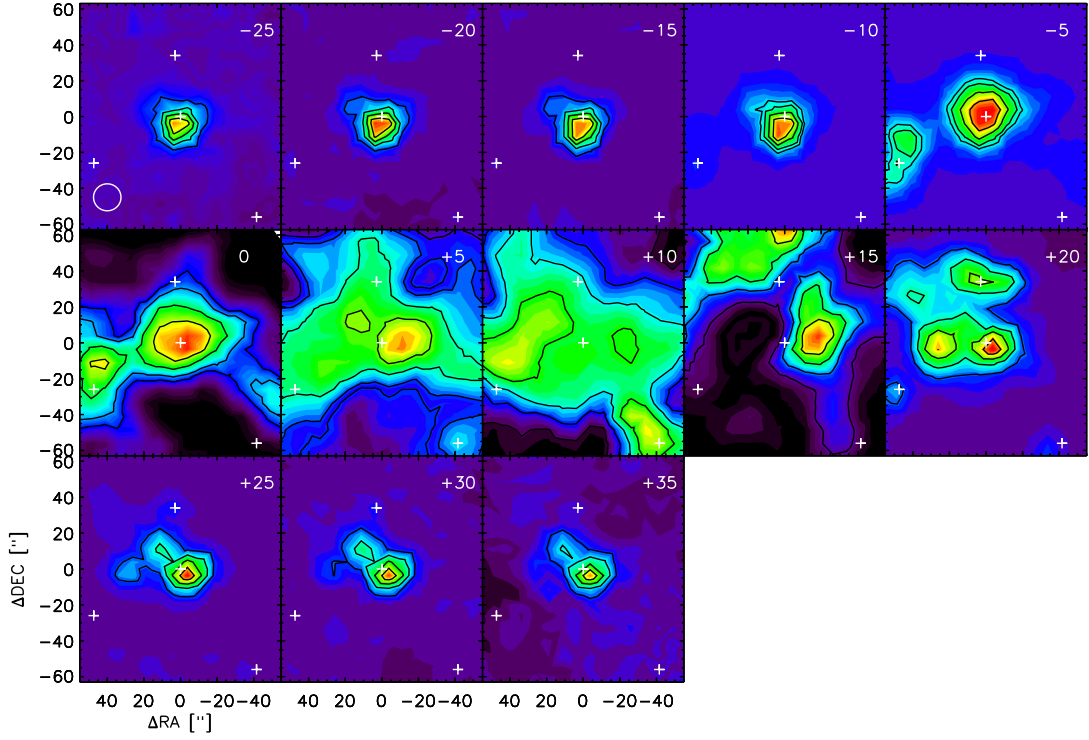}
\caption{Channel maps of CO at a resolution of $\sim$15$''$ at the given LSR velocities (in \kms).}
\label{chanmap_co}
\end{figure*}

\begin{figure*}[ht]
\centering
\includegraphics[width=15cm,trim=0.0cm 0cm 0cm 0cm,clip=true]{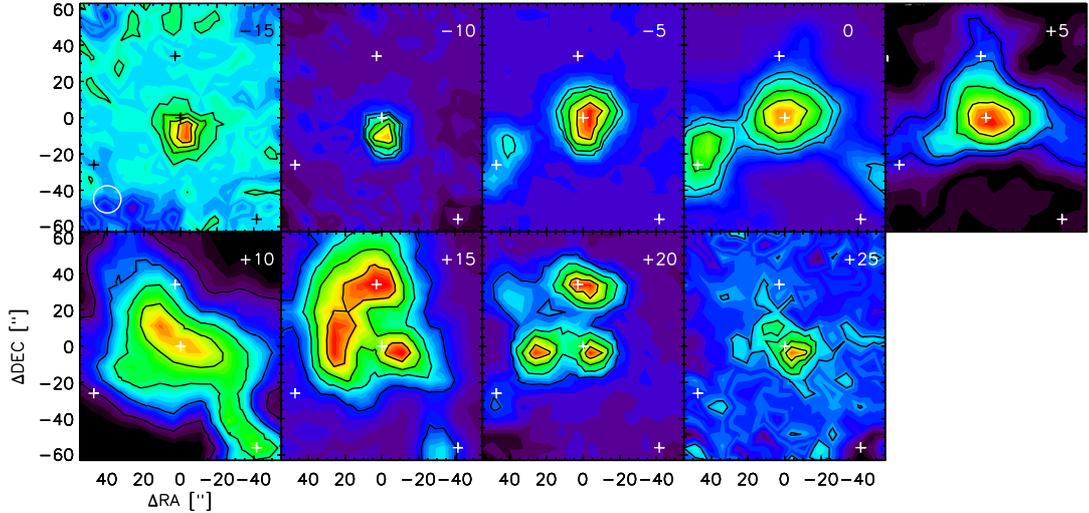}
\caption{Channel maps of $^{13}$CO at a resolution of $\sim$15$''$ at the given LSR velocities (in \kms).}
\label{chanmap_13co}
\end{figure*}
To investigate the velocity distribution of the observed line emission, we obtained channel maps for the species shown in Fig. \ref{line_profiles_w49a}. CO 3-2 shows emission in a large range of velocities between -25 and 35 km s$^{-1}$ (Fig. \ref{chanmap_co}). The most extended line emission is seen around the expected source velocity, in the range between +5 and +10 km s$^{-1}$. Toward the Eastern tail, significant blue-shifted line emission is seen at velocities between -5 and 0 km s$^{-1}$. The Northern clump is more prominent at velocities of $\sim$20 km s$^{-1}$, but also shows significant emission around the source velocity. The South-west clump is only seen around the expected source velocity (between 5 and 10 km s$^{-1}$). Significant line emission is seen at a velocity of +15 km s$^{-1}$ that does not correspond to any of the analysed subregions, including the center.
The $^{13}$CO 3-2 line shows a very similar structure (Fig \ref{chanmap_13co}) to that traced by CO 3-2 with the main difference at a velocity of +15 km s$^{-1}$, likely related to a self-absorption of the optically thick CO 3-2 lines around the center.

\begin{figure*}[h!]
\centering
\includegraphics[width=\textwidth ,trim=0.0cm 0cm 0cm 0cm,clip=true]{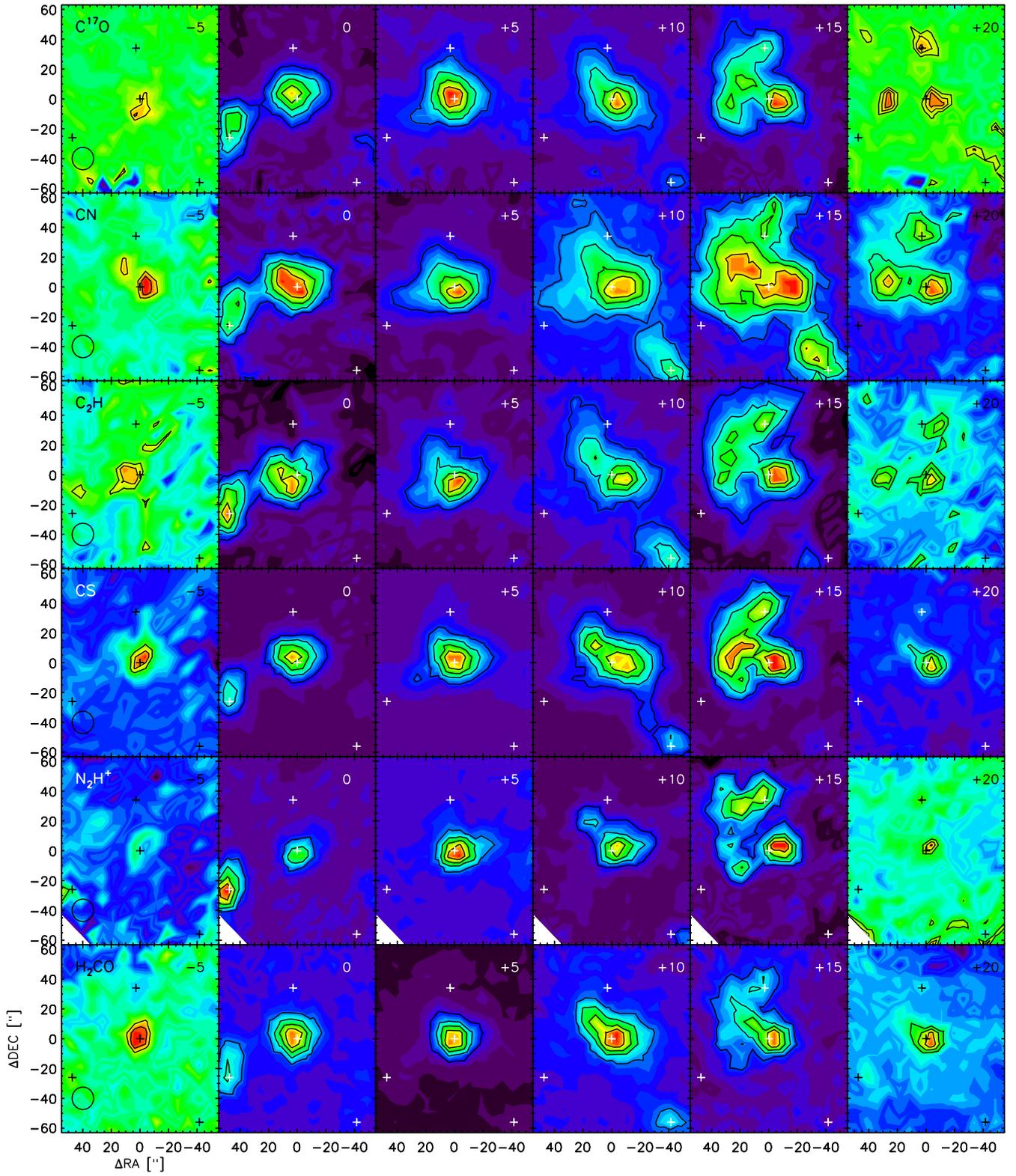}
\caption{Channel maps of the most spatially extended species in the SLS at the given LSR velocities (in \kms).}
\label{chanmaps_w49}
\end{figure*}

Lines that show spatially extended emission other than CO and $^{13}$CO are distributed over the velocity range between -5 and +20 km s$^{-1}$ (Fig. \ref{chanmaps_w49}). The strongest emission is seen in CN, especially the redshifted wings of the CN lines. While other molecules show emission toward the center and the Northern clump at a velocity of +15 km s$^{-1}$, CN emission at this velocity is also detected toward the South-west clump.

\clearpage

\subsection{Excitation and column densities}
\label{excitation}

Table \ref{lines_summary1} includes a number of species with at least three detected transitions covering a sufficient energy range to probe excitation conditions. 
We estimate the excitation conditions of these molecules using the population diagram method introduced by \citet{goldsmithlanger1999}. Column densities ($N_{\rm{tot,1}}$) and excitation temperatures ($T_{\rm{ex}}$) can be estimated based on
\begin{equation}
\label{PD}
\ln{\left( \frac{N_u}{g_u} \right)}=\ln{ \left( \frac{N_{\rm{tot,1}}}{Q_{\rm{rot}}} \right)}-\frac{E_u}{k T_{\rm{ex}}}+\ln{\left(\frac{\Omega_s}{\Omega_a}\right)}-\ln{(C_\tau)},
\end{equation}
where $N_u$ is the observed upper state column density of the molecule including line opacity and beam-source coupling effects; $g_u$ is the degeneracy of the upper state; $Q_{\rm{rot}}$ is the rotational partition function; $k$ is the Boltzmann constant; $E_u$ is the upper level energy; $f=\Omega_s/\Omega_a$ is the source filling factor with $\Omega_a\sim15''$ the beam size and $\Omega_s$ the source size; and $C_\tau=\tau/(1-e^{-\tau})$ where $\tau$ is the optical depth. 
For a uniform beam filling ($\Omega_a\sim\Omega_s$) and low optical depth, Eqn. \ref{PD} reduces to a rotational diagram with a rotational temperature $T_{\rm{rot}}$ and total column density $N_{\rm{tot}}$.
For a rotational diagram Equation \ref{PD} is simplified to only the first two terms on the right-hand side of the equation. The terms containing the correction for the opacity and the beam filling factor are not included.

We evaluate Eqn. \ref{PD} for a set of $N_{\rm{tot,1}}$, $T_{\rm{ex}}$, $f$ and $C_{\tau}$ ($N_{\rm{tot,1}}$, $T_{\rm{ex}}$). We apply $\Omega_s$ in the range between $1''$ and $15''$ (uniform beam filling) and a column density in the range between $10^{12}$ cm$^{-2}$ and $10^{18}$ cm$^{-2}$.
For molecules with just three detected transitions we use $T_{\rm{rot}}=T_{\rm{ex}}$ to derive a best fit source size and column density.

\begin{figure*}[ht!]
\centering

\includegraphics[width=3.3 cm,angle=-90,trim=0cm 0cm 0cm 0cm,clip=true]{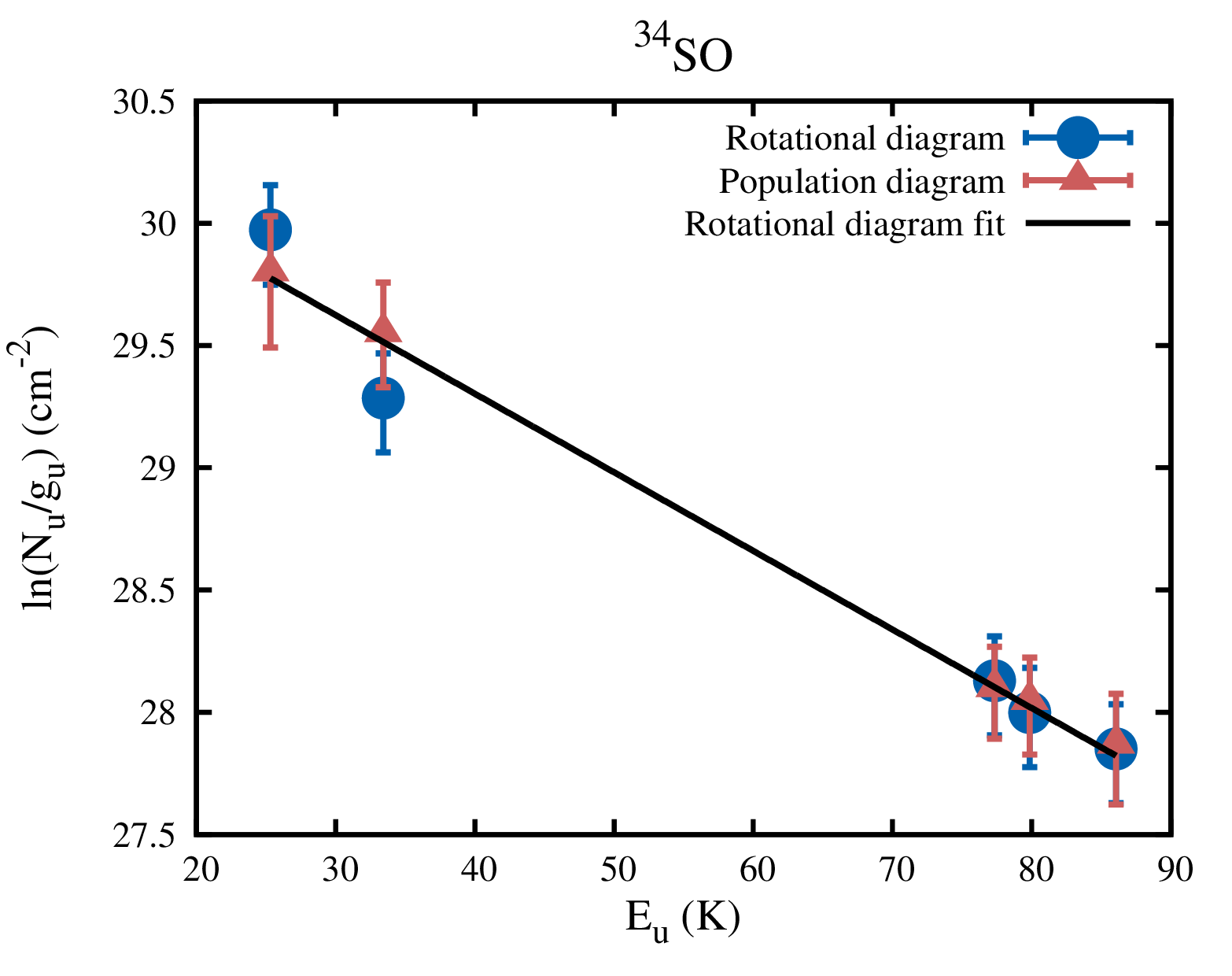}
\includegraphics[width=3.3 cm,angle=-90,trim=0cm 0cm 0cm 0cm,clip=true]{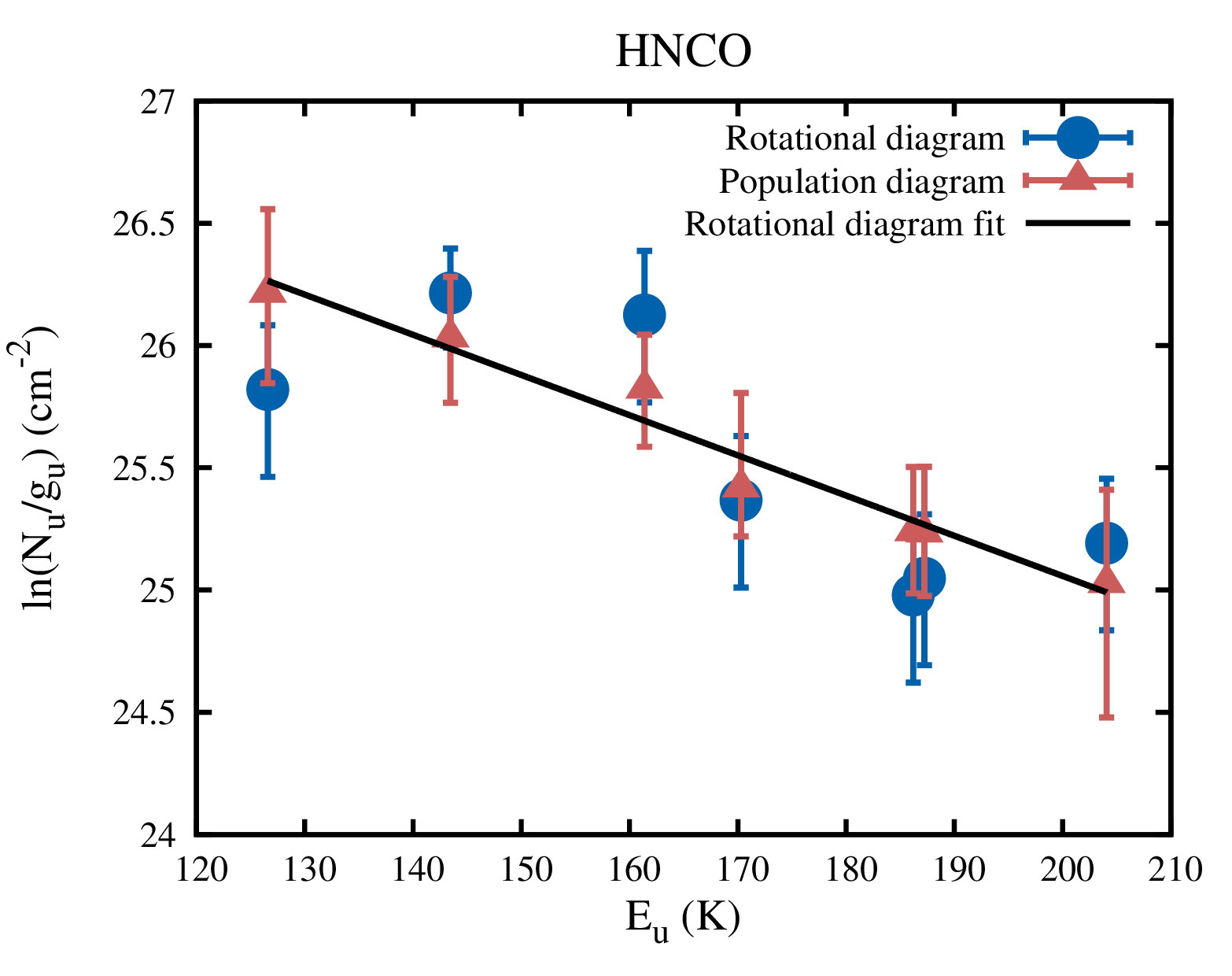}
\includegraphics[width=3.3 cm,angle=-90,trim=0cm 0cm 0cm 0cm,clip=true]{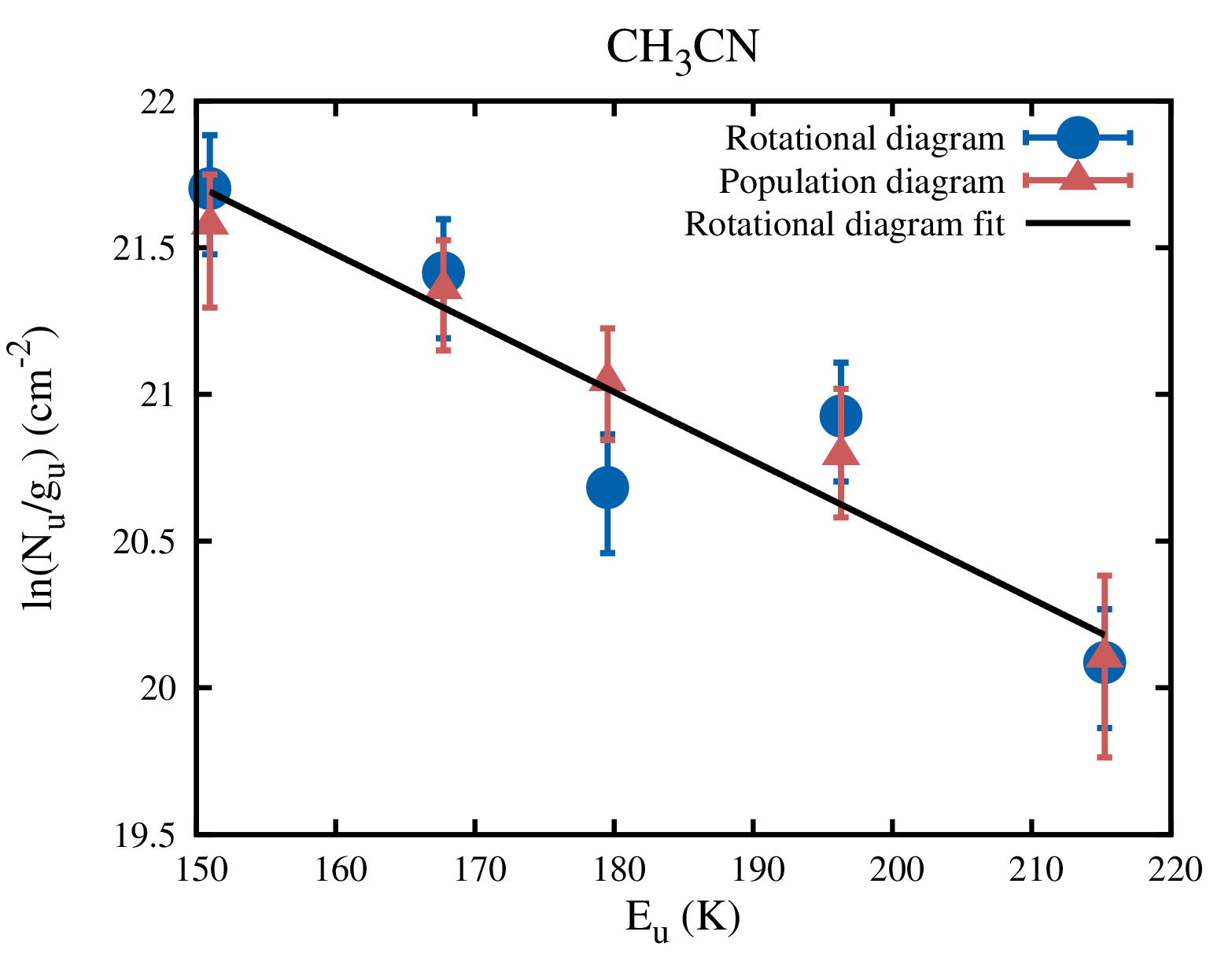}
\includegraphics[width=3.3 cm,angle=-90,trim=0cm 0cm 0cm 0cm,clip=true]{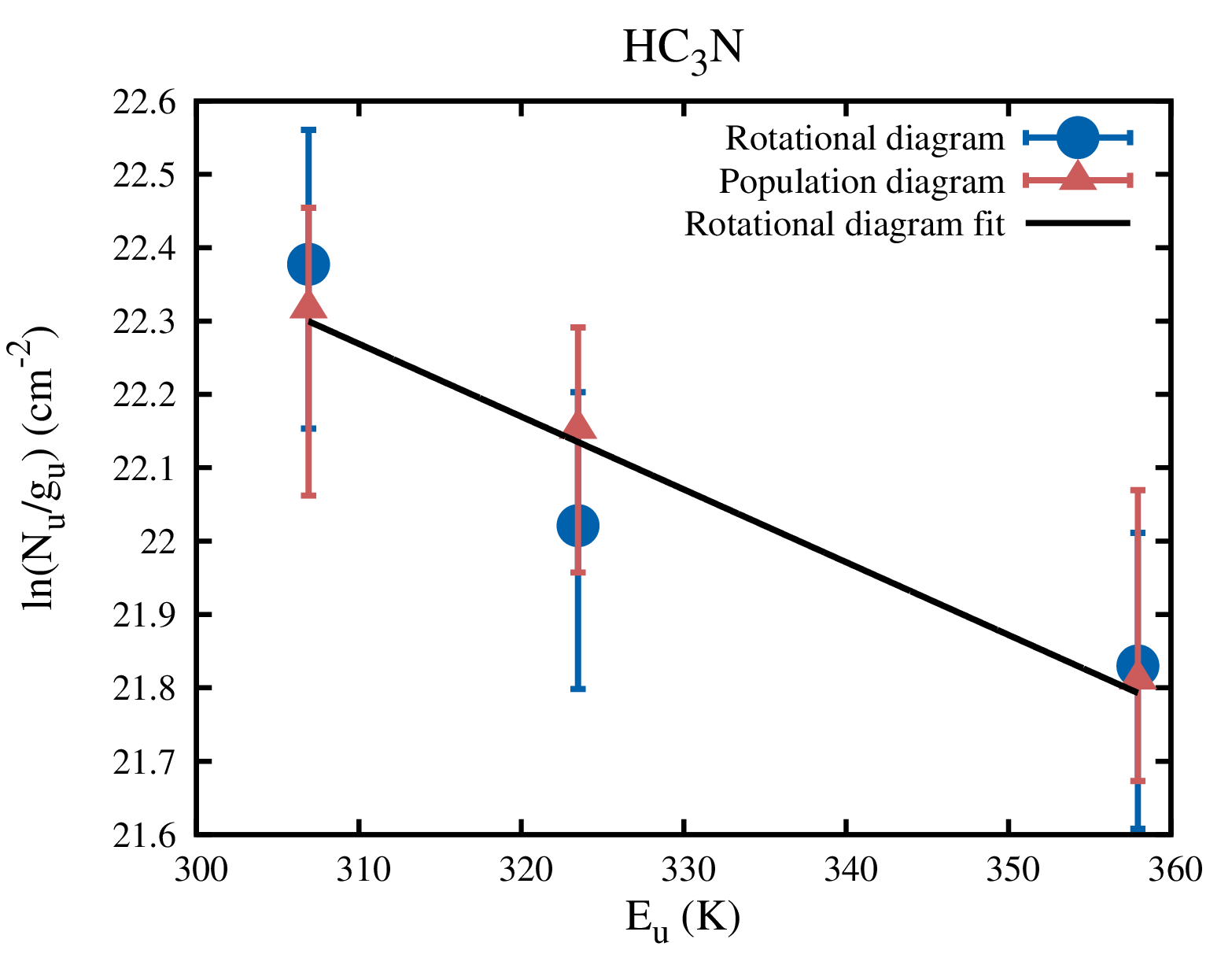}
\includegraphics[width=3.3 cm,angle=-90,trim=0cm 0cm 0cm 0cm,clip=true]{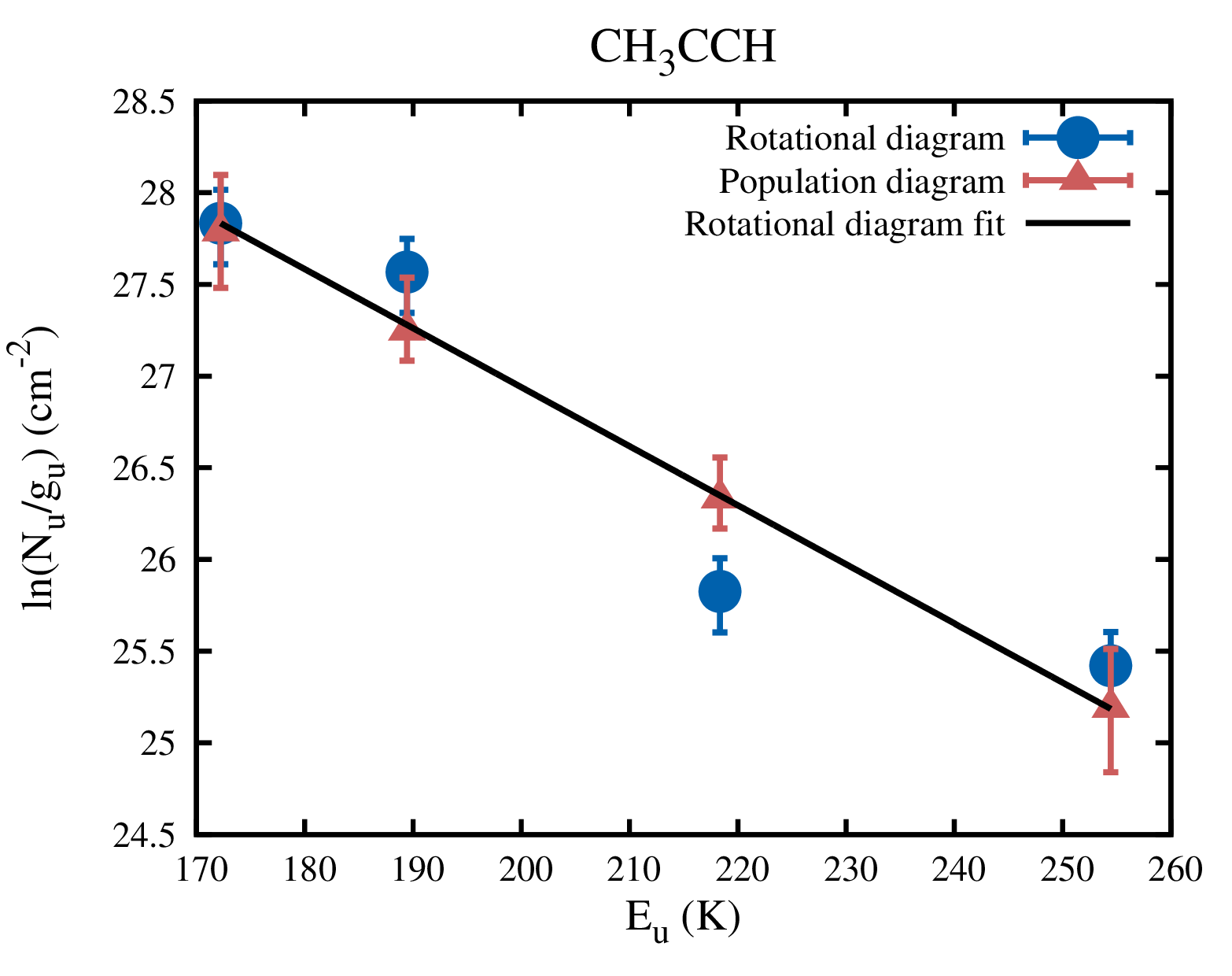}
\includegraphics[width=3.3 cm,angle=-90,trim=0cm 0cm 0cm 0cm,clip=true]{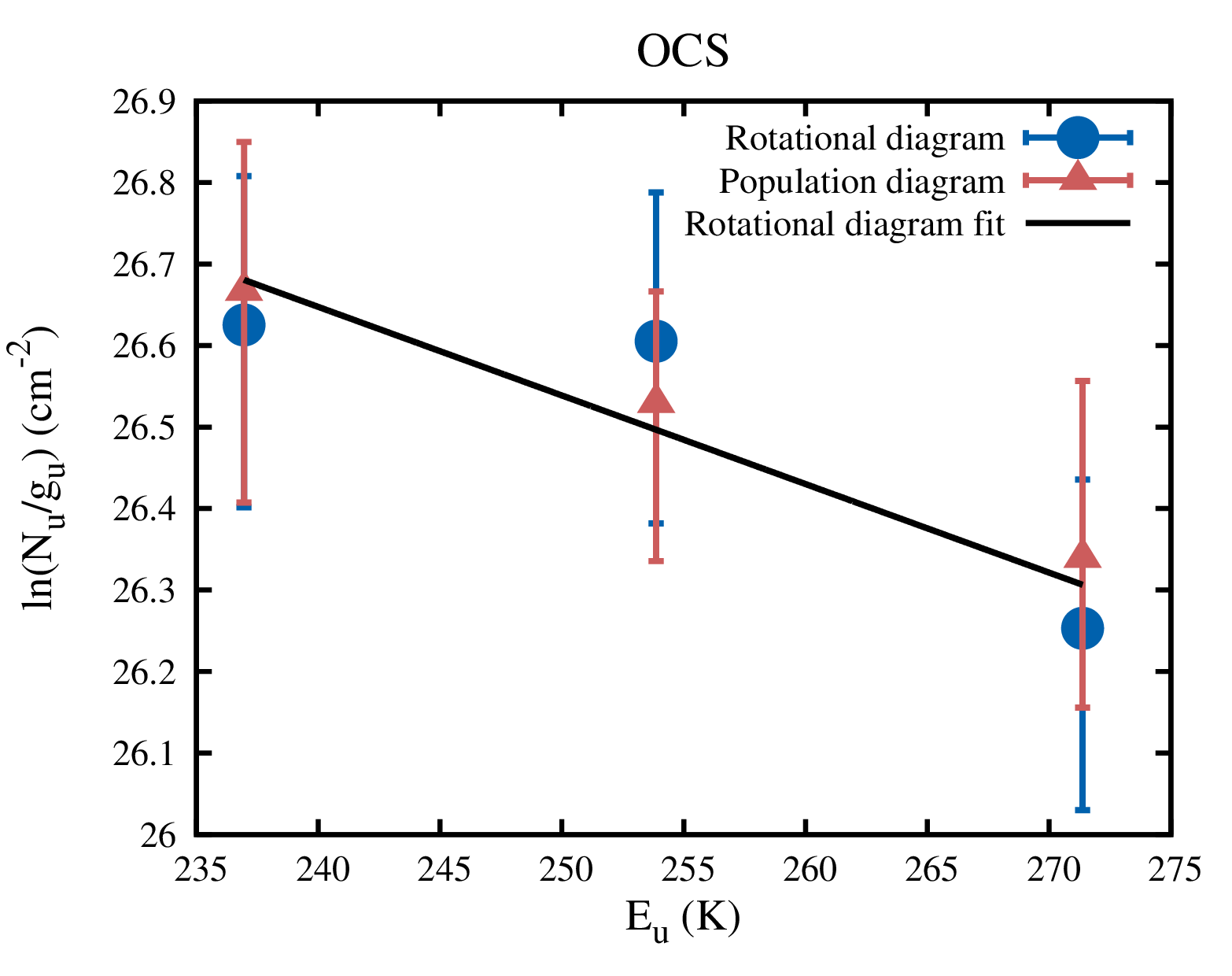}
\includegraphics[width=3.3 cm,angle=-90,trim=0cm 0cm 0cm 0cm,clip=true]{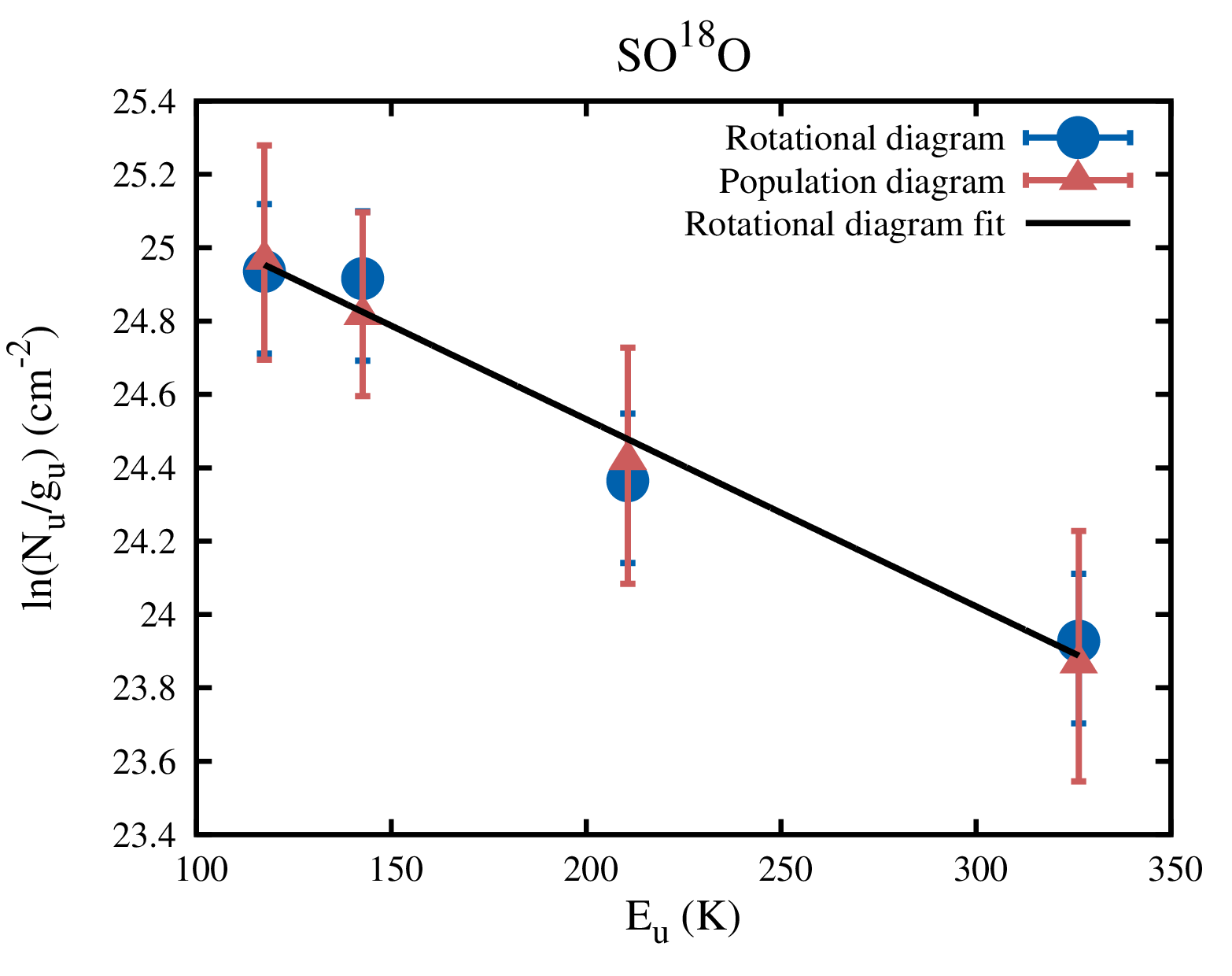}
\includegraphics[width=3.3 cm,angle=-90,trim=0cm 0cm 0cm 0cm,clip=true]{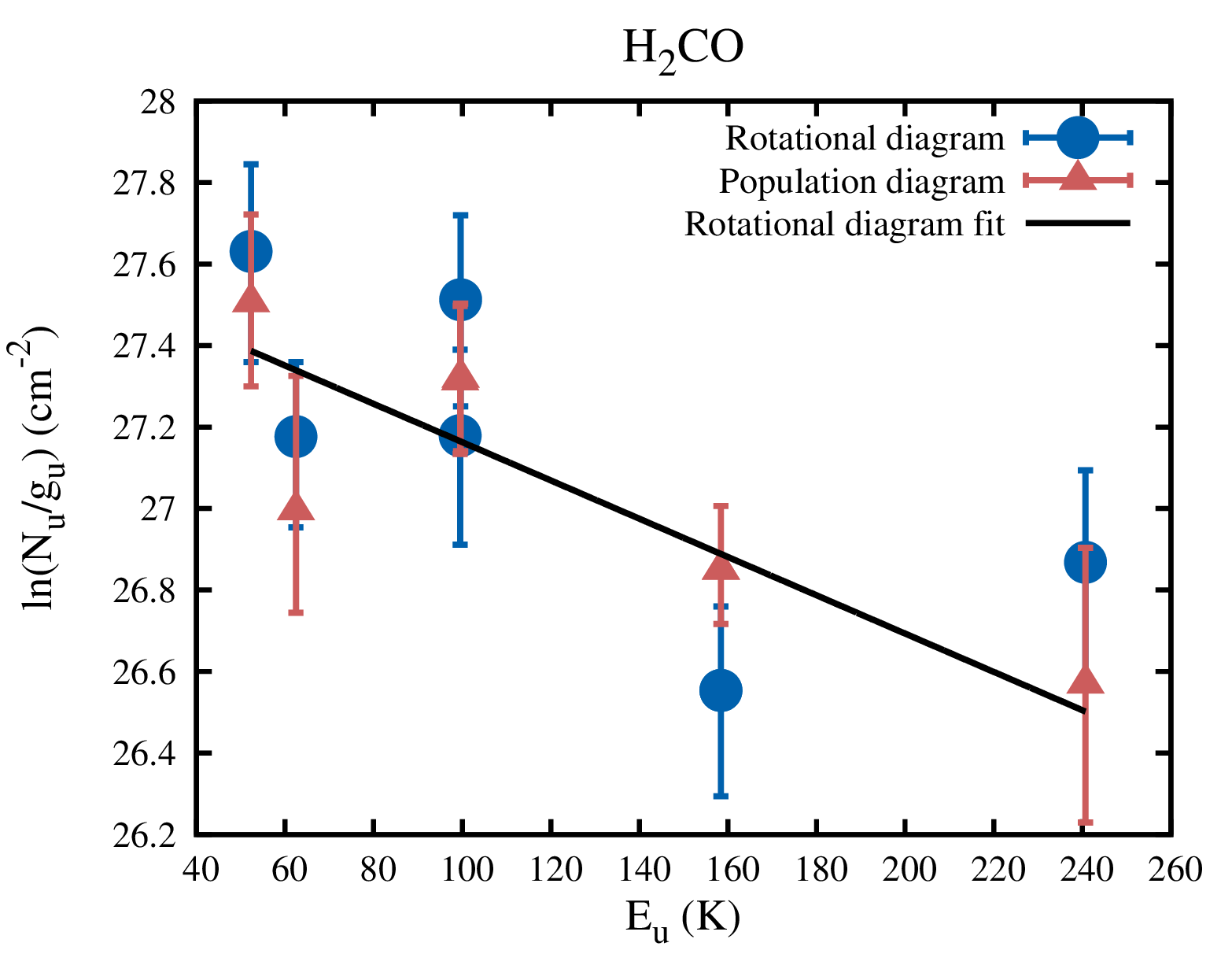}
\includegraphics[width=3.3 cm,angle=-90,trim=0cm 0cm 0cm 0cm,clip=true]{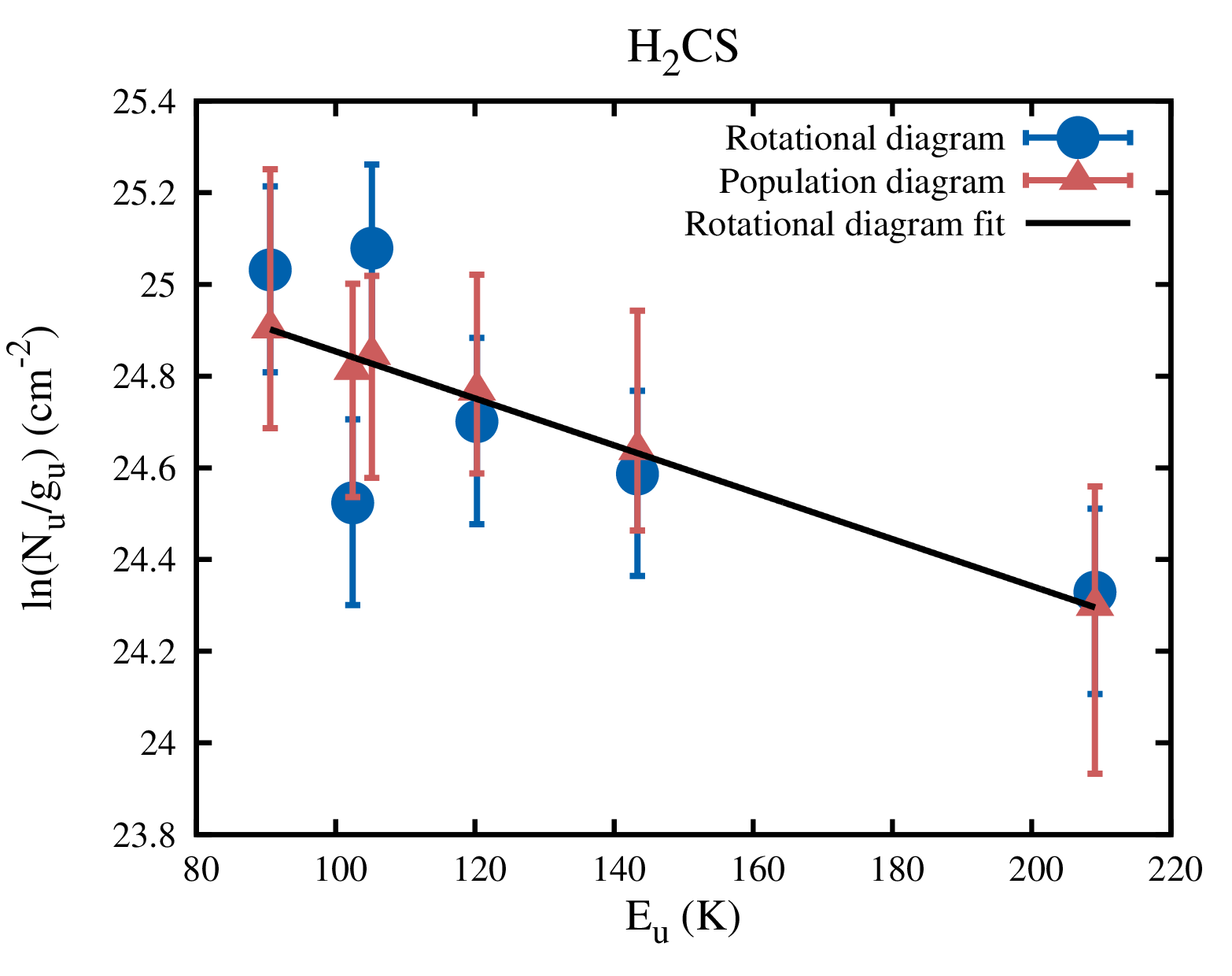}
\includegraphics[width=3.3 cm,angle=-90,trim=0cm 0cm 0cm 0cm,clip=true]{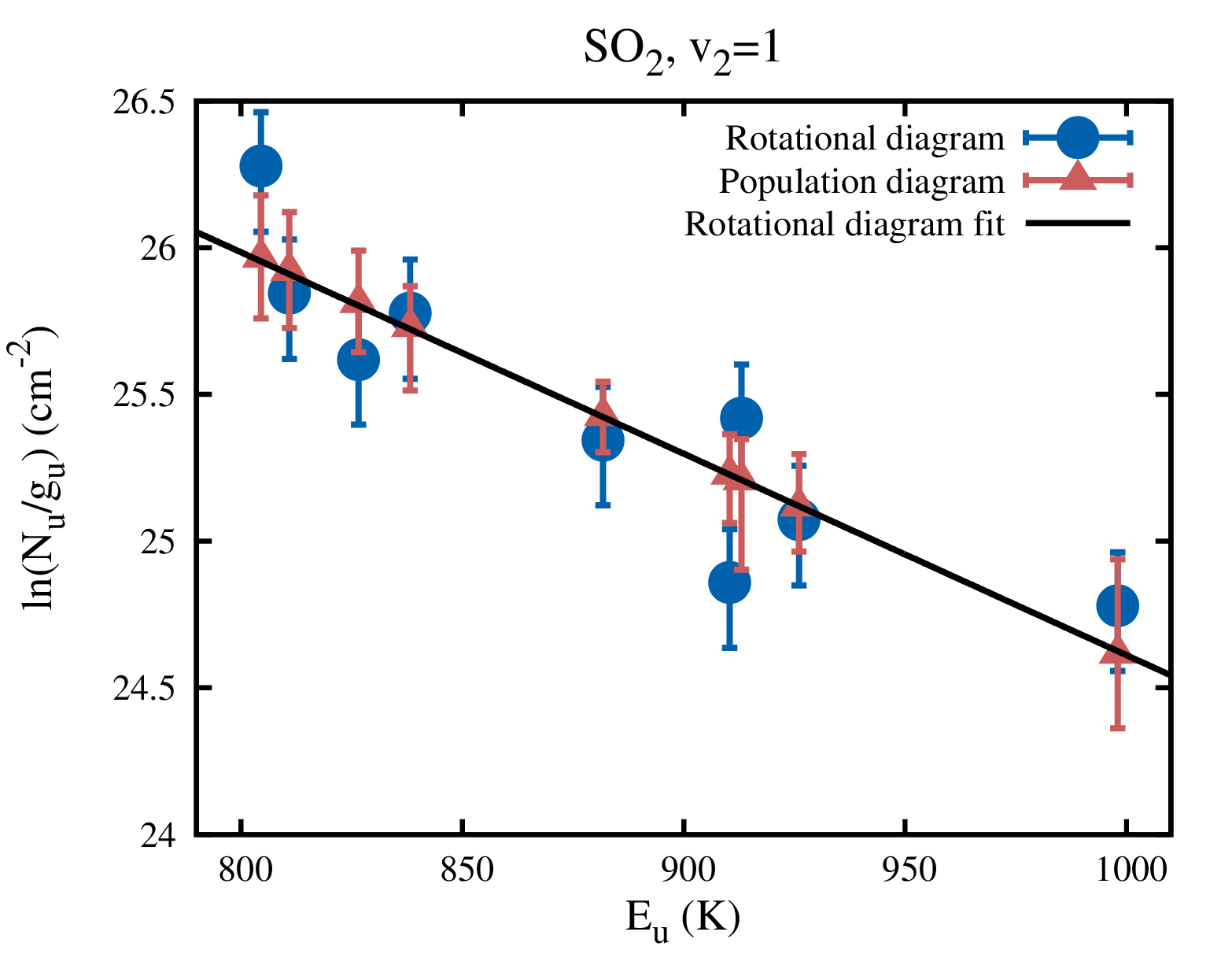}
\includegraphics[width=3.3 cm,angle=-90,trim=0cm 0cm 0cm 0cm,clip=true]{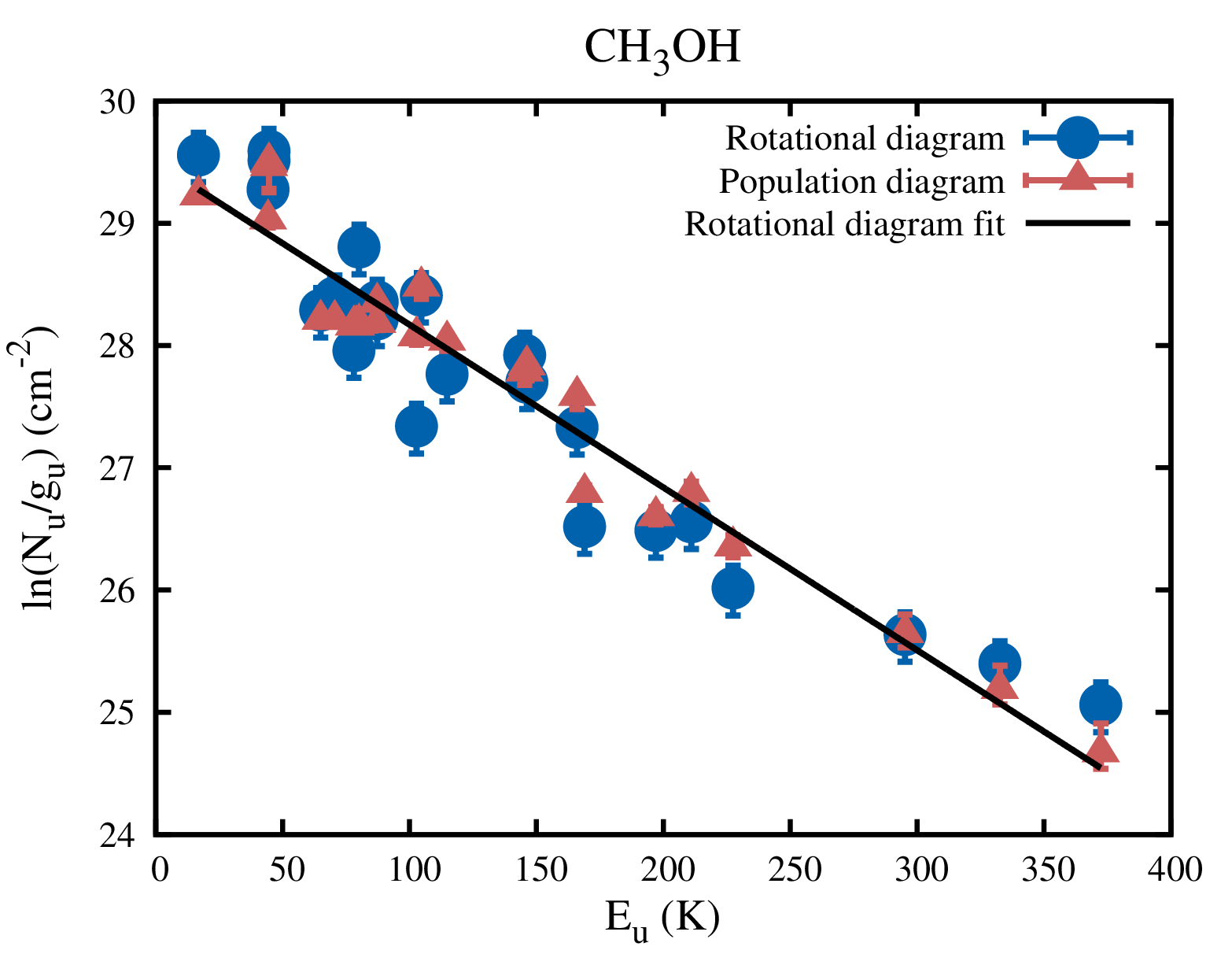}
\includegraphics[width=3.3 cm,angle=-90,trim=0cm 0cm 0cm 0cm,clip=true]{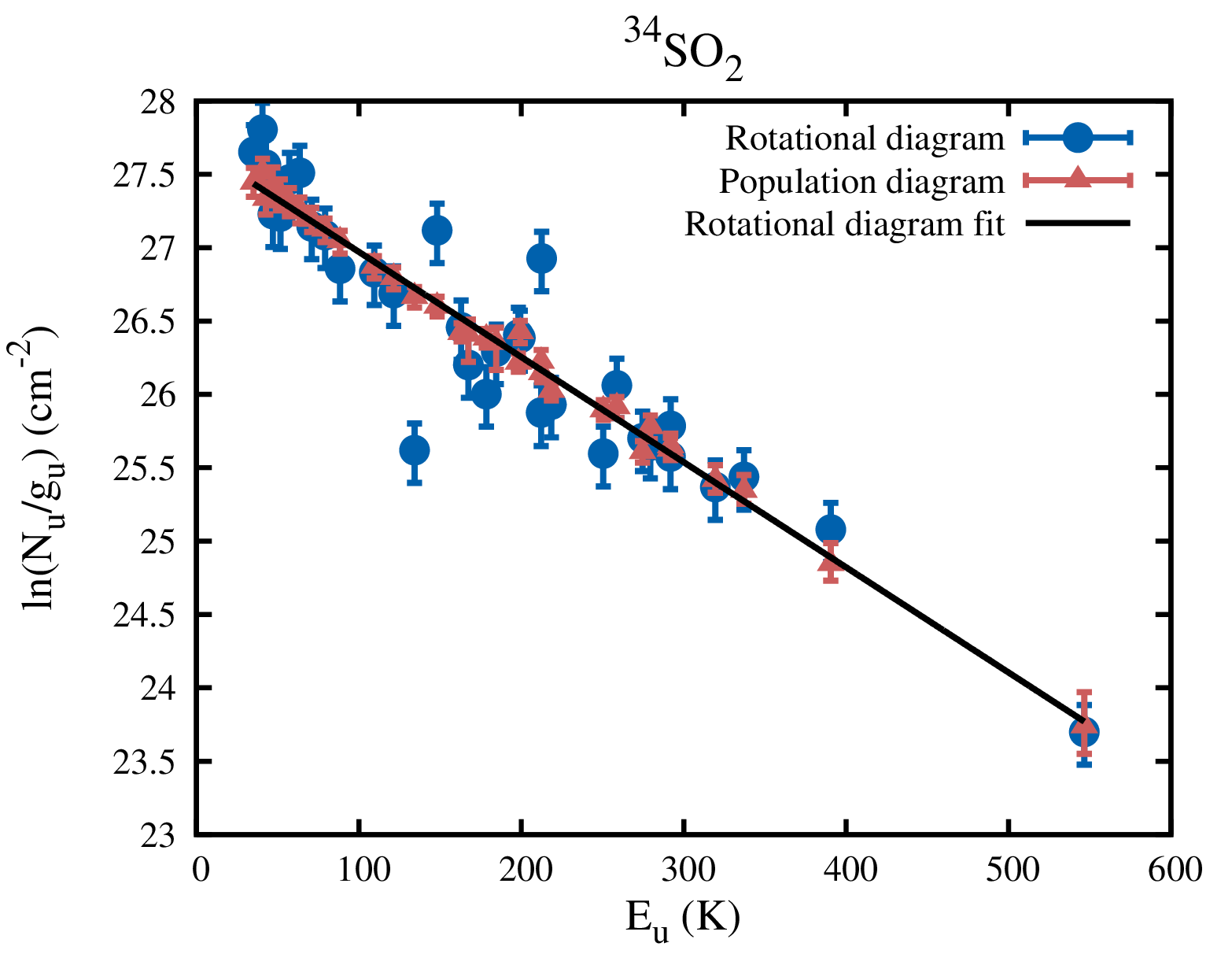}
\includegraphics[width=3.3 cm,angle=-90,trim=0cm 0cm 0cm 0cm,clip=true]{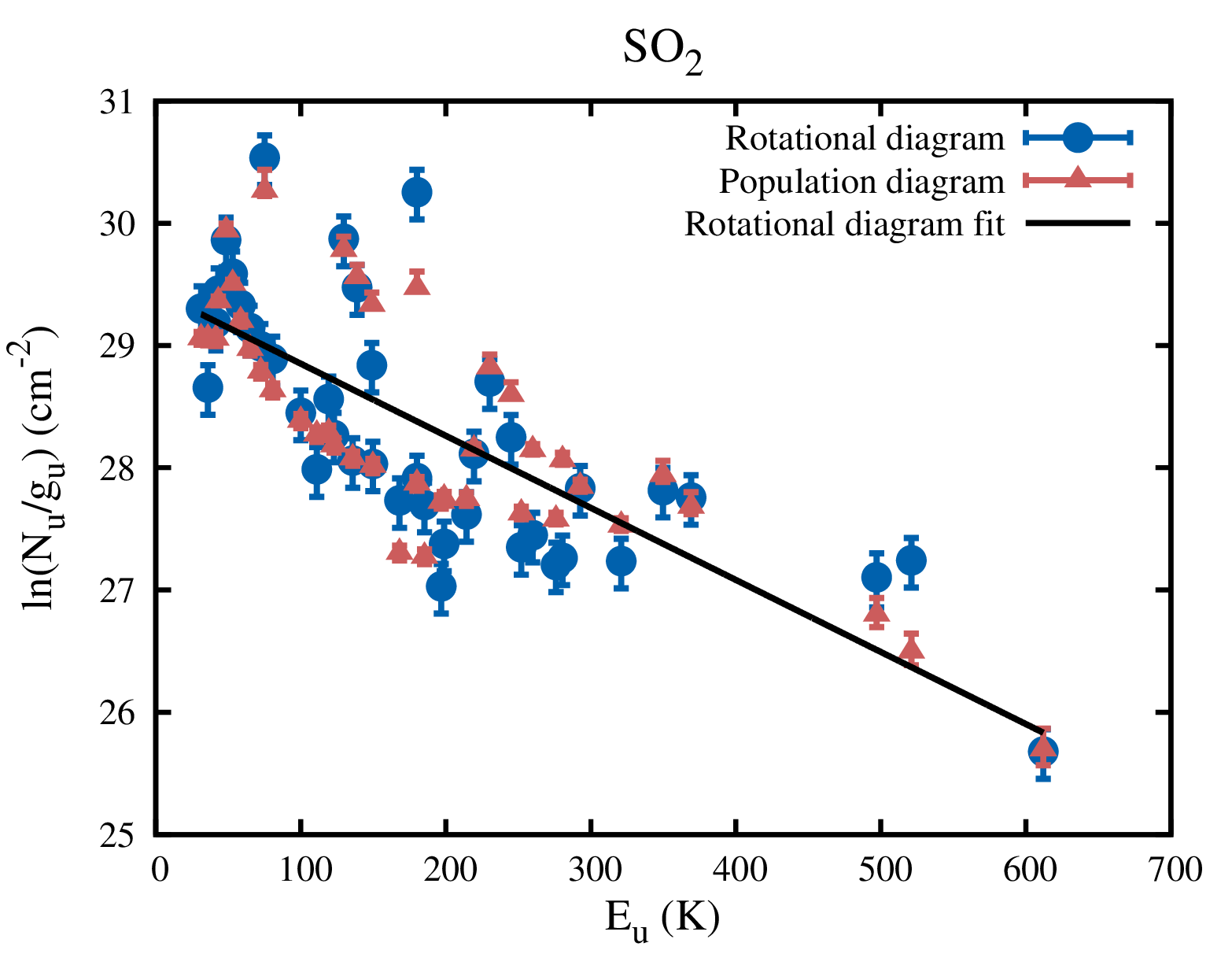}
\includegraphics[width=3.3 cm,angle=-90,trim=0cm 0cm 0cm 0cm,clip=true]{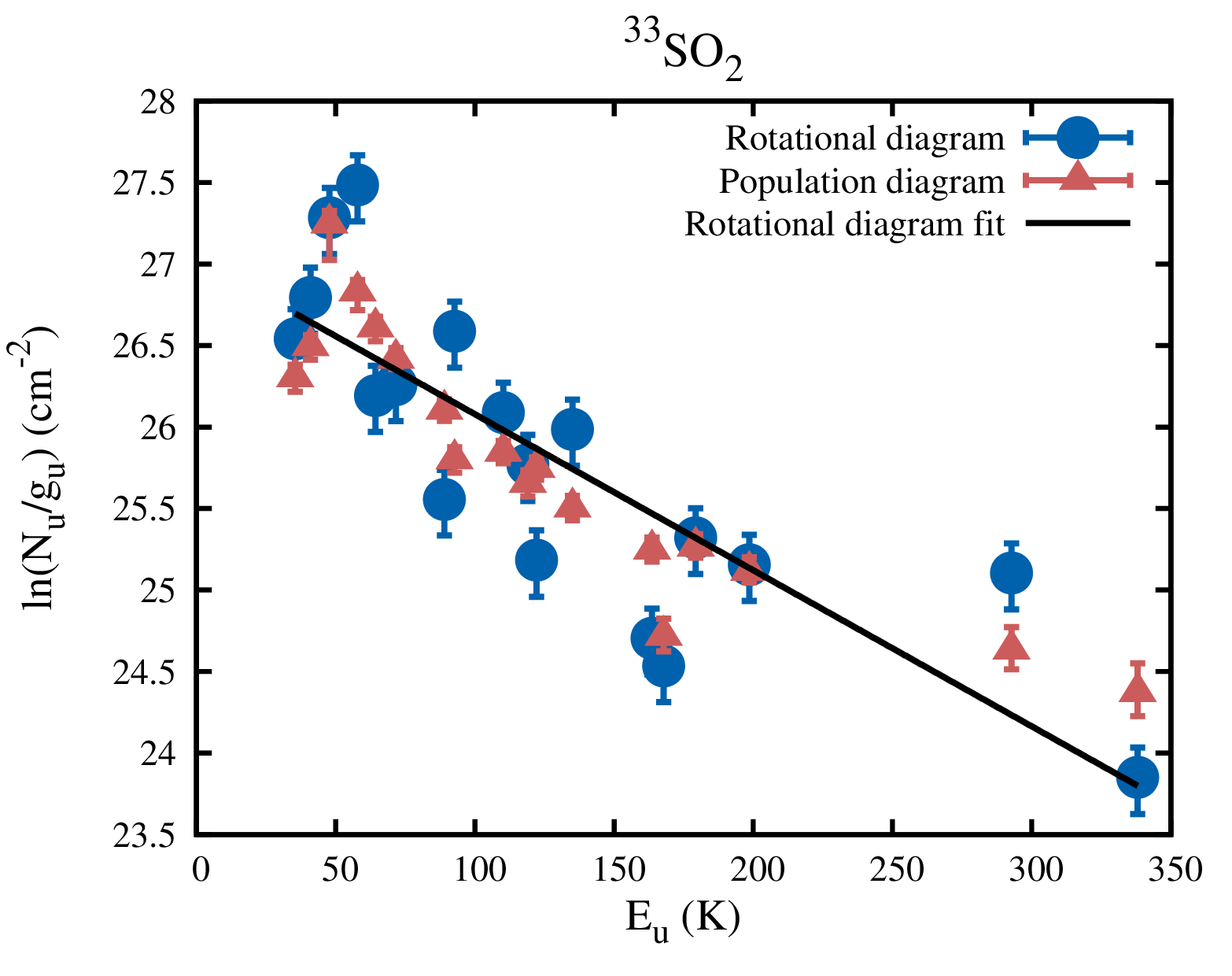}
\includegraphics[width=3.3 cm,angle=-90,trim=0cm 0cm 0cm 0cm,clip=true]{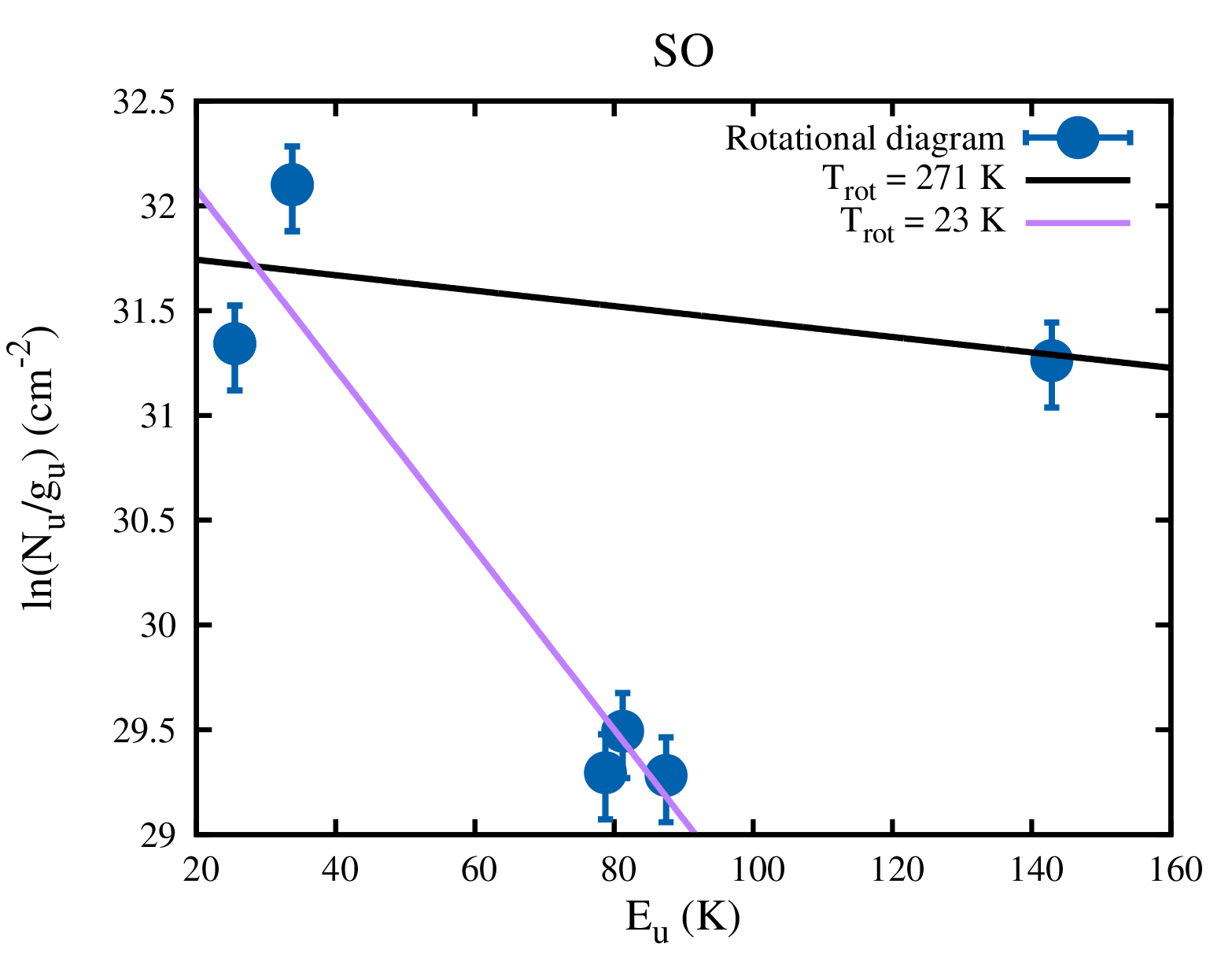}

\caption{Rotation and population diagram plots toward the center. For CH$_3$OH, H$_2$CO, and SO$_2$ the transitions from the high-frequency extension of the SLS (between 360-373 GHz) were originally analyzed in \citet{nagy2012}.}
\label{PD_plots1}
\end{figure*}

The results are summarized in Table \ref{lines_summary1}. The rotation- and population diagrams are shown on Fig. \ref{PD_plots1}. 
The $\ln(N_u/g_u)$ values corresponding to the rotational and the population diagrams are close to each other, as many of the observed lines are optically thin (with a few exceptions of some of the SO$_2$, CH$_3$OH, H$_2$CO, and $^{33}$SO$_2$ lines). 
The scatter in the population diagram values is smaller than in the rotational diagrams as the optically thicker lines have been corrected for their optical depth. In addition the lines have been corrected for the best fit size of the emitting region.

Most molecular lines have excitation temperatures between $\sim$100 and $\sim$200 K with H$_2$CO, H$_2$CS, and SO$^{18}$O showing the highest excitation. The population diagrams suggest a non-uniform beam filling in several cases and result in emission originating in regions of $\sim$1-4$''$, that are consistent with the sizes of hot cores and UCH{\sc{ii}} regions revealed by high resolution studies (e.g. \citealp{wilner2001}). Together with the spatially extended emission seen in the maps, this is evidence that the gas in W49A has a clumpy structure.

CH$_3$CCH has a very low excitation temperature toward the center ($\sim$32 K) suggesting an origin in colder material around the central cluster of UC H{\sc{ii}} regions and hot cores. CH$_3$CCH was also found to originate in a lower excitation component toward Sgr B2, Orion and DR 21 \citep{churchwellhollis1983}. 
The other species that shows a very low excitation temperature is $^{34}$SO ($T_{\rm{ex}}\sim33$ K). Results for SO are not listed in Table \ref{lines_summary1} as the detected SO transitions are not consistent with a single excitation temperature. This may suggest that some of the detected SO lines originate in the low excitation component suggested by the $^{34}$SO lines, and the others from the high-excitation ($\gtrsim$100 K) component that most detected molecular lines originate in. 
Apart from different excitation conditions, gas components with different source sizes (beam-filling factors) may contribute to the observed line emission, as suggested by the large range of source sizes corresponding to the different sulphur-bearing molecules. Distinguishing between these explanations requires observations of more SO transitions.

Differences in the excitation of the subregions (Northern clump, Eastern tail, and South-west clump) are well traced by H$_2$CO, as more of its transitions were detected with a spatial extent large enough to include the discussed subregions. More details are shown in \citet{nagy2012}. 

For molecules with no information available on their excitation we derive a column density in the optically thin LTE approximation, assuming rotation temperatures in the range between 75 K and 150 K, as most molecules with multiple detected transitions have rotation temperatures that fall into this range. Then we derive column densities using:

\begin{equation}
\label{coldens_lte}
N_{\rm{tot}}=\frac{8\pi k \nu^2}{h c^3} \frac{Q(T_{\rm{rot}})}{g_u A_{\rm{ul}}} e^{E_u/k T_{\rm{rot}}} \int{T_{\rm{mb}} d{\rm{v}}}
\end{equation}

Table \ref{table:coldens} includes a selection of molecules that have been detected in AGNs, starburst galaxies, and in Galactic star-forming regions. The line intensities were obtained using a Gaussian fit for nearly Gaussian line profiles. For double-peaked or asymmetric line profiles, the velocity-integrated intensity is used to calculate column densities.
The error of these column densities is about a factor of 2, dominated by the uncertainty in $T_{\rm{ex}}$.
For the species where the number of detected transitions was not enough to correct for the opacity, the optically thin approximation results in an under-estimate of the column densities. This is the case for some of the species detected toward the center, such as for CS and CN. The CS column density is under-estimated by a factor of $\sim$3, by applying the observed $N$(C$^{34}$S) and an isotopic ratio of $^{32}$S/$^{34}$S of 22 \citep{frerking1980}. The column density of CN cannot be constrained as no isotopologue of CN has been detected in the SLS survey. Therefore, the value derived from CN is a lower limit of the CN column density.
We discuss these column densities in more details in Sect. \ref{discussion}.

\begin{savenotes}
\begin{table*}
\begin{minipage}[h!]{\linewidth}
\caption{Results of the rotation and population diagram analysis for molecules with multiple transitions.}             
\label{lines_summary1}      
\centering           
\renewcommand{\thefootnote}{\alph{footnote}}  
\begin{tabular}{p{1.8cm}rlllll}
\hline  
\footnotetext[1]{Fixed parameter in the fit}  
\footnotetext[2]{Transitions in the frequency range between 360 and 373 GHz were included in \citet{nagy2012}.} 
\footnotetext[4]{The optical depth is calculated for each transition. Here we show the highest and lowest value for each molecule.}

Species& $T_{\rm{rot}}$& $N_{\rm{tot}}$& 
$T_{\rm{ex}}$& $N_{\rm{tot,1}}$& Source size& Optical depth\footnotemark[4]\\
       &                   (K)&      (cm$^{-2}$)&    (K)&  (cm$^{-2}$)&       ($''$)& \\  
\hline\hline\\[-0.2cm]       

SO$_2$\footnotemark[2]&
170$\pm$23& $(3.1\pm0.02)\times10^{16}$&
$118\pm7$& $2.1^{+0.3}_{-0.0}\times10^{18}$& $2.9^{+0.0}_{-0.0}$& 0.1-9.3\\[0.1cm]     

$^{33}$SO$_2$&      
105$\pm$18& $(1.3\pm0.01)\times10^{15}$& 
95$^{+11}_{-8}$& $7.6^{+1.1}_{-2.0}\times10^{17}$& 1.0$\pm$0.0& 0.8-5.3\\[0.1cm]  

$^{34}$SO$_2$&      
140$\pm$9& $(4.0\pm0.02)\times10^{15}$&
133$\pm$11& $4.8^{+11.8}_{-4.3}\times10^{16}$& $4.3^{+9.3}_{-1.9}$& 0.01-0.2\\[0.1cm] 

SO$_2$, v$_2$=1&     
146$\pm$26& $(1.9\pm0.06)\times10^{17}$&
143$^{+64}_{-31}$& $2.0^{+500}_{-1.5}\times10^{17}$& $14.5^{+0.5}_{-13.5}$& 0.001-0.004\\[0.1cm]

CH$_3$OH\footnotemark[2]&           
75$\pm$5& $(7.7\pm0.04)\times10^{15}$&
$65^{+7}_{-2}$& $5.0^{+0.0}_{-0.6}\times10^{17}$& $2.4^{+0}_{-0}$& 0.14-2.69\\[0.1cm]  

H$_2$CO\footnotemark[2]&             
213$\pm$96& $(8.6\pm0.09)\times10^{14}$&
$159^{+90}_{-51}$& $1.4^{+2.0}_{-1.2}\times10^{16}$& $3.3^{+4.9}_{-0.7}$& 0.10-1.50\\[0.1cm]

HNCO&                
61$\pm$20& $(4.9\pm0.2)\times10^{14}$&
$54^{+43}_{-34}$& $7.6^{+6599}_{-3.8}\times10^{14}$& $12.7^{+2.3}_{-9.8}$& 0.01-0.05\\[0.1cm]

H$_2$CS&             
196$\pm$85& $(1.6\pm0.02)\times10^{14}$&
$195^{+104}_{-97}$& $1.7^{+723}_{-0.5}\times10^{14}$& $14.5^{+0.7}_{-13.5}$& 0.001-0.005\\[0.1cm]

CH$_3$CN&            
70$\pm$30& $(1.1\pm0.1)\times10^{13}$&
$69^{+94}_{-27}$& $3.1^{+16.9}_{-1.1}\times10^{14}$& $3.0^{+0.0}_{-1.6}$& 0.16-0.27\\[0.1cm]

$^{34}$SO&           
31$\pm$3& $(5.9\pm0.04)\times10^{14}$&
$33^{+8}_{-6}$& $6.2^{+16.7}_{-1.2}\times10^{14}$& $15.0^{+0.0}_{-7.5}$& 0.01-0.22\\[0.1cm]

SO$^{18}$O&          
196$\pm$26& $(8.4\pm0.05)\times10^{14}$&
$173^{+126}_{-105}$& $1.91^{+7.4}_{-1.89}\times10^{17}$& $1.0^{+14.0}_{-0.0}$& 0.20-0.70\\[0.1cm]

CH$_3$CCH&           
31$\pm$7& $(2.8\pm0.1)\times10^{16}$&
$32\pm7$& $2.6^{+91}_{-1.4}\times10^{16}$& $15.0^{+0.0}_{-12.1}$& 0.01-0.03\\[0.1cm] 

HC$_3$N&             
101$\pm$39& $(4.7\pm0.2)\times10^{13}$&
101\footnotemark[1]& $4.8^{+737}_{-0.6}\times10^{13}$& $15.0^{+0.0}_{-13.5}$& 0.004-0.006\\[0.1cm]  

OCS&                 
92$\pm$46& $(1.6\pm0.08)\times10^{15}$&
92\footnotemark[1]& $1.7^{+953}_{-0.3}\times10^{15}$& $14.4^{+0.6}_{-13.2}$& 0.008-0.010\\[0.1cm]     
     
\hline                  
\end{tabular}
\end{minipage}
\end{table*}
\end{savenotes}

\begin{table*}
\begin{savenotes}
\begin{minipage}[!ht]{\linewidth}\centering
\caption{Column densities toward the four analysed subregions of W49A (in case of detections) derived in the LTE optically thin approximation. The two values correspond to adopted temperatures of 75 K and 150 K, respectively. The error of these column densities is about a factor of 2 dominated by the uncertainty in $T_{\rm{ex}}$.}             
\label{table:coldens}  
\centering       
\renewcommand{\thefootnote}{\alph{footnote}}   
\begin{tabular}{llccccc}
\hline     
\footnotetext[1]{Based on $N$(CS) and an isotopic ratio of $^{32}$S/$^{34}$S of 22 \citep{frerking1980}.}
\\[-0.3cm]
Species& Transition& \multicolumn{4}{c}{Column density (cm$^{-2}$)}\\ 
       & & Center& Eastern tail& Northern clump& South-west clump\\
\hline\hline 

C$_2$H& 4$_{9/2}$-3$_{7/2}$& 
$(1.7-2.8)\times10^{15}$& $(3.8-6.0)\times10^{14}$& $(5.7-9.0)\times10^{14}$& $(5.2-8.3)\times10^{14}$\\ 

NO& 7/2-5/2$^+$&   
$(1.0-1.8)\times10^{17}$& $(0.8-1.4)\times10^{16}$& $(1.0-1.9)\times10^{16}$& $(4.8-8.8)\times10^{15}$\\

H$^{13}$CN& 4-3& 
$(1.0-1.4)\times10^{14}$& $(6.5-9.8)\times10^{11}$& $(2.0-3.0)\times10^{12}$& $(1.6-2.5)\times10^{12}$\\

H$^{13}$CO$^+$& 4-3&  
$(2.5-3.8)\times10^{13}$& $(0.8-1.3)\times10^{12}$& $(1.3-2.0)\times10^{12}$& $(1.3-1.9)\times10^{12}$\\ 

CN& $N$=3-2, $J$=5/2-3/2& 
$(1.4-2.2)\times10^{15}$& $(3.0-4.8)\times10^{14}$& $(4.3-6.9)\times10^{14}$& $(3.7-5.9)\times10^{14}$\\

N$_2$H$^+$& 4-3& 
$(3.0-4.4)\times10^{13}$& $(0.7-1.1)\times10^{13}$& $(4.1-6.0)\times10^{12}$& $(1.4-2.0)\times10^{12}$\\

CS& 7-6&   
$(4.4-5.6)\times10^{14}$& $(4.7-6.1)\times10^{13}$& $(7.5-9.6)\times10^{13}$& $(4.8-6.2)\times10^{13}$\\

C$^{34}$S& 7-6& 
$(6.2-8.8)\times10^{13}$& $(2.2-2.8)\times10^{12}$\footnotemark[1]& $(4.2-6.1)\times10^{12}$& $(5.8-8.3)\times10^{11}$\\

H$_2$S& 3$_{2,1}$-3$_{1,2}$&    
$4.9\times10^{15}$& & & \\

HCO& 4$_{1,4}$-3$_{1,3}$&       
$(1.4-2.4)\times10^{15}$& & & \\

CO$^+$& $N$=3-2, $F$=5/2-3/2&    
$(1.4-2.2)\times10^{13}$& & & \\

HCS$^+$& 8-7&   
$(1.8-2.2)\times10^{13}$& & & \\

SO$^+$& 8-7&    
$(4.4-5.7)\times10^{14}$& & & \\

H$_3$O$^+$& 3$_{2,1}$-2$_{2,0}$& 
$(4.8-5.3)\times10^{14}$& & & \\

SiO& 8-7&      
$(0.8-1.0)\times10^{14}$& & & \\  

\hline                  
\end{tabular}
\end{minipage}
\end{savenotes}
\end{table*}

\section{Discussion}
\label{discussion}

The data reveal a complex structure both in terms of kinematics and chemical composition. The molecular line tracers detected in the SLS frequency range can be related to various physical components including shocked regions and PDRs. Some of the detected molecules can be used for a comparison between the local starburst seen in W49A and global starburst phenomenon seen for some external galaxies. In the next sections we follow-up on the simple classification of the detected species presented in section \ref{section_species} based on kinematics and a comparison to well known environments of UV-dominated and shock chemistry.

\subsection{The kinematics of chemically related species} 

\begin{figure}[ht!]
\centering
\includegraphics[width=9cm, angle=0, trim=1cm 0cm 0cm 0cm,clip=true]{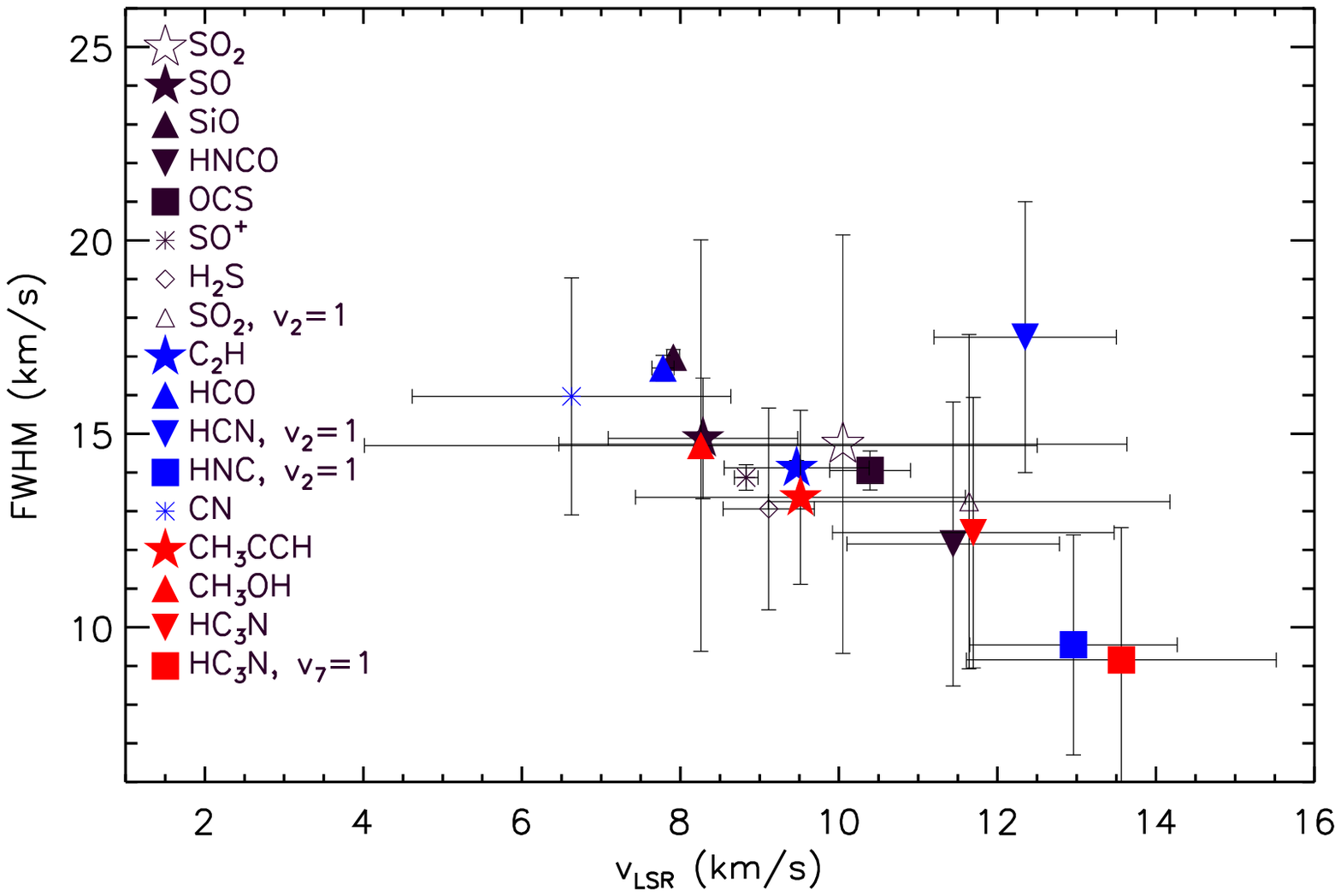}
\caption{The measured line widths vs the peak velocities of species detected toward the W49A center. The molecules classified as shock tracers are shown in black, the molecules classified as PDR tracers are shown in blue, and the molecules classified as complex organic molecules are shown in red. Molecules with double-peaked line profiles are not included in this plot.}
\label{vlsr_fwhm}
\end{figure}

\begin{figure*}[ht!]
\centering
\includegraphics[width=6cm, angle=-90, trim=0.0cm 0cm 0cm 0cm,clip=true]{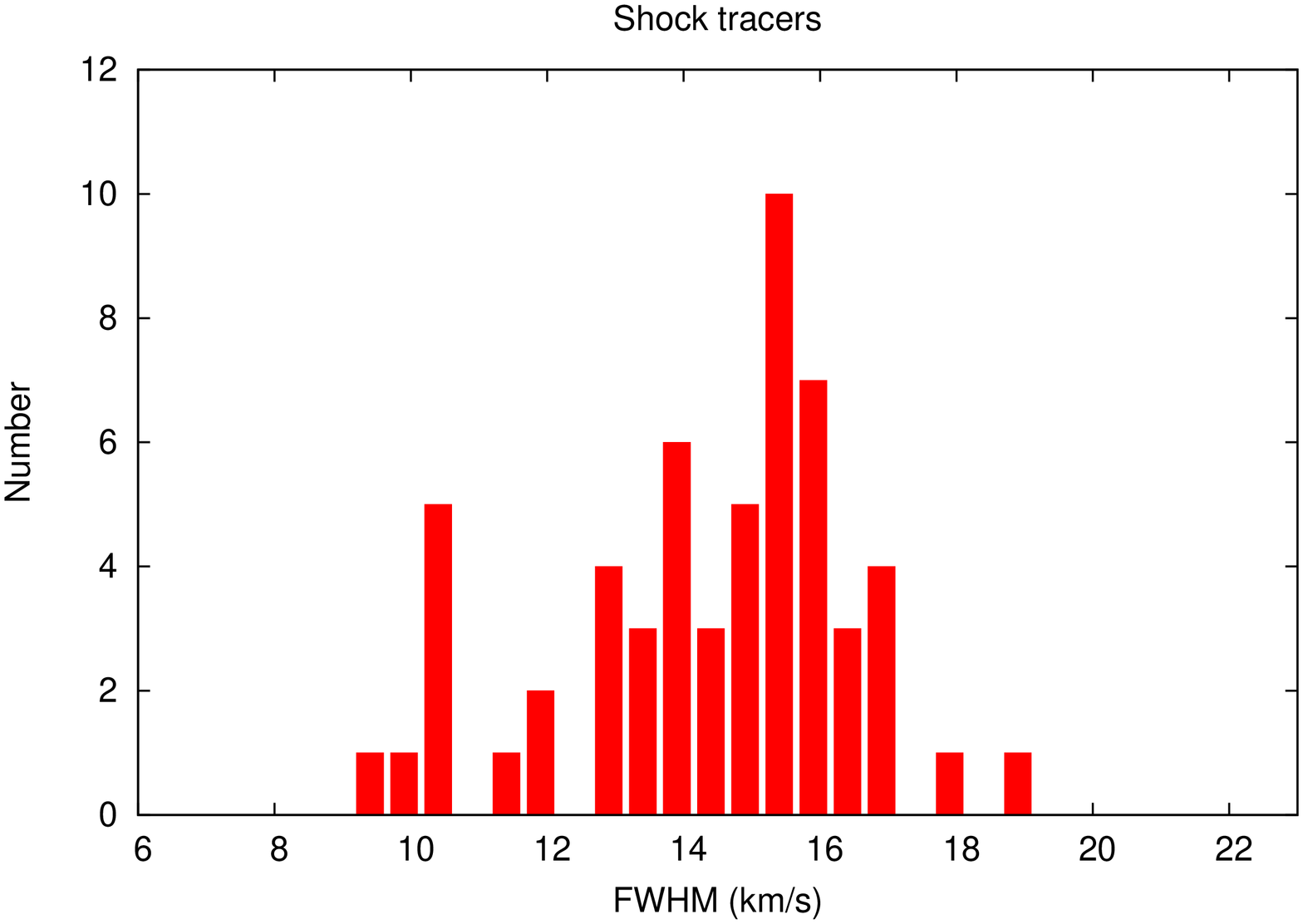}
\includegraphics[width=6cm, angle=-90, trim=0.0cm 0cm 0cm 0cm,clip=true]{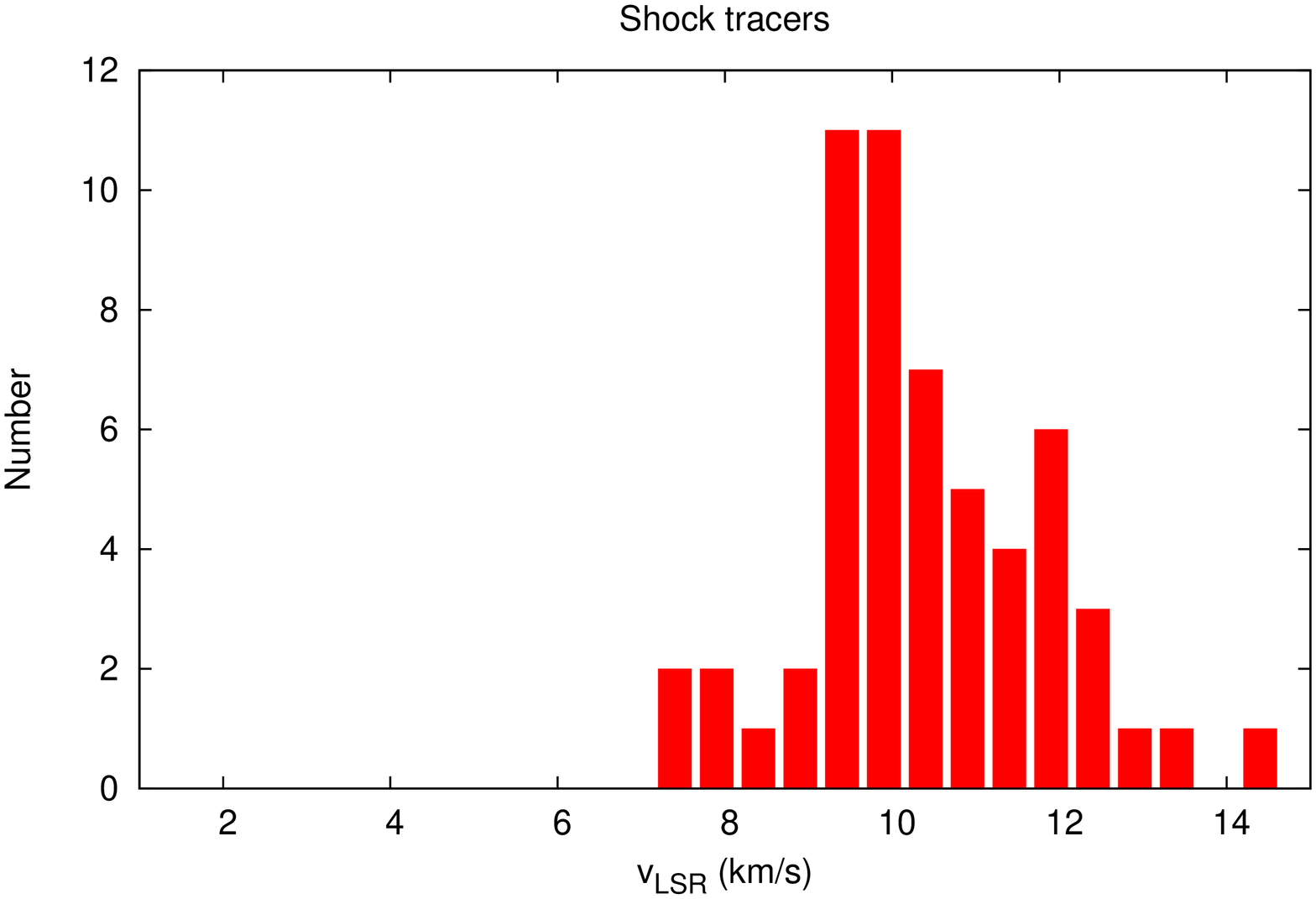}
\includegraphics[width=6cm, angle=-90, trim=0.0cm 0cm 0cm 0cm,clip=true]{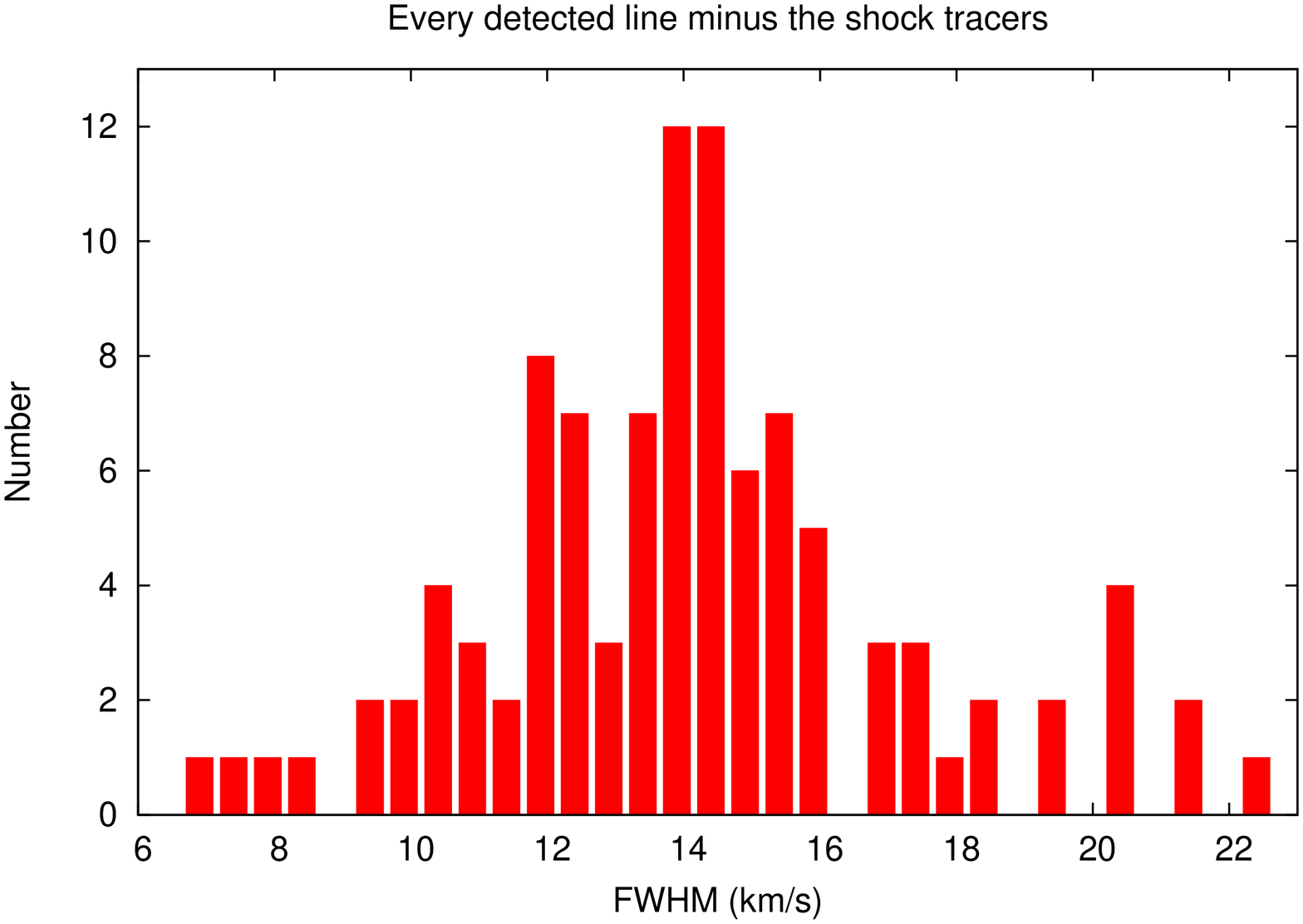}
\includegraphics[width=6cm, angle=-90, trim=0.0cm 0cm 0cm 0cm,clip=true]{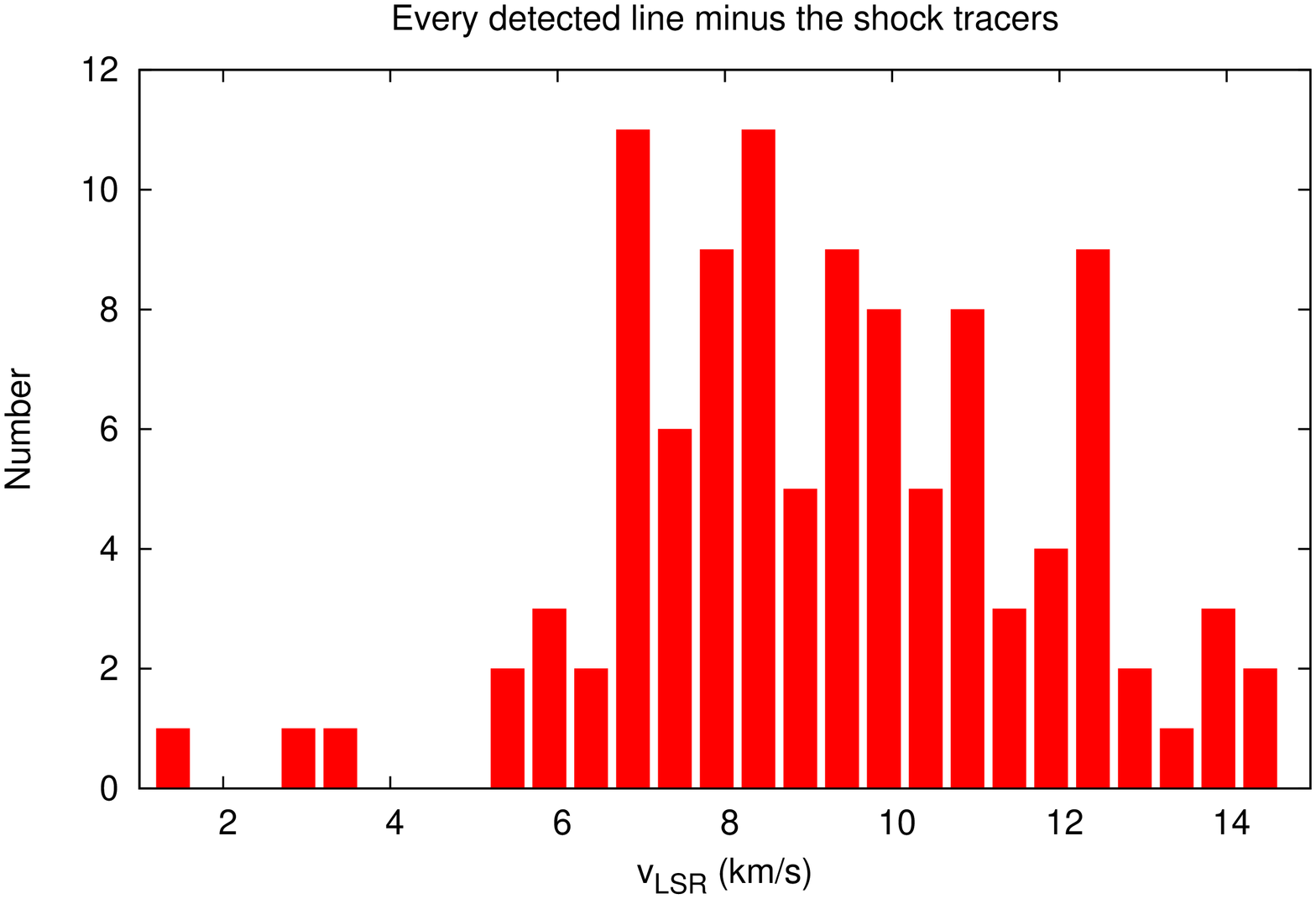}
\includegraphics[width=9cm, angle=0, trim=0.0cm 0cm 0cm 0cm,clip=true]{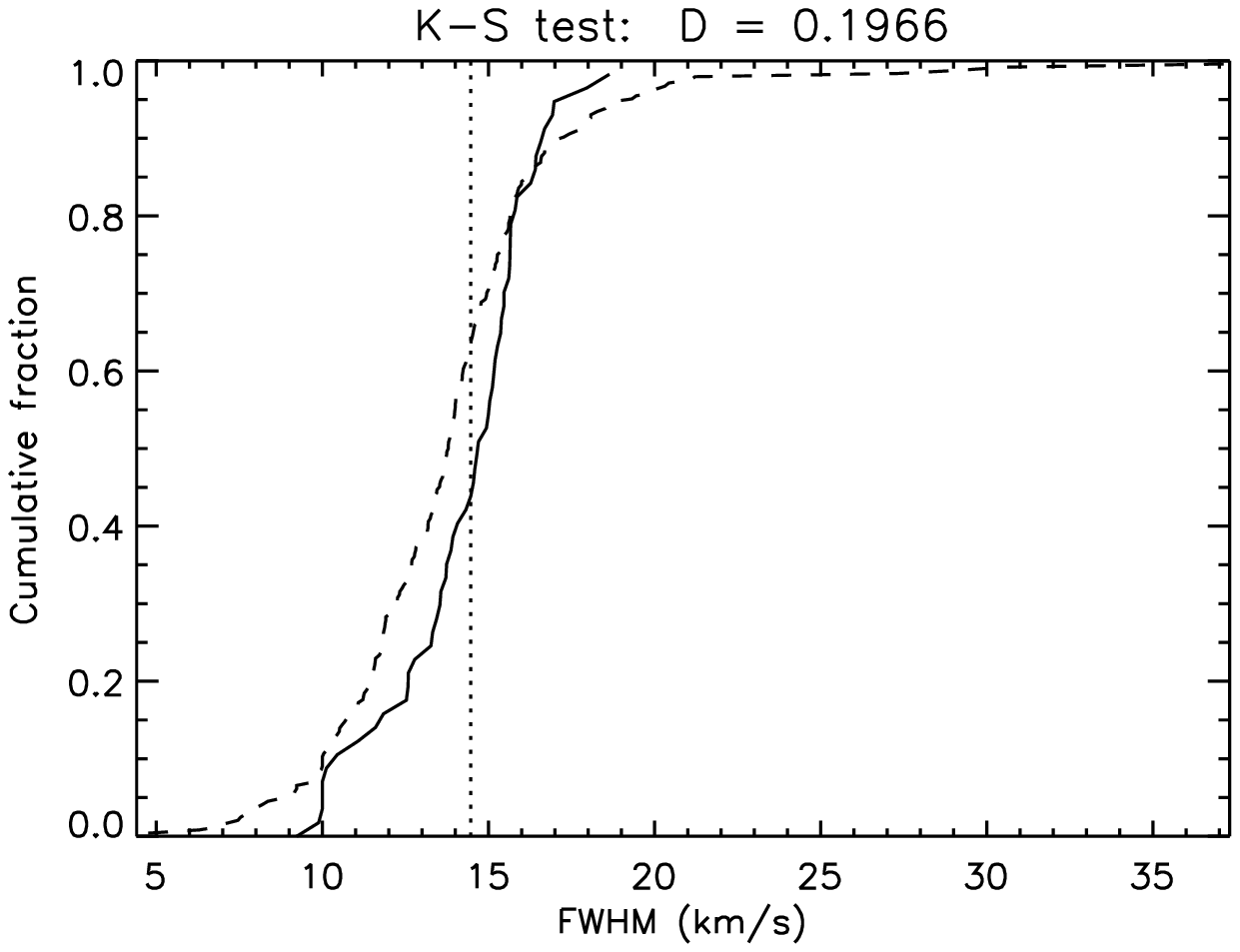}
\includegraphics[width=9cm, angle=0, trim=0.0cm 0cm 0cm 0cm,clip=true]{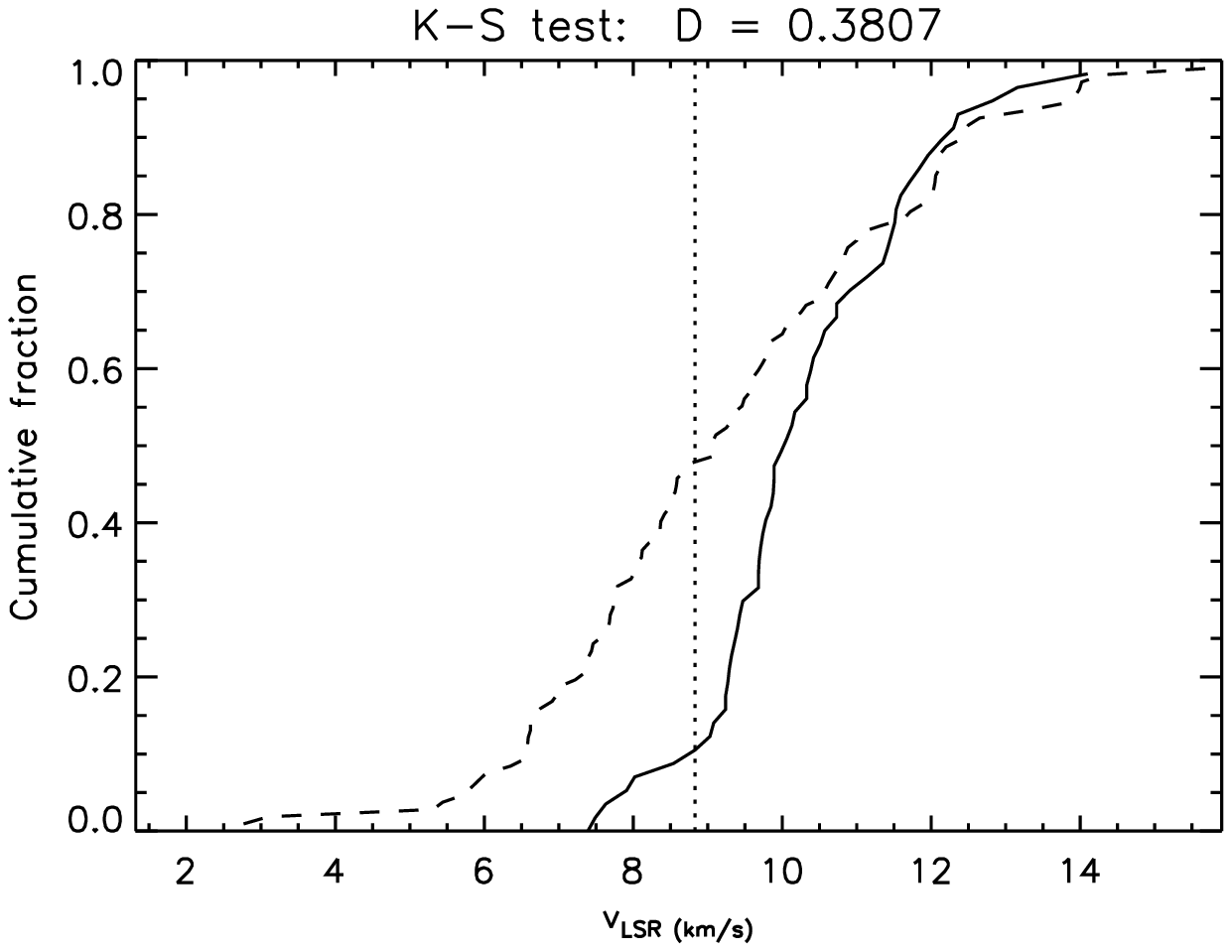}
\caption{
Top: The distribution of the fitted line widths (left) and peak velocities (right) of the lines corresponding to molecules classified as 'shock tracers'. 
Middle: The distribution of the fitted line widths (left) and peak velocities (right) of every detected line in the line survey, except for the double-peaked and very asymmetric line profiles and the shock tracers shown above.
Bottom: K-S test for the line width (left) and peak velocity (right) distributions shown above, corresponding to shock tracers only and to every other molecule that is not classified as a shock tracer. D is the maximum distance between the cumulative fraction of the shock tracers and the whole sample of lines and is related to the probability of a difference between the distributions. The part of the vertical dotted lines between the two distributions is equivalent to the D parameter.}
\label{kin_shocktracers_1}
\end{figure*}

In Table \ref{appendix_line_parameters} in Appendix \ref{line_ident_details} we list the basic line parameters derived from fitting a single profile Gaussian to each line.
Here we use the velocities and FWHM line widths to obtain information on the kinematics of the previously discussed species. This is a simplification, as the fitting of most line profiles requires more components than a single Gaussian fit (e.g. \citealp{galvanmadrid2013}, \citealp{roberts2011}). In this simple comparison fitting the line profiles accurately is beyond our scope, which is a comparison of the line widths and velocities of the observed lines which gives information on the regions where the bulk of emission arises from. 
For the most asymmetric lines we use only the line width derived using a Gaussian fit, and for its $v_{\rm{LSR}}$ we use the velocity of the peak intensity.
Most of the observed lines correspond to species associated with shock tracers (as explained in Sect. \ref{shock_tracers}) including sulphur-bearing molecules. Fig. \ref{kin_shocktracers_1} includes a summary of the fitted FWHM and peak velocity values observed for the molecules associated with shock tracers. The lines of molecules with a possible origin in shocks covers a large range of line widths (9-19 \kms) and peak velocities (7-14 \kms), but most species peak at a velocity of 9-10 \kms and have a width of 15-16 \kms. The large range in the observed parameters is possibly related to the large region ($\sim$0.8 pc) covered by the JCMT beam in the 345 GHz band, which covers multiple sources, which are only fully resolved with interferometers, such as in the maps presented in \citet{galvanmadrid2013}.
Figure \ref{vlsr_fwhm} shows the fitted peak velocities vs the line widths for species corresponding to shock tracers (black color), PDR tracers (blue), and complex organic molecules (red).
The error bars represent the range of the parameters for the different transitions.
All three groups cover a substantial fraction of the parameter space for both velocities and line widths that are discussed in this paper. 
The highest velocities are found for the vibrationally excited molecules. For HCN, v$_2$=1, the high peak velocities are combined with large line widths (FWHM$\sim$20 \kms).
The fact that the molecules which trace different chemistries do not correlate with kinematics may be related to the fact that different individual sources are covered by the JCMT beam. The kinematical signatures are likely to be dependent on the source that they correspond to, and not on the different chemical groups to which they can be related.

Another method to test whether the assumed chemically related species indeed trace different regions is a comparison of the distributions of the observed line parameters. As most observed transitions correspond to shock tracers, we compare the distribution of the observed v$_{\rm{LSR}}$ and FWHM line widths to those of every other transition. Fig. \ref{kin_shocktracers_1} shows the v$_{\rm{LSR}}$ and FWHM line width distributions of the species classified as shock tracers and of all the other transitions. 
We compare the distribution of the fitted parameters of these groups of species using Kolmogorov-Smirnov tests, as shown in Fig. \ref{kin_shocktracers_1}. The probability that the two distributions are different is about 20\% for the line widths and about 38\% for the peak velocities. 
This supports our conclusion that the classification of chemically related species is not confirmed by the measured line parameters, likely, because on the scale probed in this paper they correspond to many different sources, which are not resolved by the JCMT beam. The larger difference seen in the velocity distributions compared to the distributions of the line widths is likely due to the slightly asymmetric line profiles included in the samples.  

\subsection{The importance of UV irradiation}
\label{comparison_pdrs}

We probe the importance of UV-irradiation in the chemistry of W49A by a comparison of the observed column density ratios to values observed toward well-known Galactic PDRs, and by a comparison to predictions of PDR models. 

For the comparison of column density ratios to those observed toward PDRs we select a sample of lines that have been observed both in the SLS survey and in multiple PDRs (Table \ref{table:line_ratios}).
One of the most commonly used tracers of PDRs is the [CN]/[HCN] abundance ratio, which was found to be larger than unity in NGC 7023 \citep{fuente1993}.
An increase in the abundance ratio of [CN]/[HCN] in PDRs is a result of the production of CN via the photodissociation of HCN. Using the HCN column densities presented in \citep{nagy2012} derived in a non-LTE approximation and CN column densities from Table \ref{coldens_lte}, we estimate a [CN]/[HCN] abundance ratio of $\sim$2.0-4.6 toward the Northern clump, $\sim$3.8-6.9 toward the Eastern tail, $\sim$0.4-0.95 toward the South-West clump, and [CN]/[HCN]=0.04-0.07 toward the Center.
The low [CN]/[HCN] ratio toward the center suggests that the chemistry is only partially affected by UV-irradiation.

Though SiO is a tracer of shocks, it has been detected toward several PDRs as well (e.g. \citealp{schilke2001}, \citealp{rizzo2005}). Typical [SiO]/[H$^{13}$CO$^+$] abundance ratios of 0.03-0.37 were found toward Mon R2 \citep{rizzo2005}, 0.54 toward NGC 7023 \citep{schilke2001}, and 3-14 toward the Orion Bar \citep{schilke2001}, based on detections toward multiple positions. We measure an [SiO]/[H$^{13}$CO$^+$] abundance ratio of 2.7-3.3 toward the W49A center. 
The HCO abundance with respect to the SiO abundance was found to be 50-720 in Mon R2 \citep{rizzo2005}, while is in the range between 16 and 23 toward the W49A center. The [HCO]/[SiO] abundance ratio of $>$4.6-10 measured toward NGC 7023 \citep{schilke2001} is based on an upper limit for the SiO column density. The [HCO]/[SiO] abundance ratio toward the Orion Bar was measured to be 2-9, based on detections toward multiple positions. Though the large ($\sim$17 \kms) line width of SiO detected toward the W49A center (Table \ref{appendix_line_parameters}) suggests an origin in shocks rather than PDRs, the [HCO]/[SiO] and [SiO]/[H$^{13}$CO$^+$] abundance ratios do not confirm it.

The CO$^+$ ion is a tracer of warm ($T_{\rm{gas}}\gtrsim500$ K) PDR surfaces, corresponding to the region of $A_V<2$ mag (\citealp{fuente2003}, \citealp{rizzo2003}). We measure an abundance ratio of [CO$^+$]/[H$^{13}$CO$^+$] of 0.54-0.56. This is an upper limit on the [CO$^+$]/[H$^{13}$CO$^+$] ratio as the CO$^+$ line may be blended with SO$^{17}$O. \citet{ginard2012} measure [CO$^+$]/[H$^{13}$CO$^+$]=0.53 in Mon R2, similar to that in the Orion Bar (0.52, based on \citealp{fuente1996} and \citealp{fuente2003}) and to that in NGC 7023 (0.6, based on \citealp{fuente1993} and \citealp{fuente2003}). 
The detection of CO$^+$ at several positions toward the center suggests the existence of a hot gas component directly exposed to UV-irradiation by a radiation field of $G_0\sim3\times10^5$ \citep{vastel2001}. 

HCS$^+$ was found to be abundant in PDRs \citep{ginard2012}. We measure a CS abundance with respect to that of HCS$^+$ of $\sim$76-108. This is between the lower ratios of 11 and 25 found toward the ionization front and MP2 positions in Mon R2 \citep{ginard2012} and the higher [CS]/[HCS$^+$] ratio of 175 measured toward the Horsehead nebula \citep{goicoechea2006}. This suggests that the [CS]/[HCS$^+$] ratio is not a good tracer to probe the contribution of UV irradiation in the chemistry of W49A.

The sulphur-bearing species SO$^+$ and SO$_2$ can be related to shock chemistry, but have also been detected in several PDRs. In PDRs SO$^+$ is formed near the PDR surface (at $A_{\rm{V}}<2$) through the reaction S$^+$ + OH $\rightarrow$ SO$^+$ + H and destroyed by dissociative recombination into S and O. SO and SO$_2$ form in the more shielded regions in the PDR. The [SO$^+$]/[SO$_2$] abundance ratio of $<$0.02 that we measured toward the center of W49A (based on the beam-averaged estimate for SO$_2$ in Table \ref{lines_summary1}) is at least an order of magnitude below the value measured in several PDRs, such as in the Orion Bar ([SO$^+$]/[SO$_2$]$\sim$0.4-1, \citealp{fuente2003}). This difference may also be interpreted as the effect of shocks on the SO$_2$ and SO$^+$ the abundances toward the W49 center.

\begin{table}

\begin{minipage}[h!]{\linewidth}

\caption{Summary of observed column density ratios in W49A (Center, Eastern Tail, Northern Clump and South-west clump, when available) and in Galactic PDRs. 
References: (a) \citet{fuente1996}; (b) \citet{fuente1993}; (c) \citet{ginard2012}; (d) \citet{schilke2001}; (e) \citet{rizzo2005}; (f) \citet{fuente2003}; (g) \citet{goicoechea2006}
}         
\label{table:line_ratios}      
\centering          
\renewcommand{\thefootnote}{\alph{footnote}}   
\resizebox{\textwidth}{!}{%
\begin{tabular}{lccccc}
\hline
Ratio&          W49A&  Orion&  NGC&  Mon&  Horse-\\ 
     &              &  Bar&   7023&   R2&  head\\ 
\hline\hline       
CN/HCN center&  0.55&  3.0$^{({\rm{a}})}$& 4.5$^{({\rm{b}})}$& 2-12$^{({\rm{c}})}$\\
CN/HCN ET&      3.8-6.9& \\ 
CN/HCN NC&      2-4.6& \\
CN/HCN SWC&     0.4-0.95& \\

SiO/H$^{13}$CO$^+$&    2.7-3.3&               3-14$^{({\rm{d}})}$&  0.54$^{({\rm{d}})}$& 0.03-0.37$^{({\rm{e}})}$\\
HCO/SiO&                 16-23&  2-9$^{({\rm{d}})}$& $>$4.6-10$^{({\rm{d}})}$& 50-720$^{({\rm{e}})}$& \\
CO$^+$/H$^{13}$CO$^+$& $<$0.56&  0.52$^{({\rm{a,f}})}$& 0.6$^{({\rm{b,f}})}$& 0.53$^{({\rm{c}})}$& \\
CS/HCS$^+$&             76-108&  & & 11-25$^{({\rm{c}})}$& 175$^{({\rm{g}})}$\\
SO$^+$/SO$_2$&         $<$0.02&  0.4-1$^{({\rm{f}})}$& $>$0.4$^{({\rm{f}})}$& 0.2-0.7$^{({\rm{c}})}$& \\
\hline    
\end{tabular}
}
\end{minipage}
\end{table}

The importance of FUV-irradiation in the chemistry of W49A can be probed by comparing the observed column density ratios to predictions of PDR models. \citet{vastel2001} derive a radiation field toward the W49A center that is equivalent to $\chi=3.5\times10^5 \chi_0$ in \citet{draine1978} units, with $\chi_0 = 2.7 \times 10^{-3}$ erg s$^{-1}$ cm$^{-2}$. We use the  1.4.4 version of the Meudon code (\citealp{lepetit2006}, \citealp{goicoechealebourlot2007}, \citealp{lebourlot2012}) to calculate column densities for the radiation field derived by \citet{vastel2001} for expected densities of 10$^5$ cm$^{-3}$ and 10$^6$ cm$^{-3}$ and temperatures of order 100 K derived by previous studies, corresponding to isobaric models with pressures of 10$^7$ cm$^{-3}$ K and 10$^{8}$ cm$^{-3}$ K. We apply a cosmic ray ionization rate of 2$\times$10$^{-16}$ s$^{-1}$ / H$_2$ molecule \citep{hollenbach2012}. 
Predictions of column density ratios of selected species are summarized in Table \ref{table:pdr_model_comparison}. 

The [HCO]/[H$^{13}$CO$^+$] and [CO$^+$]/[H$^{13}$CO$^+$] line ratios may be consistent with the 10$^7$ cm$^{-3}$ K pressure PDR model, showing that the abundance of some species may be described by PDR chemistry.
The importance of PDR chemistry toward the region in W49A analysed in this paper is further confirmed by Spitzer 8 $\mu$m observations analysed by \citet{peng2010}, showing emission along arcs and other structures toward most of the region where extended emission is seen in PDR tracers such as C$_2$H and CN.
However, the [CN]/[HCN], [CS]/[HCS$^+$], and [SO$^+$]/[SO$_2$] column density ratios clearly show, that the chemistry of the W49A center cannot be explained by a PDR component only, and different effects such as shocks also play an important role. 
The orders of magnitude difference seen in the [CS]/[HCS$^+$] and [SO$^+$]/[SO$_2$] column density ratios may suggest the importance of shock chemistry.
The result that the abundances (and abundance ratios) of sulphur-bearing molecules cannot be explained by a predominantly PDR chemistry is not surprising. Sulphur is released from grains into the gas phase in high temperature regions (such as PDRs and hot cores), but can also be evaporated by shocks. The regions with different chemistries (and different evaporation processes of sulphur from the grains) result in very different abundances for the sulphur-bearing species. This has been observationally confirmed toward regions, where the different types of chemistries can more easily be disentangled. One example is Orion KL, where very different abundance ratios of sulphur-bearing molecules have been measured toward the Plateau region (dominated by shock chemistry) and the Extended ridge (dominated by PDR chemistry) as presented in \citet{tercero2010}.

Based on this comparison we conclude that UV irradiation contributes to the chemistry of the W49A center, but it is not the dominant effect.

\begin{table}
\begin{minipage}[h!]{\linewidth}

\caption{
Comparison of the column density ratios measured toward the center of W49A to PDR models with pressures of 10$^7$ cm$^{-3}$ K and 10$^{8}$ cm$^{-3}$ K illuminated by a radiation field of $\chi=3.5\times10^5 \chi_0$.
}         
\label{table:pdr_model_comparison}      
\centering          
\renewcommand{\thefootnote}{\alph{footnote}}   
\resizebox{\textwidth}{!}{%
\begin{tabular}{lccc}
\hline
     &                     &  \multicolumn{2}{c}{Models}\\
Ratio&          W49A center&  $P=10^7$ cm$^{-3}$ K& $P=10^8$ cm$^{-3}$ K\\ 
\hline\hline       
CN/HCN&                     0.55&    42.4&   7.1\\

HCO/H$^{13}$CO$^+$&    43.2-75.9&    31.5&   5.0\\              

CO$^+$/H$^{13}$CO$^+$& $<$0.56&      0.81&   0.86\\

CS/HCS$^+$&             76-108&      7151&   5696\\

SO$^+$/SO$_2$&         $<$0.02&      6564&   4892\\

\hline    
\end{tabular}
}
\end{minipage}
\end{table}

\subsection{Comparison to regions with shock chemistry}

As we have shown above, the chemistry toward the W49A center is mostly related to effects other than FUV-irradiation. Another possible effect that has an impact on the chemistry in the W49 center is shocks related to outflows and winds of the young massive stars and protostars in the central stellar cluster. The importance of shocks in W49A has previously been studied by \citet{peng2010} through their relation to expanding shells identified in $^{13}$CO data, and by \citet{nagy2012} as a possible contribution to the gas heating. Here we select a sample of regions with shock chemistry for a comparison to the observed column density ratios (Table \ref{table:shocked_regions}).

The Orion KL is a well studied high-mass star-forming region at a distance of $\sim$414 pc \citep{menten2007}. Spectral line surveys toward Orion KL show a component (Plateau) related to outflows and shocks (e.g. \citealp{tercero2010} and references therein). We use the OCS/CS, CS/HCS$^+$, CS/HCO$^+$, CS/H$_2$CO, HCO$^+$/HCS$^+$, and H$_2$CO/H$_2$CS column density ratios measured for the Orion KL plateau by \citet{tercero2010}. The SiO/H$_2$CO and SO$_2$/H$_2$CO ratios correspond to the low velocity flow component reported by \citet{persson2007}.

The low-mass Class 0 protostar L1157 drives a strong molecular outflow (e.g. \citealp{bachillerperezgutierrez1997} and references therein). Its blue lobe has a heating rate related to shocks comparable to that in the shocked region of Orion KL. The column density ratios shown in Table \ref{table:shocked_regions} correspond to the B1 position of the blue lobe measured by \citet{bachillerperezgutierrez1997}.

W3 IRS5 is a bright infra-red source located in the W3 star-forming region at a distance of 2.0 kpc (e.g. \citealp{chavarria2010} and references therein). The embedded young massive stars drive several outflows, which affect the chemistry of the region. In Table \ref{table:shocked_regions} we show the column density ratios measured by \citet{helmichvandishoeck1997}.

The NGC 1333 star forming region is located at a distance of 253 pc \citep{hirota2008} and contains several low-mass protostars including IRAS 4A, which is also a source of outflows. The column density ratios toward this source in Table \ref{table:shocked_regions} are from \citet{blake1995}.

The column density ratios observed toward the W49A center are closest to those measured toward the W3 IRS5 source, and match within a factor of 3. 
The lower limit that we derive for the CS/HCS$^+$ column density ratio is also close to those toward the Orion KL Plateau and the L1157 regions. While the CS/HCS$^+$ column density is not explained by any of the PDR models shown above, they match those observed toward regions with shock chemistry reasonably well.
The CS/H$_2$CO column density ratios are also consistent with those measured toward L1157 (within a factor of 2) and NGC 1333 IRAS 4A (within a factor of 3). The SiO/H$_2$CO and H$_2$CO/H$_2$CS column density ratios are very similar to that measured toward L1157. The largest differences between the regions compared in Table \ref{table:shocked_regions} is seen in the SO$_2$/H$_2$CO line ratios. As we have shown in the population diagram analysis, the SO$_2$ lines probably originate in clumps smaller than the beam-size (Table \ref{lines_summary1}). Therefore, the large differences in the SO$_2$/H$_2$CO column density ratio between the regions may be related to the combination of the different spatial scales that correspond to the observations and to the clumpy structure of SO$_2$.

Even this simple comparison of observed column density ratios shows the importance of shocks in the chemistry of the center of W49A. Further evidence could be given by a comparison to shock models and by smaller scale observations of sulphur-bearing molecules in particular.

\begin{table}

\begin{minipage}[h!]{\linewidth}

\caption{
Comparison of the column density ratios measured toward the center of W49A to values measured toward regions with shock chemistry. References: Orion KL Plateau SiO/H$_2$CO and SO$_2$/H$_2$CO \citep{persson2007}, Orion KL Plateau other column density ratios \citep{tercero2010}; L1157 position B1 \citep{bachillerperezgutierrez1997}; W3 IRS5 \citep{{helmichvandishoeck1997}}; NGC 1333 IRAS 4A outflow \citep{blake1995}.
}         
\label{table:shocked_regions}      
\centering          
\renewcommand{\thefootnote}{\alph{footnote}}   
\resizebox{\textwidth}{!}{%
\begin{tabular}{lccccc}
\hline
Ratio&                 W49A&   Orion KL&   L1157&   W3 IRS5&   NGC 1333\\ 
     &               center&   Plateau&         &          &   IRAS 4A\\ 
\hline\hline       

OCS/CS&             0.9-1.3&    5&           0.2&       0.4&          \\
CS/HCS$^+$&          76-108&   50&            63&       321\\
CS/HCO$^+$&         0.5-1.0&  5.6&           3.7&       1.1&      19.5\\              
CS/H$_2$CO&         1.6-2.3&   0.1&       0.3-0.9&       2.4&       0.6\\
HCO$^+$/HCS$^+$&     90-165&    9&             17&       304&          \\
H$_2$CO/H$_2$CS&    5.0-5.1&   18&          2-5.3&       6.2&          \\
SiO/H$_2$CO&            0.1&  0.8&        0.1-0.3&      0.06&      0.04\\
SO$_2$/H$_2$CO&   35.4-36.7&  140&          0.4-1&      64.5&     $<$0.55\\
\hline    
\end{tabular}
}
\end{minipage}
\end{table}

\subsection{Comparison to starburst galaxies and AGNs}
\label{comparison_starburst}

\begin{figure*}[ht]
\centering
\includegraphics[width=7 cm,angle=-90,trim=0cm 0cm 0cm 0cm,clip=true]{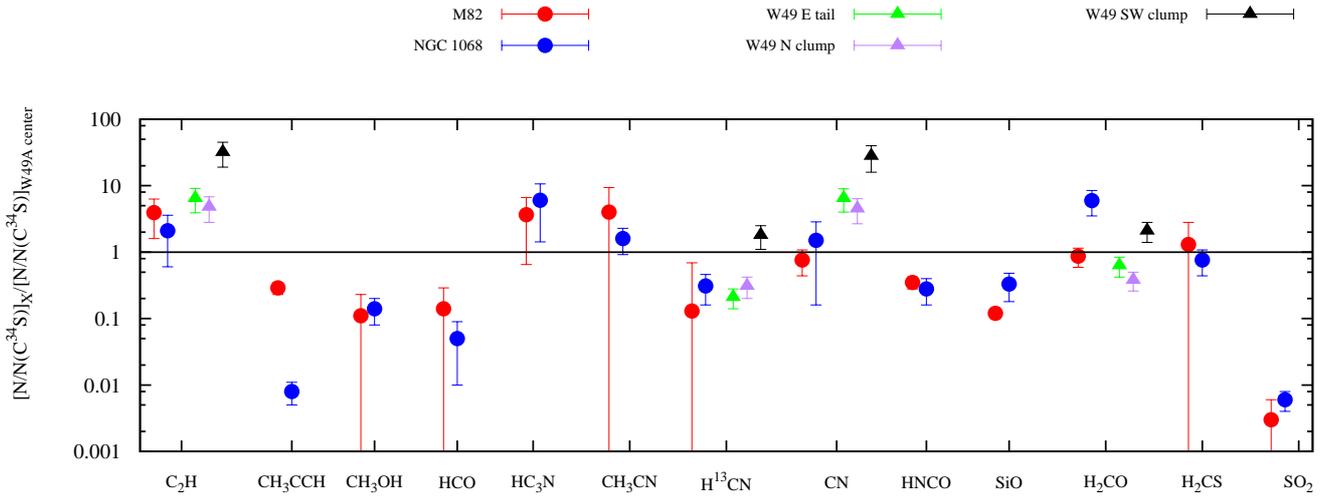}
\caption{Comparison of column density ratios (with respect to that of C$^{34}$S) estimated in W49A and its subregions to those measured in the starburst galaxy M82 \citep{aladro2011b} and in the AGN NGC 1068 \citep{aladro2013}. The error bars for HNCO and SiO toward M82 and those for H$_2$CO, H$_2$CS, and SO$_2$ toward NGC 1068 are under-estimate of the error bars for the ratios, as these values were calculated based on upper limits for the column densities.}
\label{w49_starburst_agn}
\end{figure*}

Several molecules seen in our line survey have been detected in starburst galaxies and AGNs. A comparison between these regions and W49A is shown on Fig. \ref{w49_starburst_agn}, based on line surveys toward the starburst galaxy M82 (\citealp{aladro2011b}, carried out using the IRAM-30m with beam sizes of $14''-19''$ at $\sim$2~mm and $9''-10''$ at $\sim$1.3~mm) and the AGN NGC 1068 (\citealp{aladro2013}, carried out using the IRAM-30m with a $\sim$21$''-29''$ beam). 
For the column density ratios toward the center the beam-averaged column densities are used in the case of species with multiple detected transitions, and not the value corresponding to the best fit source size indicated by the population diagrams (see Sect. \ref{excitation}).
We use C$^{34}$S as a reference for the column density ratios to be consistent with \citet{aladro2011b} and \citet{aladro2013}. C$^{34}$S is also a dense gas tracer and is expected to be optically thin.
The error bars of the ratios presented in Fig. \ref{w49_starburst_agn} are based on the error bars for $N$(X)/$N$(C$^{34}$S) for M82 \citep{aladro2011b}, the column densities for NGC 1068 \citep{aladro2013}, and the column densities presented in this paper and in \citet{nagy2012}.
For some of the abundance ratios this is only a lower limit of the error, as those were calculated using column density upper limits given for M82 (HNCO and SiO, \citealp{aladro2011b}) and for NGC 1068 (H$_2$CO, H$_2$CS, and SO$_2$, \citealp{aladro2013}).
Though a comparison of Galactic scales to those probed in external galaxies, such as a scale of $\sim$0.8 pc for W49A to the 1.5-2 kpc for NGC 1068 \citep{aladro2013} and 158-333 pc for M82 \citep{aladro2011b} is an over-simplification, a few conclusions can be drawn.

The most similar abundances with respect to that of C$^{34}$S between W49A and the AGN and starburst environments are seen for C$_2$H, HC$_3$N, CH$_3$CN, CN, H$_2$CS, and for H$_2$CO in the case of M82. In the case of H$_2$CO, the Northern clump, Eastern tail, and southwest clump regions also show similar abundances w.r.t. C$^{34}$S as M82 and the W49A center. 
C$_2$H, CN, and H$_2$CO are detected with a large spatial extent in W49A. The similarity of the fractional abundance of these species between M82 and W49A, and for C$_2$H and CN also to that toward NGC 1068 suggests that they are good tracers to compare physical and chemical properties of in Galactic star-forming regions at scales of $<$1 pc and global star-formation seen in external galaxies on scales of~$>$1 kpc. 

Among the four subregions of W49A, the largest differences compared to the other regions in the observed fractional abundances w.r.t. C$^{34}$S are seen for the South-west clump. As C$^{34}$S is a dense-gas tracer, the explanation is possibly related to the lower average H$_2$ volume density of the South-west clump compared to the other regions, which is traced by HCN 3-2 and 4-3 line intensity ratios \citep{nagy2012}.

The largest difference compared both to the AGN and starburst examples is seen in the SO$_2$ fractional abundance. This is probably related to the difference between the spatial scales of the observations. SO$_2$ and most sulphur-bearing species in W49A are detected in a $\sim$20$''\times20''$ region around the center and therefore are least likely to trace 'global' properties of $>$kpc regions such as the more spatially extended species C$_2$H, CN, and H$_2$CO. 
Other species with a similar spatial extent to SO$_2$ also show significant differences in Fig. \ref{w49_starburst_agn}: CH$_3$CCH, HCO, and SiO.

\begin{figure*}[ht]
\centering
\includegraphics[width=7 cm,angle=-90,trim=0cm 0cm 0cm 0cm,clip=true]{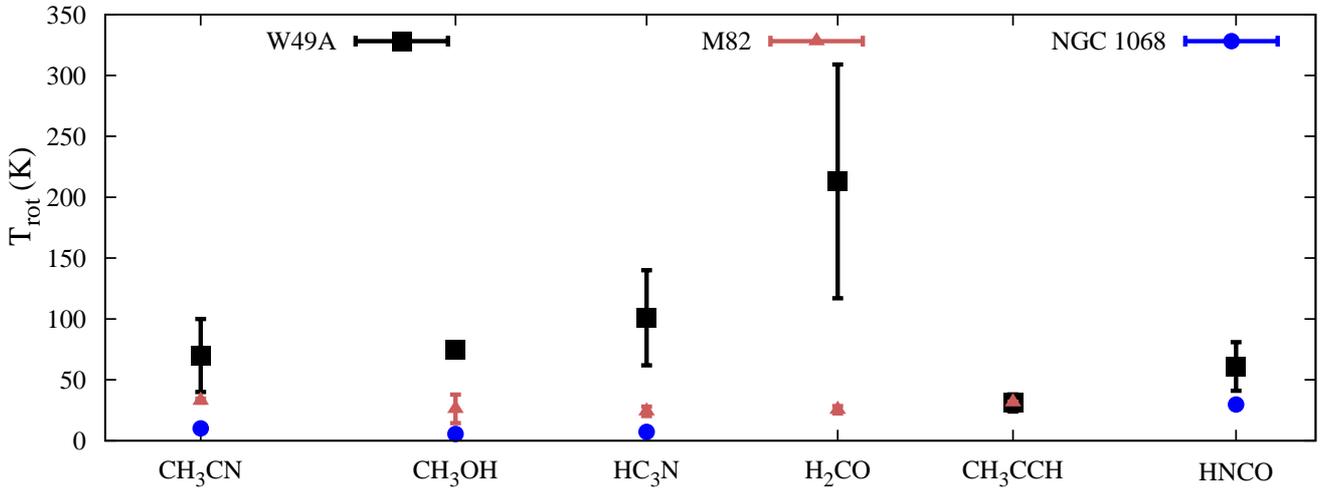}
\caption{A comparison of the rotational temperatures measured toward the W49A center and those measured toward M82 \citep{aladro2011b} and toward NGC 1068 \citep{aladro2013}.}
\label{w49_starburst_agn_temp}
\end{figure*}

Apart from the abundances, the excitation conditions can also be used to compare W49A to external galaxies. We presented an excitation analysis for 14 molecules, and estimated rotational and excitation temperatures. These temperatures can also be compared to those found toward M82 and NGC 1068. In the following, we use the derived rotational temperatures to be consistent with the values derived toward these two galaxies, and to use the values that correspond to uniform beam filling equivalent to emitting regions of $\sim$0.8 pc (the JCMT beam at the observed frequencies). Fig. \ref{w49_starburst_agn_temp} shows the comparison of the species that have a measured excitation temperature toward the W49A center and toward M82 \citep{aladro2011b} and toward NGC 1068 \citep{aladro2013}. Most of the rotational temperatures measured toward the W49A center are significantly higher (factors of 2-10) compared to those measured toward M82 and NGC 1068. This is not surprising due to the very different spatial cales that the measurements correspond to.
CH$_3$CCH is an exception as the rotational temperature of this molecule very close to the value derived toward M82. Apart from CH$_3$CCH toward M82, CH$_3$CN toward M82 and HNCO toward NGC 1068 have rotational temperatures close to the value measured toward the W49A center. The largest difference is seen for H$_2$CO, which is not the case when comparing the abundances toward W49A, M82, and NGC 1068. This comparison toward W49A is biased toward the highest density region with the highest excitation. For a better comparison, more transitions toward the off-center regions (such as the Northern clump, Eastern tail, and South-west clump) could be compared to similar observations in starburst galaxies and AGNs.

\section{Summary}

We presented results from the SLS survey carried out using the JCMT at an angular resolution of $\sim$15$''$ in the 330-373 GHz frequency range toward the luminous and massive star-forming region W49A. Our maps cover a 2$\times$2 arcminutes region around the center of W49A, including the regions referred to as the center, the Eastern tail, the Northern clump, and the Southwest clump, which were selected based on a combination of the morphology and the kinematics of the detected molecular lines and were previously used in \citet{roberts2011}.

\begin{itemize}

\item[-] The detected 255 transitions correspond to 60 species including shock-, and PDR-tracers, and complex organic molecules. Excitation conditions can be probed using multiple detected transitions for 16 species. 

\item[-] The species detected with the largest spatial extent include CO (and its $^{13}$CO and C$^{17}$O isotopologues), N$_2$H$^+$, and H$_2$CO; the dense gas tracers including CS, HCN, HNC, and HCO$^+$; PDR tracers such as CN and C$_2$H.

\item[-] The most spatially extended species reveal a complex spatial and kinematic structure, covering the velocity range of [-5, +20] km s$^{-1}$. Most of the observed lines toward the center are double-peaked or asymmetric, with signatures of infall and outflow motions. Mostly blue-shifted emission is detected toward the Eastern tail region, mostly red-shifted emission toward the Northern clump region, while emission toward the South-west clump is detected around the source velocity.

\item[-] Based on column density ratios of characteristic species, a combination of shock and PDR chemistry affects the physical and chemical structure of the central 17$''\times17''$ region, while UV-irradiation dominates outside of this region including the Northern clump, Eastern tail, and South-west clump regions.  
A preliminary comparison to a starburst galaxy and to an AGN suggests similar C$_2$H, CN, and H$_2$CO abundances between the $\sim$0.8 pc scale probed for W49A and the $>$1 kpc regions in external galaxies with global star-formation, suggesting them to be best tracers for such a comparison of physical and chemical properties of star-formation seen on different scales.

\end{itemize}
In the future, observations with interferometers, such as with ALMA, will probe the chemistry of the W49A region on the smaller scales, indicating the differences between the individual components of the source that are not resolved by single dish telescopes.

\begin{acknowledgements}
We thank the referee for the useful suggestions and the editor Malcolm Walmsley for additional comments. We thank Kuo-Song Wang for providing his population diagram code and John Black for useful comments on an earlier version of this manuscript.
\end{acknowledgements}


\Online

\onecolumn
\begin{appendix}

\section{The detected lines toward the center of W49A}
\label{line_ident_details}

\begin{small}
\setlength\LTcapwidth{18cm}
\begin{longtable}{lllllllll}

\caption[]{The detected species toward the central position of the SLS field. The transitions marked with $^\star$ have been detected toward AFGL 2591 in the SLS survey \citep{vanderwiel2011}. The transitions marked with $^\blacklozenge$ have a non-gaussian line profile. The FWHM line widths of these asymmetric lines are from a Gaussian fitting. Their intensity is obtained by integrating over the whole line and their $V_{\rm{LSR}}$ is the velocity at $T_{\rm{peak}}$. The parameters without error bars are fixed parameters in the fit.}  
\label{appendix_line_parameters}      \\
 
\hline\hline       

Molecule& Transition& Freq.& $E_{\rm{up}}$& $A_\mathrm{ij}$& $\int T_\mathrm{mb} {\rm{d}}v$& $V_{\rm{LSR}}$& FWHM& $T_{\rm{peak}}$\\
&           &     (MHz)&  (K)& (s$^{-1}$)& (K km s$^{-1}$)& (km s$^{-1}$)& (km s$^{-1}$)& (K)\\

\hline

\endhead

\hline \multicolumn{9}{|r|}{{Continued on next page}} \\ \hline
\endfoot

\hline \hline
\endlastfoot

\hline

\textbf{HNCO}& 15$_{0,15}$-14$_{0,14}$& 
329664.4& 126.6& $5.04\times10^{-4}$& 
12.09$\pm$0.95& 9.68$\pm$0.64& 16.57$\pm$1.50& 0.69$\pm$0.14\\

\textbf{$^{34}$SO$_2$}& 8$_{2,6}$-7$_{1,7}$& 
330191.1& ~~42.8& $1.19\times10^{-4}$& 
8.90$\pm$0.58& 10.70$\pm$0.42& 13.31$\pm$1.02& 0.63$\pm$0.12\\

\textbf{$^{13}$CO$^\star$}& 3-2& 
330588.0& ~~31.7& $2.19\times10^{-6}$& 
616.28$\pm$1.01& 5.31$\pm$0.01& 15.88$\pm$0.03& 36.46$\pm$1.80\\

\textbf{$^{34}$SO$_2$}& 21$_{2,20}$-21$_{1,21}$& 
330667.6& 218.9& $1.45\times10^{-4}$& 
5.39$\pm$0.61& 10.70$\pm$0.87& 14.77$\pm$1.66& 0.34$\pm$0.07\\

\textbf{CH$_3$OH$^\star$}&  8$_{-3}$-9$_{-2}$ $E$& 
330793.9& 146.3& $5.39\times10^{-5}$& 
4.62$\pm$0.45& 9.85$\pm$0.86& 17.44$\pm$1.80&  0.25$\pm$0.09\\

\textbf{HNCO}& 15$_{1,14}$-14$_{1,13}$& 
330848.6& 170.3& $5.01\times10^{-4}$& 
7.77$\pm$0.51& 11.35$\pm$0.49& 14.65$\pm$1.08& 0.50$\pm$0.10\\

\textbf{CH$_3$CN$^\star$}& 18$_3$-17$_3$& 
331014.3& 215.2& $3.07\times10^{-3}$& 
2.59$\pm$0.25& 9.68$\pm$1.00& 10.51$\pm$1.00& 0.23$\pm$0.11\\

\textbf{CH$_3$CN$^\star$}& 18$_2$-17$_2$& 
331046.1& 179.5& $3.12\times10^{-3}$& 
3.53$\pm$0.77& 11.88$\pm$1.32& 13.05$\pm$2.96& 0.25$\pm$0.10\\

\textbf{CH$_3$CN$^\star$}& 18$_0$-17$_0$& 
331071.5& 151.0& $3.16\times10^{-3}$& 
9.50$\pm$0.26& 11.91$\pm$1.00& 17.90$\pm$1.00& 0.50$\pm$0.11\\

\textbf{CH$_3$OH$^\star$}& 11$_1$-11$_0$ $E^\mp$& 
331502.4& 169.0& $3.93\times10^{-4}$& 
15.60$\pm$0.64& 8.11$\pm$0.31& 15.01$\pm$0.72& 0.98$\pm$0.12\\

\textbf{SO$_2$}& 11$_{6,6}$-12$_{5,7}$& 
331580.2& 149.0& $4.35\times10^{-5}$& 
17.80$\pm$0.61& 10.77$\pm$0.24& 14.73$\pm$0.61& 1.14$\pm$0.12\\

\textbf{SO$_2~^\star$}& 21$_{2,20}$-21$_{1,21}$& 
332091.4& 219.5& $1.51\times10^{-4}$& 
52.14$\pm$0.69& 10.08$\pm$0.10& 16.28$\pm$0.26& 3.01$\pm$0.17\\

\textbf{$^{34}$SO$_2$}& 23$_{3,21}$-23$_{2,22}$& 
332173.6& 275.1& $2.54\times10^{-4}$& 
9.10$\pm$0.63& 11.94$\pm$0.47& 14.16$\pm$1.21& 0.60$\pm$0.13\\

\textbf{SO$_2~^\star$}& 4$_{3,1}$-3$_{2,2}$& 
332505.2& ~~31.3& $3.29\times10^{-4}$& 
75.04$\pm$0.77& 9.11$\pm$0.08& 15.99$\pm$0.20& 4.41$\pm$0.15\\

\textbf{$^{34}$SO$_2$}& 16$_{4,12}$-16$_{3,13}$& 
332836.2& 163.1& $3.02\times10^{-4}$& 
17.88$\pm$1.02& 12.26$\pm$0.44& 16.58$\pm$1.23& 1.01$\pm$0.13\\

\textbf{S$^{17}$O}& 8$_9$-7$_8$& 
333121.6& ~~75.9& $4.50\times10^{-4}$& 
3.89$\pm$0.67& 12.38$\pm$1.49& 15.91$\pm$2.71& 0.23$\pm$0.12\\

\textbf{$^{34}$SO$^\star$}& 7$_8$-6$_7$& 
333901.0& ~~79.9& $4.69\times10^{-4}$& 
47.01$\pm$0.42& 9.57$\pm$0.06& 14.00$\pm$0.15& 3.15$\pm$0.11\\

\textbf{SO$_2$~$^\star$}& 8$_{2,6}$-7$_{1,7}$& 
334673.4& ~~43.1& $1.27\times10^{-4}$& 
42.19$\pm$0.69& 10.38$\pm$0.12& 15.60$\pm$0.30& 2.54$\pm$0.09\\

\textbf{CH$_3$OH$^\star$}& 2$_2$-3$_1$ A$^-$& 
335133.7& ~~44.7& $2.69\times10^{-5}$& 
6.95$\pm$0.46& 12.41$\pm$0.66& 19.32$\pm$1.30& 0.34$\pm$0.09\\

\textbf{H}& 33 $\beta$&
335207.3& & &
5.19$\pm$0.65& 6.62$\pm$1.91& 30.50$\pm$4.37& 0.16$\pm$0.07\\

\textbf{CH$_3$CHO}& $18_{0,18,2}$-$17_{0,17,2}$&
335318.1& 154.9& $1.30\times10^{-3}$& 
6.53$\pm$0.41& 11.02$\pm$0.61& 20.02$\pm$1.44& 0.31$\pm$0.06\\

\textbf{CH$_3$OH$^\star$}& 7$_1$-6$_1$ A$^+$& 
335582.0& ~~79.0& $1.63\times10^{-4}$& 
20.67$\pm$0.32& 7.68$\pm$0.11& 14.28$\pm$0.24& 1.36$\pm$0.11\\

\textbf{SO$_2$~$^\star$}& 23$_{3,21}$-23$_{2,22}$& 
336089.2& 276.0& $2.67\times10^{-4}$& 
37.35$\pm$0.41& 10.02$\pm$0.08& 14.77$\pm$0.20& 2.38$\pm$0.10\\

\textbf{HC$_3$N}& 37-36& 
336520.1& 306.9& $3.05\times10^{-3}$& 
7.16$\pm$0.62& 10.74$\pm$0.60& 14.57$\pm$1.64& 0.46$\pm$0.09\\

\textbf{SO}& 11$_{10}$-10$_{10}$& 
336553.8& 142.9& $6.12\times10^{-6}$& 
22.03$\pm$1.01& 9.40$\pm$0.31& 13.57$\pm$0.69& 1.52$\pm$0.08\\

\textbf{SO$_2$}& 16$_{7,9}$-17$_{6,12}$& 
336669.6& 245.1& $5.84\times10^{-5}$& 
16.22$\pm$0.11& 11.05$\pm$0.17& 11.46$\pm$0.34& 1.33$\pm$0.14\\

\textbf{CH$_3$OH$^\star$}& 12$_1$-12$_0$ A$^\mp$& 
336865.1& 197.1& $4.07\times10^{-4}$& 
14.71$\pm$0.44& 7.69$\pm$0.22& 15.21$\pm$0.53& 0.91$\pm$0.10\\

\textbf{C$^{17}$O$^\star$}& 3-2& 
337062.0& ~~32.4& $4.30\times10^{-7}$& 
84.53$\pm$0.39& 7.39$\pm$0.03& 13.62$\pm$0.07& 5.83$\pm$0.32\\

\textbf{$^{33}$SO}& 7$_8$-6$_7$& 
337199.4& ~~80.5& $4.65\times10^{-4}$& 
19.13$\pm$0.64& 10.11$\pm$0.24& 14.68$\pm$0.59& 1.22$\pm$0.10\\

\textbf{HC$_3$N, v$_7$=1}& $J$=37-36&
337344.7& 628.5& $3.05\times10^{-3}$&
3.19$\pm$0.55& 14.02$\pm$1.04& 11.72$\pm$2.43& 0.26$\pm$0.09\\

\textbf{C$^{34}$S$^\star$}& 7-6& 
337396.5& ~~50.2& $8.00\times10^{-4}$& 
24.98$\pm$0.56& 8.02$\pm$0.15& 13.79$\pm$0.35& 1.70$\pm$0.15\\

\textbf{$^{34}$SO$^\star$}& 8$_8$-7$_7$& 
337580.1& ~~86.1& $4.89\times10^{-4}$& 
46.74$\pm$0.43& 9.08$\pm$0.06& 13.66$\pm$0.14& 3.21$\pm$0.10\\

\textbf{CH$_3$OH, $v_t$=1}& $7_{7,1}-6_{6,1}$ A$^+$& 
337643.9& 365.4& $1.69\times10^{-4}$& 
3.54$\pm$0.50& 9.49$\pm$1.20& 15.61$\pm$2.04& 0.21$\pm$0.06\\

\textbf{CH$_3$OCH$_3$}& $7_{4,3,0}-6_{3,4,0}$& 
337787.2& ~~48.0& $1.94\times10^{-4}$& 
1.80$\pm$0.53& 11.72$\pm$1.38& 10.13$\pm$3.81& 0.17$\pm$0.04\\

\textbf{$^{34}$SO}& 3$_3$-2$_3$& 
337892.2& ~~25.3& $1.40\times10^{-5}$& 
4.59$\pm$0.54& 10.81$\pm$0.70& 12.32$\pm$1.81& 0.35$\pm$0.08\\

\textbf{H$_2$CS}& 10$_{1,10}$-9$_{1,9}$& 
338083.2& 102.4& $5.77\times10^{-4}$& 
7.31$\pm$0.57& 7.36$\pm$0.54& 13.65$\pm$1.17& 0.50$\pm$0.07\\

\textbf{CH$_3$OH$^\star$}& 7$_0$-6$_0$ $E$& 
338124.5& ~~78.1& $1.70\times10^{-4}$& 
15.87$\pm$0.88& 6.64$\pm$0.37& 13.31$\pm$0.85& 1.12$\pm$0.10\\

\textbf{SO$_2$~$^\star$}& 18$_{4,14}$-18$_{3,15}$& 
338306& 196.8& $3.27\times10^{-4}$& 
29.82$\pm$0.39& 7.47$\pm$0.11& 18.52$\pm$0.28& 1.51$\pm$0.10\\

\textbf{CH$_3$OH$^\star$}& 7$_{-1}$-6$_{-1}$ $E$& 
338344.6& ~~70.6& $1.67\times10^{-4}$& 
23.99$\pm$0.31& 6.59$\pm$0.09& 13.53$\pm$0.21& 1.67$\pm$0.10\\

\textbf{CH$_3$OH$^\star$}& 7$_0$-6$_0$ A$^+$& 
338408.7& ~~65.0& $1.70\times10^{-4}$& 
22.13$\pm$0.30& 6.62$\pm$0.08& 12.26$\pm$0.19& 1.70$\pm$0.10\\

\textbf{CH$_3$OH$^\star$}& $7_{4,4}-6_{4,3}$ A$^-$& 
338512.6& 145.3& $1.15\times10^{-4}$& 
10.32$\pm$0.48& 7.35$\pm$0.32& 13.83$\pm$0.72& 0.70$\pm$0.09\\

\textbf{CH$_3$OH}& $7_{3,4}-6_{3,3}$ A$^-$& 
338543.2& 114.8& $1.39\times10^{-4}$& 
10.68$\pm$0.50& 7.97$\pm$0.33& 13.87$\pm$0.75& 0.72$\pm$0.09\\

\textbf{SO$_2$~$^\star$}& 20$_{1,19}$-19$_{2,18}$& 
338611.8& 198.9& $2.87\times10^{-4}$& 
40.93$\pm$0.50& 6.95$\pm$0.09& 15.73$\pm$0.22& 2.44$\pm$0.10\\

\textbf{CH$_3$OH}& $7_2-6_2$ $A^+$& 
338639.9& 102.7& $1.57\times10^{-4}$& 
7.93$\pm$0.90& 6.75$\pm$1.00& 11.79$\pm$1.00& 0.63$\pm$0.10\\

\textbf{CH$_3$OH$^\star$}& 7$_2$-6$_2$ $E$& 
338721.6& ~~87.3& $1.55\times10^{-4}$& 
21.52$\pm$0.27& 5.92$\pm$0.09& 13.16$\pm$0.20& 1.54$\pm$0.17\\

\textbf{$^{34}$SO$_2$}& 14$_{4,10}$-14$_{3,11}$& 
338785.7& 134.5& $3.08\times10^{-4}$& 
5.35$\pm$0.32& 11.23$\pm$0.38& 12.57$\pm$0.83& 0.40$\pm$0.04\\

\textbf{SO$^\star$}& 3$_3$-3$_2$& 
339341.5& ~~25.5& $1.45\times10^{-5}$& 
18.61$\pm$0.34& 8.02$\pm$0.12& 13.74$\pm$0.28& 1.27$\pm$0.08\\

\textbf{$^{33}$SO$_2$}& 16$_{4,12}$-16$_{3,13}$& 
339482.3& 166.1& $3.18\times10^{-4}$& 
2.72$\pm$0.31& 10.53$\pm$0.71& 12.33$\pm$1.53& 0.21$\pm$0.05\\

\textbf{CN$^\star$}& $N$=3-2, $J$=5/2-5/2& 
339516.6& ~~32.6& $2.54\times10^{-5}$& 
3.15$\pm$0.50& 7.46$\pm$1.40& 17.27$\pm$3.49& 0.17$\pm$0.07\\

\textbf{$^{34}$SO$^\star$}& 9$_8$-8$_7$& 
339857.3& ~~77.3& $5.08\times10^{-4}$& 
70.69$\pm$0.33& 9.15$\pm$0.03& 14.43$\pm$0.08& 4.60$\pm$0.15\\

\textbf{CN$^{\star \blacklozenge}$}& $N$=3-2, $J$=5/2-3/2& 
340035.4& ~~32.6& $2.89\times10^{-4}$& 
51.50$\pm$1.01& 6.98$\pm$1.0& 17.38$\pm$0.68& 3.07$\pm$0.06\\ 

\textbf{CH$_3$OH$^\star$}& 2$_2$-3$_1$ A$^+$& 
340141.2& ~~44.7& $2.78\times10^{-5}$& 
4.38$\pm$0.35& 8.03$\pm$0.60& 14.91$\pm$1.39& 0.28$\pm$0.05\\

\textbf{CN$^{\star  \blacklozenge}$}& $N$=3-2, $J$=7/2-5/2& 
340248.5& ~~32.7& $3.67\times10^{-4}$& 
51.14$\pm$0.99& 5.44$\pm$1.0& 13.25$\pm$1.42& 3.61$\pm$0.07\\

\textbf{$^{13}$CH$_3$OH}& $2_{2,1}-3_{1,2}$ A$^-$& 
340313.9& ~~44.6& $2.82\times10^{-5}$& 
31.79$\pm$0.32& 8.21$\pm$0.07& 14.43$\pm$0.17& 2.07$\pm$0.08\\

\textbf{OCS$^\star$}& 28-27& 
340449.3& 237.0& $1.15\times10^{-4}$& 
10.68$\pm$0.41& 10.38$\pm$0.29& 14.55$\pm$0.58&  0.69$\pm$0.08\\

\textbf{$^{33}$SO$_2$}& 20$_{1,19}$-19$_{2,18}$& 
340526.1& 201.5& $2.96\times10^{-4}$& 
4.61$\pm$0.32& 11.27$\pm$0.40& 11.24$\pm$0.86& 0.39$\pm$0.06\\

\textbf{HC$^{18}$O$^+$}& 4-3& 
340630.7& ~~40.9& $3.11\times10^{-3}$& 
6.98$\pm$0.36& 7.58$\pm$0.29& 11.26$\pm$0.68& 0.58$\pm$0.07\\

\textbf{SO$^\star$}& 7$_8$-6$_7$& 
340714.2& ~~81.2& $4.99\times10^{-4}$&  
213.26$\pm$1.09& 7.49$\pm$0.04& 15.61$\pm$0.10& 12.83$\pm$0.52\\

\textbf{$^{33}$SO}& 8$_8$-7$_7$& 
340839.6& ~~86.8& $5.03\times10^{-4}$&  
22.64$\pm$0.26& 10.49$\pm$0.08& 13.84$\pm$0.18& 1.54$\pm$0.09\\

\textbf{SO$_2$}& $21_{8,14}-22_{7,15}$&
341275.5& 369.1& $6.86\times10^{-5}$&
14.77$\pm$0.36& 11.60$\pm$0.15& 12.69$\pm$0.37& 1.09$\pm$0.10\\

\textbf{HCS$^+$}& 8-7& 
341350.2& ~~73.7& $8.35\times10^{-4}$& 
5.68$\pm$0.37& 9.40$\pm$0.46& 13.38$\pm$0.87& 0.40$\pm$0.08\\

\textbf{CH$_3$OH$^\star$}& 7$_1$-6$_1$ A$^-$& 
341415.6& ~~80.1& $1.71\times10^{-4}$& 
36.71$\pm$0.48& 10.62$\pm$0.14& 20.43$\pm$0.31& 1.69$\pm$0.19\\

\textbf{HCO}& $4_{1,4}-3_{1,3}$&
341671.6& ~~73.8& $3.46\times10^{-4}$&
18.60$\pm$0.31& 7.78$\pm$0.14& 16.70$\pm$0.33& 1.05$\pm$0.09\\

\textbf{$^{33}$SO$_2$}& $13_{2,12}-12_{1,11}$& 
341721.7& ~~94.1& $2.30\times10^{-4}$& 
10.70$\pm$1.45& 10.51$\pm$0.95& 17.67$\pm$2.28& 0.57$\pm$0.09\\

\textbf{CH$_3$CCH}& $20-19$& 
341741.0& 172.2& $1.39\times10^{-4}$& 
9.38$\pm$1.43& 9.82$\pm$0.86& 14.53$\pm$1.58& 0.61$\pm$0.09\\

\textbf{$^{34}$SO$_2$}& $5_{3,3}-4_{2,2}$& 
342208.9& ~~35.1& $3.10\times10^{-4}$& 
15.29$\pm$0.51& 11.77$\pm$0.19& 11.83$\pm$0.46& 1.21$\pm$0.10\\

\textbf{$^{34}$SO$_2$}& $20_{1,19}-19_{2,18}$& 
342231.6& 198.5& $3.06\times10^{-4}$& 
16.20$\pm$0.52& 11.60$\pm$0.21& 14.02$\pm$0.55& 1.09$\pm$0.10\\

\textbf{$^{34}$SO$_2$}& $12_{4,8}-12_{3,9}$& 
342332.0& 109.7& $3.06\times10^{-4}$& 
15.07$\pm$0.47& 11.57$\pm$0.20& 12.77$\pm$0.46& 1.11$\pm$0.10\\

\textbf{CH$_3$OH$^\star$}& 13$_1$-13$_0$ A$^\mp$& 
342729.8& 227.5& $4.23\times10^{-4}$& 
9.95$\pm$0.38& 9.06$\pm$0.26& 13.71$\pm$0.57& 0.68$\pm$0.09\\

\textbf{O$^{13}$C$^{34}$S}& 29-28& 
342759.8& 246.8& $1.18\times10^{-4}$& 
17.47$\pm$0.35& 9.66$\pm$0.14& 13.45$\pm$0.34& 1.22$\pm$0.09\\

\textbf{CS$^\star$}& $7-6$& 
342882.9& ~~65.8& $8.40\times10^{-4}$& 
155.79$\pm$0.66& 6.35$\pm$0.03& 14.21$\pm$0.08& 10.30$\pm$0.80\\

\textbf{H$_2$CS}& $10_{0,10}-9_{0,9}$& 
342946.4& ~~90.6& $6.08\times10^{-4}$& 
4.15$\pm$0.35& 9.11$\pm$0.52& 11.58$\pm$0.99& 0.34$\pm$0.06\\

\textbf{HO$^{13}$C$^+$}& $4-3$&
342983.3& ~~41.2& $3.34\times10^{-3}$&
2.49$\pm$0.35& 10.83$\pm$0.82& 11.91$\pm$2.29& 0.20$\pm$0.07\\

\textbf{$^{33}$SO}& $9_8-8_7$, $F=\frac{17}{2}-\frac{15}{2}$& 
343087.3& ~~78.0& $5.09\times10^{-4}$& 
30.65$\pm$0.53& 9.66$\pm$0.12& 14.21$\pm$0.29& 2.03$\pm$0.10\\

\textbf{H$_2^{13}$CO}& 5$_{15}$-4$_{14}$& 
343325.7& ~~61.3& $1.12\times10^{-3}$& 
8.76$\pm$0.58& 10.88$\pm$0.60& 18.27$\pm$1.50& 0.45$\pm$0.07\\

\textbf{H$_2$CS}& $10_{3,7}-9_{3,6}$& 
343414.1& 209.1& $5.56\times10^{-4}$& 
5.62$\pm$0.33& 12.12$\pm$0.48& 15.75$\pm$0.89& 0.33$\pm$0.09\\

\textbf{H$_2$CS}& $10_{2,8}-9_{2,7}$& 
343813.2& 143.4& $5.88\times10^{-4}$& 
2.56$\pm$0.41& 10.25$\pm$0.83& 11.91$\pm$2.68& 0.20$\pm$0.10\\

\textbf{HC$^{15}$N$^\star$}& 4-3& 
344200.1& ~~41.3& $1.88\times10^{-3}$& 
13.85$\pm$0.46& 8.49$\pm$0.24& 14.51$\pm$0.56& 0.90$\pm$0.09\\

\textbf{$^{34}$SO$_2$}& $10_{4,6}-10_{3,7}$& 
344245.3& ~~88.5& $2.96\times10^{-4}$& 
12.41$\pm$0.36& 10.85$\pm$0.17& 11.56$\pm$0.39& 1.01$\pm$0.07\\

\textbf{SO$^\star$}& 8$_8$-7$_7$& 
344310.6& ~~87.5& $5.19\times10^{-4}$& 
199.46$\pm$1.96& 7.39$\pm$0.07& 15.37$\pm$0.17& 12.19$\pm$0.50\\

\textbf{$^{34}$SO$_2$~$^\star$}& 19$_{1,19}$-18$_{0,18}$& 
344581.0& 167.7& $5.16\times10^{-4}$& 
20.86$\pm$0.43& 11.35$\pm$0.12& 12.20$\pm$0.30& 1.61$\pm$0.15\\

\textbf{$^{34}$SO$_2$~$^\star$}& 13$_{4,10}$-13$_{3,11}$& 
344807.9& 121.6& $3.17\times10^{-4}$& 
14.45$\pm$0.54& 11.54$\pm$0.25& 13.69$\pm$0.64& 0.99$\pm$0.12\\

\textbf{$^{34}$SO$_2$}& 15$_{4,12}$-15$_{3,13}$& 
344987.6& 148.3& $3.27\times10^{-4}$& 
26.22$\pm$0.64& 6.18$\pm$0.22& 17.92$\pm$0.49& 1.37$\pm$0.14\\

\textbf{$^{33}$SO$_2$}& $14_{4,10}-4_{3,11}$& 
345134.4& 137.1& $3.19\times10^{-6}$& 
7.92$\pm$1.14& 12.86$\pm$0.93& 15.36$\pm$2.61& 0.48$\pm$0.12\\

\textbf{SO$_2$}& 5$_{5,1}$-6$_{4,2}$& 
345149.0& ~~75.1& $9.81\times10^{-6}$& 
8.51$\pm$1.20& 11.36$\pm$0.41& 9.70$\pm$1.13& 0.82$\pm$0.12\\

\textbf{$^{34}$SO$_2$}& 8$_{4,4}$-8$_{3,5}$& 
345168.7& ~~71.0& $2.75\times10^{-4}$& 
12.42$\pm$0.68& 12.25$\pm$0.28& 11.39$\pm$0.83& 1.02$\pm$0.12\\

\textbf{$^{34}$SO$_2$}& 9$_{4,6}$-9$_{3,7}$& 
345285.6& ~~79.3& $2.88\times10^{-4}$& 
13.67$\pm$0.66& 11.93$\pm$0.25& 10.22$\pm$0.56& 1.26$\pm$0.17\\

\textbf{H$^{13}$CN$^\star$}& 4-3& 
345339.8& ~~41.4& $1.90\times10^{-3}$& 
110.90$\pm$0.72& 8.59$\pm$0.05& 15.70$\pm$0.12& 6.64$\pm$0.46\\

\textbf{SO$_2$}& $26_{9,17}-27_{8,20}$& 
345449.0& 521.0& $7.63\times10^{-5}$& 
11.82$\pm$0.65& 11.12$\pm$0.35& 13.25$\pm$0.90& 0.84$\pm$0.17\\

\textbf{$^{34}$SO$_2$~$^\star$}& 7$_{4,4}$-7$_{3,5}$& 
345519.7& ~~63.7& $2.59\times10^{-4}$& 
14.84$\pm$0.63& 11.70$\pm$0.24& 11.80$\pm$0.62& 1.18$\pm$0.13\\

\textbf{$^{34}$SO$_2$}& 6$_{4,2}$-6$_{3,3}$& 
345553.1& ~~57.3& $2.35\times10^{-4}$& 
11.10$\pm$2.13& 12.08$\pm$0.96& 10.04$\pm$2.15& 1.04$\pm$0.13\\

\textbf{$^{33}$SO$_2$}& $19_{1,19}-18_{0,18}$& 
345584.7& 170.2& $5.19\times10^{-4}$& 
4.22$\pm$0.60& 12.85$\pm$0.53& 7.71$\pm$1.28& 0.51$\pm$0.11\\

\textbf{HC$_3$N}& 38-37& 
345609.0& 323.5& $3.30\times10^{-3}$& 
6.83$\pm$0.41& 12.06$\pm$0.35& 13.19$\pm$1.07& 0.49$\pm$0.09\\

\textbf{$^{34}$SO$_2$}& 5$_{4,2}$-5$_{3,3}$& 
345651.3& ~~51.8& $1.96\times10^{-4}$& 
6.12$\pm$0.78& 11.99$\pm$0.51& 7.46$\pm$0.98& 0.77$\pm$0.11\\

\textbf{$^{34}$SO$_2$}& $4_{4,0}-4_{3,1}$& 
345678.8& ~~47.2& $1.31\times10^{-4}$& 
3.38$\pm$0.39& 11.30$\pm$0.61& 10.38$\pm$1.30& 0.31$\pm$0.09\\

\textbf{CO$^\star$}& 3-2& 
345796.0& ~~33.2& $2.50\times10^{-6}$& 
1161.6$\pm$4.43& 2.77$\pm$0.001& 21.22$\pm$0.10& 51.42$\pm$7.3\\

\textbf{$^{34}$SO$_2$}& 17$_{4,14}$-17$_{3,15}$& 
345929.3& 178.8& $3.37\times10^{-4}$& 
9.96$\pm$0.53& 12.02$\pm$0.32& 12.24$\pm$0.77& 0.76$\pm$0.09\\

\textbf{NS$^\star$}& 15/2-13/2& 
346220.1& ~~71.0& $7.38\times10^{-4}$& 
7.35$\pm$0.50& 10.84$\pm$0.68& 20.31$\pm$1.54& 0.34$\pm$0.06\\

\textbf{SO$_2$, v$_2$=1}& $19_{1,19}-18_{0,18}$& 
346379.2& 930.6& $5.16\times10^{-4}$&
9.82$\pm$0.41& 9.03$\pm$0.38& 18.64$\pm$0.91& 0.49$\pm$0.09\\


\textbf{HC$_3$N, $v_7$=1}& $J$=38-37&
346455.7& 645.1& $3.31\times10^{-3}$&
2.10$\pm$0.34& 13.99$\pm$0.96& 9.77$\pm$2.07& 0.20$\pm$0.07\\

\textbf{SO$^\star$}& 9$_8$-8$_7$& 
346528.5& ~~78.8& $5.38\times10^{-4}$& 
231.50$\pm$0.69& 7.63$\pm$0.02& 16.69$\pm$0.06& 13.03$\pm$0.68\\

\textbf{$^{33}$SO$_2$}& 5$_{3,3}$-4$_{2,2}$& 
346589.8& ~~36.0& $1.76\times10^{-5}$& 
6.41$\pm$0.33& 10.91$\pm$0.30& 11.76$\pm$0.71& 0.51$\pm$0.06\\

\textbf{SO$_2$~$^\star$}& 19$_{1,19}$-18$_{0,18}$& 
346652.2& 168.1& $5.22\times10^{-4}$& 
96.43$\pm$0.32& 9.42$\pm$0.03& 15.47$\pm$0.06& 5.85$\pm$0.14\\

\textbf{HC$_3$N $v_7$=1}& $J$=38-37&
346949.1& 645.6& $3.32\times10^{-3}$&
1.55$\pm$0.30& 13.93$\pm$0.73& 6.88$\pm$1.66& 0.21$\pm$0.09\\

\textbf{H$^{13}$CO$^+$~$^\star$}& 4-3& 
346998.3& ~~41.6& $3.29\times10^{-3}$& 
50.66$\pm$0.41& 6.58$\pm$0.05& 11.86$\pm$0.12& 4.01$\pm$0.19\\

\textbf{SiO}& 8-7& 
347330.6& ~~75.0& $2.20\times10^{-3}$& 
68.67$\pm$0.66& 7.91$\pm$0.08& 16.98$\pm$0.20& 3.80$\pm$0.16\\

\textbf{$^{34}$SO$_2$}& 28$_{2,26}$-28$_{1,27}$& 
347483.1& 391.2& $2.65\times10^{-4}$& 
5.01$\pm$0.37& 11.48$\pm$0.45& 11.89$\pm$1.03& 0.40$\pm$0.07\\

\textbf{SO$^+$}& 8-7& 
347740& ~~70.1& $2.28\times10^{-4}$&
19.44$\pm$0.41& 8.83$\pm$0.15& 13.87$\pm$0.33& 1.32$\pm$0.09\\

\textbf{SO$_2$, v$_2$=1}& $13_{2,12}-12_{1,11}$& 
347991.8& 854.3& $2.41\times10^{-4}$&
4.48$\pm$0.47& 9.98$\pm$0.88& 16.43$\pm$1.81& 0.26$\pm$0.08\\

\textbf{$^{34}$SO$_2$}& 19$_{4,16}$-19$_{3,17}$& 
348117.5& 212.9& $3.50\times10^{-4}$& 
28.62$\pm$0.38& 11.22$\pm$0.09& 14.24$\pm$0.22& 1.89$\pm$0.10\\

\textbf{HC$^{17}$O$^+$}& 4-3& 
348211.2& ~~41.8& $3.32\times10^{-3}$& 
1.64$\pm$0.33& 8.75$\pm$1.12& 10.73$\pm$2.59& 0.14$\pm$0.05\\

\textbf{HN$^{13}$C}& 4-3& 
348340.8& ~~41.8& $2.03\times10^{-3}$& 
5.66$\pm$1.05& 11.00$\pm$1.00& 11.88$\pm$1.00& 0.45$\pm$0.11\\

\textbf{SO$_2$~$^\star$}& 24$_{2,22}$-23$_{3,21}$& 
348387.8& 292.7& $1.91\times10^{-4}$& 
48.45$\pm$1.14& 10.12$\pm$1.00& 13.99$\pm$1.00& 3.25$\pm$0.10\\

\textbf{$^{33}$SO$_2$}& 12$_{4,8}$-12$_{3,9}$& 
348490.9& 111.9& $2.96\times10^{-6}$& 
6.28$\pm$0.78& 8.90$\pm$1.17& 18.08$\pm$3.31& 0.33$\pm$0.10\\

\textbf{o-H$_2$CS$^\star$}& 10$_{1,9}$-9$_{1,8}$& 
348534.4& 105.2& $6.32\times10^{-4}$& 
13.13$\pm$0.62& 9.32$\pm$0.36& 15.55$\pm$0.86& 0.79$\pm$0.10\\

\textbf{C$_2$H$^\star$}& 4$_{9/2}$-3$_{5/2}$, $F$=4-3& 
349108.5& ~~41.9& $5.76\times10^{-8}$& 
9.05$\pm$0.46& 9.59$\pm$0.37& 14.35$\pm$0.80& 0.59$\pm$0.08\\

\textbf{SO$^{18}$O}& 15$_{4,12}$-15$_{3,13}$& 
349209.6& 149.8& $3.37\times10^{-4}$& 
2.84$\pm$0.62& 10.65$\pm$1.58& 12.26$\pm$3.04& 0.22$\pm$0.09\\

\textbf{SO$^{18}$O}& 13$_{4,10}$-13$_{3,11}$& 
349224.9& 123.3& $3.27\times10^{-4}$& 
2.44$\pm$0.53& 11.95$\pm$0.28& 4.76$\pm$1.27& 0.48$\pm$0.09\\

\textbf{C$_2$H$^\star$}& 4$_{9/2}$-3$_{7/2}$& 
349338.3& ~~41.9& $1.28\times10^{-4}$& 
33.15$\pm$0.46& 8.49$\pm$0.10& 13.97$\pm$0.22& 2.23$\pm$0.17\\

\textbf{C$_2$H$^\star$}& 4$_{7/2}$-3$_{5/2}$& 
349400.7& ~~41.9& $1.20\times10^{-4}$& 
28.34$\pm$1.14& 10.32$\pm$0.24& 14.04$\pm$0.46& 1.90$\pm$0.10\\

\textbf{CH$_3$CN$^\star$}& 19$_2$-18$_2$& 
349426.8& 196.3& $3.68\times10^{-3}$& 
5.91$\pm$3.13& 10.57$\pm$2.57& 19.45$\pm$12.17& 0.29$\pm$0.10\\

\textbf{CH$_3$CN$^\star$}& 19$_0$-18$_0$& 
349453.7& 167.7& $3.72\times10^{-3}$& 
9.74$\pm$1.63& 12.05$\pm$1.12& 15.03$\pm$1.74& 0.61$\pm$0.10\\

\textbf{$^{33}$SO$_2$}& 10$_{4,6}$-10$_{3,7}$& 
350302.4& ~~90.4& $4.02\times10^{-6}$& 
2.87$\pm$0.34& 10.39$\pm$0.73& 11.41$\pm$1.29& 0.24$\pm$0.08\\

\textbf{HNCO$^\star$}& 16$_{1,16}$-15$_{1,15}$& 
350333.1& 186.2& $5.97\times10^{-4}$& 
5.82$\pm$0.32& 12.30$\pm$0.26& 9.89$\pm$0.64& 0.55$\pm$0.08\\

\textbf{NO$^\star$}& 7/2-5/2$^+$& 
350694.8& ~~36.1& $4.81\times10^{-6}$& 
35.27$\pm$0.44& 12.03$\pm$0.09& 14.78$\pm$0.21& 2.24$\pm$0.06\\

\textbf{$^{33}$SO$_2$}& 13$_{4,10}$-13$_{3,11}$& 
350787.0& 124.0& $2.64\times10^{-6}$& 
2.84$\pm$0.36& 11.42$\pm$0.89& 14.02$\pm$2.08& 0.19$\pm$0.08\\

\textbf{SO$_2$~$^\star$}& 10$_{6,4}$-11$_{5,7}$& 
350862.8& 138.8& $4.40\times10^{-5}$& 
24.44$\pm$0.51& 11.46$\pm$0.13& 12.58$\pm$0.30& 1.83$\pm$0.12\\

\textbf{CH$_3$OH$^\star$}& 1$_1$-0$_0$ $A^+$& 
350905.1& ~~16.8& $3.31\times10^{-4}$& 
28.53$\pm$0.52& 5.82$\pm$0.11& 12.22$\pm$0.28& 2.19$\pm$0.24\\

\textbf{SO$^{17}$O}& 12$_{4,8}$-12$_{3,9}$& 
350993.2& 112.1& $3.27\times10^{-4}$& 
5.55$\pm$0.54& 11.15$\pm$0.46& 9.97$\pm$1.17& 0.52$\pm$0.12\\

\textbf{NO$^\star$}& 7/2-5/2$^-$& 
351049.0& ~~36.1& $4.99\times10^{-6}$& 
35.03$\pm$0.61& 9.79$\pm$0.14& 16.80$\pm$0.34& 1.96$\pm$0.13\\

\textbf{$^{33}$SO$_2$}& 8$_{4,4}$-8$_{3,5}$& 
351177.4& ~~72.7& $5.65\times10^{-6}$& 
4.14$\pm$0.43& 12.48$\pm$0.38& 8.01$\pm$1.05& 0.49$\pm$0.09\\

\textbf{SO$_2$~$^\star$}& 5$_{3,3}$-4$_{2,2}$& 
351257.2& ~~~~35.9& $3.36\times10^{-4}$& 
42.96$\pm$0.59& 9.27$\pm$0.11& 16.27$\pm$0.28& 2.48$\pm$0.13\\

\textbf{$^{33}$SO$_2$}& 7$_{4,4}$-7$_{3,5}$& 
351508.5& ~~~~65.2& $6.76\times10^{-6}$& 
3.09$\pm$0.34& 11.35$\pm$0.47& 8.20$\pm$0.96& 0.35$\pm$0.09\\

\textbf{$^{33}$SO$_2$}& 6$_{4,2}$-6$_{3,3}$& 
351542.1& ~~~~58.7& $8.06\times10^{-6}$& 
8.37$\pm$0.55& 13.22$\pm$0.61& 18.63$\pm$1.30& 0.42$\pm$0.09\\

\textbf{HNCO}& $16_{0,16}-15_{0,15}$& 
351633.3& ~~143.5& $6.13\times10^{-4}$&
20.41$\pm$0.79& 12.36$\pm$0.24& 12.78$\pm$0.57& 1.50$\pm$0.12\\

\textbf{$^{33}$SO$_2$}& 4$_{4,0}$-4$_{3,1}$& 
351661.8& ~~~~48.5& $1.18\times10^{-5}$& 
3.39$\pm$0.53& 9.73$\pm$0.85& 10.67$\pm$1.90& 0.30$\pm$0.11\\

\textbf{$^{33}$SO$_2$}& 17$_{4,14}$-17$_{3,15}$& 
351743.7& ~~181.9& $1.78\times10^{-6}$& 
5.17$\pm$0.53& 13.39$\pm$0.56& 11.22$\pm$1.08& 0.43$\pm$0.18\\

\textbf{o-H$_2$CO$^\star$}& 5$_{1,5}$-4$_{1,4}$& 
351768.6& ~~~~62.5& $1.20\times10^{-3}$& 
104.71$\pm$1.35& 6.54$\pm$0.08& 13.75$\pm$0.22& 7.15$\pm$0.65\\

\textbf{SO$_2$~$^\star$}& 14$_{4,10}$-14$_{3,11}$& 
351873.9& ~~135.9& $3.43\times10^{-4}$& 
63.49$\pm$0.71& 10.17$\pm$0.08& 14.94$\pm$0.19& 3.99$\pm$0.17\\

\textbf{$^{34}$SO$_2$}& 21$_{4,18}$-21$_{3,19}$& 
352082.9& ~~250.8& $3.65\times10^{-4}$& 
8.53$\pm$0.59& 11.65$\pm$0.37& 11.33$\pm$0.96& 0.71$\pm$0.07\\

\textbf{OCS$^\star$}& 29-28& 
352599.6& ~~253.9& $1.28\times10^{-4}$& 
11.23$\pm$0.33& 10.91$\pm$0.20& 13.54$\pm$0.44& 0.78$\pm$0.08\\

\textbf{HNCO$^\star$}& 16$_{1,15}$-15$_{1,14}$& 
352897.6& ~~187.2& $6.10\times10^{-4}$& 
6.28$\pm$0.33& 11.71$\pm$0.27& 10.12$\pm$0.63& 0.58$\pm$0.07\\

\textbf{$^{34}$SO$_2$}& 14$_{7,7}$-15$_{6,10}$& 
353002.4& ~~212.6& $5.70\times10^{-5}$& 
1.18$\pm$0.20& 11.24$\pm$0.98& 10.59$\pm$1.51& 0.11$\pm$0.05\\

\textbf{SO$^{18}$O}& 26$_{3,24}$-26$_{2,25}$& 
353195.8& ~~342.7& $2.94\times10^{-4}$& 
1.59$\pm$0.25& 12.54$\pm$0.89& 11.07$\pm$1.87& 0.13$\pm$0.03\\

\textbf{H}& 26$\alpha$&
353622.8& & &
18.97$\pm$0.46& 11.63$\pm$0.44& 37.31$\pm$1.07& 0.48$\pm$0.06\\

\textbf{CO$^+$}\footnote{May be blended with SO$^{17}$O}& $N$=3-2, $F$=$\frac{5}{2}$-$\frac{3}{2}$& 
353741.3& ~~~~33.9& $2.06\times10^{-4}$&
5.55$\pm$0.45& 12.11$\pm$0.82& 21.09$\pm$2.06& 0.25$\pm$0.06\\  
     
\textbf{HCN,v$_2$=1$^{\rm{blend/1}}$~$^\blacklozenge$}& 4-3&
354460.4& 1066.9& $1.87\times10^{-3}$& 
9.78$\pm$0.95& 12.50$\pm$1.0& 15.00$\pm$1.00& 0.62$\pm$0.06\\

\textbf{HCN$^{\rm{blend/2}}$~$^{\star \blacklozenge}$}& 4-3& 
354505.5& ~~~~42.5& $2.05\times10^{-3}$& 
184.9$\pm$1.30& 1.32$\pm$1.0& 22.00$\pm$2.00& 8.55$\pm$0.06\\

\textbf{SO$_2$, v$_2$=1}& $16_{4,12}-16_{3,13}$& 
354800& ~~927.8& $3.60\times10^{-4}$&
3.16$\pm$0.32& 10.73$\pm$0.91& 15.39$\pm$1.89& 0.19$\pm$0.08\\

\textbf{SO$_2$~$^\star$}& $12_{4,8}-12_{3,9}$& 
355045.5& ~~111.0& $3.40\times10^{-4}$& 
49.46$\pm$0.32& 9.74$\pm$0.05& 15.04$\pm$0.12& 3.09$\pm$0.09\\

\textbf{SO$_2$}& $17_{4,14}-18_{1,17}$& 
355186.5& ~~180.1& $2.62\times10^{-6}$& 
5.16$\pm$0.36& 9.08$\pm$0.57& 16.41$\pm$1.47& 0.30$\pm$0.05\\

\textbf{H$^{15}$NC}& $4-3$&
355439.5& ~~~~42.6& $1.69\times10^{-3}$&
1.25$\pm$0.23& 10.68$\pm$0.83& 8.36$\pm$1.71& 0.14$\pm$0.05\\

\textbf{S$^{18}$O}& $8_9-7_8$& 
355571.1& ~~~~93.1& $5.74\times10^{-4}$& 
8.35$\pm$0.26& 11.21$\pm$0.24& 16.00$\pm$0.58& 0.49$\pm$0.06\\

\textbf{CH$_3$OH$^\star$}& $13_{0,13}$-$12_{1,12}$ A$^+$& 
355603.1& ~~211.0& $2.53\times10^{-4}$& 
9.52$\pm$0.27& 9.61$\pm$0.22& 15.04$\pm$0.47& 0.59$\pm$0.06\\

\textbf{CH$_3$OH$^\star$}& $15_{1,14}$-$15_{0,15}$ A$^+$& 
356007.2& ~~295.3& $4.60\times10^{-4}$& 
7.87$\pm$0.62& 8.70$\pm$0.58& 14.19$\pm$1.13& 0.52$\pm$0.10\\

\textbf{SO$_2$}& $15_{7,9}-16_{6,10}$& 
356040.6& ~~230.4& $6.40\times10^{-5}$& 
23.58$\pm$0.67& 10.51$\pm$0.19& 13.27$\pm$0.45& 1.67$\pm$0.10\\

\textbf{$^{34}$SO$_2$}& $25_{3,23}-25_{2,24}$& 
356222.2& ~~320.0& $2.97\times10^{-4}$& 
6.39$\pm$0.51& 10.72$\pm$0.45& 10.87$\pm$0.92& 0.55$\pm$0.08\\

\textbf{HCN, $v_2=1$}& $J=4-3$&
356255.6& 1067.12& $5.33\times10^{-3}$&
11.42$\pm$0.71& 12.20$\pm$0.57& 20.00$\pm$1.61& 0.54$\pm$0.07\\

\textbf{HCO$^+$~$^\star$}& 4-3& 
356734.2& ~~~~42.8& $3.57\times10^{-3}$& 
503.55$\pm$0.55& 3.10$\pm$0.01& 27.35$\pm$0.03& 17.30$\pm$1.30\\

\textbf{$^{34}$SO$_2$~$^\star$}& $20_{0,20}-19_{1,19}$& 
357102.2& ~~184.8& $5.81\times10^{-4}$& 
25.19$\pm$0.53& 11.25$\pm$0.14& 13.36$\pm$0.33& 1.77$\pm$0.14\\

\textbf{SO$_2$~$^\star$}& $13_{4,10}-13_{3,11}$& 
357165.4& ~~123.0& $3.51\times10^{-4}$& 
72.28$\pm$0.41& 9.24$\pm$0.04& 15.20$\pm$0.10& 4.47$\pm$0.15\\

\textbf{SO$_2$~$^\star$}& $15_{4,12}-15_{3,13}$& 
357241.2& ~~149.7& $3.62\times10^{-4}$& 
67.49$\pm$0.68& 9.24$\pm$0.07& 14.70$\pm$0.17& 4.31$\pm$0.14\\

\textbf{SO$_2$~$^\star$}& $11_{4,8}-11_{3,9}$& 
357387.6& ~~100.0& $3.38\times10^{-4}$& 
71.00$\pm$0.84& 9.85$\pm$0.09& 15.00$\pm$0.21& 4.45$\pm$0.13\\

\textbf{$^{34}$SO$_2$}& $32_{5,27}-32_{4,28}$& 
357497.8& ~~547.5& $4.55\times10^{-4}$& 
2.34$\pm$0.24& 12.21$\pm$0.47& 9.23$\pm$1.22& 0.24$\pm$0.05\\

\textbf{SO$_2$~$^\star$}& $8_{4,4}-8_{3,5}$& 
357581.4& ~~~~72.4& $3.06\times10^{-4}$& 
81.60$\pm$0.61& 9.32$\pm$0.06& 17.96$\pm$0.16& 4.27$\pm$0.24\\

\textbf{SO$_2$~$^\star$}& $9_{4,6}-9_{3,7}$& 
357671.8& ~~~~80.6& $3.20\times10^{-4}$& 
86.02$\pm$0.30& 9.88$\pm$0.03& 16.93$\pm$0.07& 4.77$\pm$0.23\\

\textbf{SO$_2$}$^{\rm{blend/1}}$~$^\star$& $7_{4,4}-7_{3,5}$& 
357892.4& ~~~~65.0& $2.87\times10^{-4}$& 
78.52$\pm$2.36& 9.36$\pm$1.00& 15.65$\pm$1.00& 4.71$\pm$0.15\\

\textbf{SO$_2$}$^{\rm{blend/2}}$~$^\star$& $6_{4,2}-6_{3,3}$& 
357925.8& ~~~~58.6& $2.60\times10^{-4}$& 
74.40$\pm$0.36& 9.47$\pm$0.04& 15.64$\pm$0.09& 4.47$\pm$0.14\\

\textbf{SO$_2$}$^{\rm{blend/3}}$& $17_{4,14}-17_{3,15}$& 
357962.9& ~~180.1& $3.73\times10^{-4}$& 
69.61$\pm$0.38& 10.06$\pm$0.04& 15.86$\pm$0.10& 4.12$\pm$0.14\\

\textbf{SO$_2$}$^{\rm{blend/4}}$~$^\star$& $5_{4,2}-5_{3,3}$& 
358013.2& ~~~53.1& $2.18\times10^{-4}$& 
68.11$\pm$0.36& 9.71$\pm$0.01& 15.47$\pm$0.10& 4.14$\pm$0.14\\

\textbf{SO$_2$}$^{\rm{blend/5}}$~$^\star$& $4_{4,0}-4_{3,1}$& 
358037.9& ~~~48.5& $1.45\times10^{-4}$& 
48.82$\pm$2.39& 9.68$\pm$1.00& 13.32$\pm$1.00& 3.44$\pm$0.14\\

\textbf{SO$_2$~$^\star$}& $20_{0,20}-19_{1,19}$& 
358215.6& 185.3& $5.83\times10^{-4}$& 
102.21$\pm$0.56& 9.29$\pm$0.04& 15.47$\pm$0.10& 6.21$\pm$0.17\\

\textbf{$^{34}$SO$_2$}& $23_{4,20}-23_{3,21}$& 
358347.3& 292.4& $3.86\times10^{-4}$& 
9.32$\pm$0.30& 11.67$\pm$0.20& 12.70$\pm$0.48& 0.69$\pm$0.08\\

\textbf{CH$_3$OCH$_3$}& $5_{5,1,1}-4_{4,1,1}$& 
358454.0& ~~48.8& $2.94\times10^{-4}$& 
3.32$\pm$0.30& 8.56$\pm$0.68& 14.95$\pm$1.40& 0.21$\pm$0.05\\

\textbf{CH$_3$OH$^\star$}& $4_1-3_0$ $E$& 
358605.8& ~~44.3& $1.32\times10^{-4}$& 
24.58$\pm$1.35& 7.66$\pm$0.38& 13.84$\pm$0.83& 1.67$\pm$0.14\\ 

\textbf{S$^{18}$O}& $9_9-8_8$& 
358645.7& ~~99.3& $5.92\times10^{-4}$& 
9.23$\pm$3.28& 9.38$\pm$2.27& 12.81$\pm$5.15& 0.68$\pm$0.05\\

\textbf{CH$_3$CCH}$^{\rm{blend/1}}$& $21_3-20_3$& 
358756.5& 254.5& $1.58\times10^{-4}$& 
3.63$\pm$0.18& 9.34$\pm$0.30& 12.03$\pm$0.66& 0.28$\pm$0.04\\

\textbf{CH$_3$CCH}$^{\rm{blend/2}}$& $21_2-20_2$& 
358790.6& 218.3& $1.60\times10^{-4}$& 
2.75$\pm$0.28& 8.37$\pm$0.67& 12.40$\pm$1.45& 0.21$\pm$0.06\\

\textbf{CH$_3$CCH}$^{\rm{blend/3}}$& $21_0-20_0$& 
358817.9& 189.5& $1.61\times10^{-4}$&  
7.92$\pm$0.43& 10.53$\pm$0.36& 14.48$\pm$0.93& 0.51$\pm$0.06\\

\textbf{SO$^{17}$O}\footnote{Probably blended with $^{34}$SO$_2$ $15_{2,14}-14_{1,13}$ at 358988 MHz}& $15_{2,14}$-$14_{1,13}$&
358986.2& 119.9& $2.84\times10^{-4}$& 
20.64$\pm$0.25& 9.95$\pm$0.08& 13.18$\pm$0.19& 1.47$\pm$0.10\\

\textbf{SO$_2$~$^\star$}& $25_{3,23}-25_{2,24}$&
359151.2& 320.9& $3.10\times10^{-4}$& 
42.63$\pm$0.21& 10.57$\pm$0.04& 14.47$\pm$0.09& 2.77$\pm$0.08\\

\textbf{$^{34}$SO$_2$}& $24_{2,22}-23_{3,21}$& 
359651.7& 292.4& $2.20\times10^{-4}$& 
6.76$\pm$0.34& 12.02$\pm$0.26& 11.01$\pm$0.71& 0.58$\pm$0.07\\

\textbf{SO$_2$~$^\star$}& $19_{4,16}-19_{3,17}$&
359770.7& 214.3& $3.85\times10^{-4}$& 
58.89$\pm$0.28& 10.13$\pm$0.03& 15.16$\pm$0.08& 3.65$\pm$0.10\\

\textbf{SO$_2$, $v_2=1$}& $14_{4,10}-14_{3,11}$&
360133.2& 898.7& $1.43\times10^{-4}$&
4.42$\pm$0.37& 11.53$\pm$0.58& 13.44$\pm$1.23& 0.31$\pm$0.06\\
 
\textbf{SO$_2$}& $34_{5,29}-34_{4,30}$&
360290.4& 612.0& $4.80\times10^{-4}$&
18.63$\pm$0.32& 11.14$\pm$0.11& 13.93$\pm$0.28& 1.26$\pm$0.06\\

\textbf{S$^{18}$O}& $10_9-9_8$& 
360637.9& ~~90.6& $6.10\times10^{-4}$& 
11.30$\pm$0.52& 9.65$\pm$0.32& 13.98$\pm$0.71& 0.76$\pm$0.09\\

\textbf{SO$_2$~$^\star$}& $20_{8,12}-21_{7,15}$&
360721.8& 349.8& $7.72\times10^{-5}$& 
15.07$\pm$0.55& 11.51$\pm$0.22& 12.59$\pm$0.56& 1.12$\pm$0.12\\

\textbf{CH$_3$OH$^\star$}& $11_{0,11}$-$10_{1,9}$ $E$& 
360848.9& 166.0& $1.21\times10^{-4}$& 
8.12$\pm$0.44& 8.42$\pm$0.42& 15.26$\pm$0.87& 0.50$\pm$0.08\\
 
\textbf{NO}& $J$=7/2-5/2&
360948.3& 209.4& $4.52\times10^{-6}$& 
15.30$\pm$0.55& 11.31$\pm$0.52& 29.18$\pm$1.19& 0.49$\pm$0.06\\

\textbf{CH$_3$OH$^\star$}& $8_1$-$7_2$ $E$&
361852.3& 104.6& $7.70\times10^{-5}$&
11.25$\pm$0.38& 6.58$\pm$0.27& 16.60$\pm$0.68& 0.64$\pm$0.07\\

\textbf{DCN$^\star$}& 5-4&
362045.8& ~~52.1& $2.25\times10^{-3}$&
2.26$\pm$0.33& 6.99$\pm$0.75& 9.17$\pm$1.22& 0.23$\pm$0.09\\

\textbf{$^{34}$SO$_2$~$^\star$}& $6_{3,3}-5_{2,4}$&
362158.2& ~~40.7& $3.29\times10^{-4}$&
19.93$\pm$0.27& 11.17$\pm$0.08& 12.88$\pm$0.21& 1.45$\pm$0.09\\

\textbf{$^{33}$SO$_2$}& $15_{2,14}-14_{1,13}$&
362487.6& 120.8& $2.98\times10^{-4}$& 
6.11$\pm$0.25& 10.16$\pm$0.26& 13.16$\pm$0.61& 0.44$\pm$0.04\\

\textbf{HNC, v$_2$=1}& 4-3& 
362554.4& 709.3& $2.15\times10^{-3}$& 
4.20$\pm$0.29& 12.65$\pm$0.35& 11.39$\pm$1.06& 0.35$\pm$0.05\\

\textbf{HNC$^\star$}& 4-3&
362630.3& ~~43.5& $2.30\times10^{-3}$&
52.96$\pm$0.56& 5.72$\pm$0.06& 12.93$\pm$0.18& 3.85$\pm$0.42\\

\textbf{p-H$_2$CO$^\star$}& $5_{0,5}-4_{0,4}$&
362736& ~~52.3& $1.37\times10^{-3}$&
60.08$\pm$0.26& 6.91$\pm$0.03& 12.45$\pm$0.06& 4.53$\pm$0.25\\

\textbf{$^{34}$SO$_2$}& $23_{2,22}-23_{1,23}$&
362834.1& 259.2& $1.77\times10^{-4}$& 
6.75$\pm$0.31& 11.91$\pm$0.26& 11.59$\pm$0.65& 0.55$\pm$0.06\\

\textbf{SO$_2$~$^\star$}& $21_{4,18}-21_{3,19}$&
363159.3& 252.1& $4.00\times10^{-4}$&
50.66$\pm$0.33& 10.33$\pm$0.05& 15.12$\pm$0.12& 3.15$\pm$0.10\\

\textbf{$^{33}$SO$_2$}& $23_{4,20}-23_{3,21}$&
363286.4& 297.1& $4.01\times10^{-4}$&
6.25$\pm$0.35& 9.08$\pm$0.35& 13.79$\pm$1.02& 0.43$\pm$0.07\\

\textbf{CH$_3$OH$^\star$}& $16_1-16_0$ $A^\mp$&
363440.4& 332.6& $2.40\times10^{-4}$&
6.63$\pm$0.39& 9.74$\pm$0.46& 15.26$\pm$0.94& 0.41$\pm$0.06\\

\textbf{CH$_3$OH$^\star$}& $7_2-6_1$ $E$&
363739.8& ~~87.3& $1.70\times10^{-4}$&
17.91$\pm$0.26& 8.36$\pm$0.10& 13.80$\pm$0.22& 1.22$\pm$0.07\\

\textbf{HC$_3$N}& 40-39&
363785.4& 358.0& $3.85\times10^{-3}$&
3.67$\pm$0.22& 12.39$\pm$0.28& 9.71$\pm$0.70& 0.35$\pm$0.05\\

\textbf{SO$_2$}& $24_{1,23}-24_{0,24}$&
363890.9& 280.5& $1.78\times10^{-4}$& 
23.45$\pm$0.24& 10.42$\pm$0.07& 14.59$\pm$0.18& 1.51$\pm$0.07\\

\textbf{SO$_2$~$^\star$}& $23_{2,22}-23_{1,23}$& 
363925.8& 259.9& $1.83\times10^{-4}$&
27.90$\pm$0.34& 10.33$\pm$0.08& 15.27$\pm$0.22& 1.72$\pm$0.07\\

\textbf{p-H$_2$CO$^\star$}& $5_{2,4}-4_{2,3}$&
363945.9& ~~99.5& $1.16\times10^{-3}$&
25.65$\pm$0.28& 7.73$\pm$0.07& 12.59$\pm$0.15& 1.91$\pm$0.07\\
 
\textbf{p-H$_2$CO}& $5_{4,2}-4_{4,1}$&
364103.2& 240.7& $4.99\times10^{-4}$&
11.15$\pm$0.28& 8.12$\pm$0.17& 13.31$\pm$0.39& 0.79$\pm$0.06\\

\textbf{o-H$_2$CO$^\star$}& $5_{3,3}-4_{3,2}$&
364275.1& 158.4& $8.88\times10^{-4}$&
39.52$\pm$0.32& 7.79$\pm$0.07& 12.00& 3.09$\pm$0.14\\

\textbf{o-H$_2$CO$^\star$}& $5_{3,2}-4_{3,1}$&
364288.9& 158.4& $8.88\times10^{-4}$&
39.70$\pm$0.38& 7.22$\pm$0.08& 12.00& 3.11$\pm$0.14\\

\textbf{Atmospheric}& & 
364403.7\\

\textbf{HC$_3$N v$_7$=1}& $J$=40-39&
364676.3& 679.7& $3.86\times10^{-3}$&
3.05$\pm$0.37& 15.90$\pm$1.10& 18.08$\pm$2.55& 0.16$\pm$0.06\\

\textbf{OCS$^\star$}& $30-29$&
364749& 271.4& $1.42\times10^{-4}$&
8.45$\pm$0.31& 9.89$\pm$0.27& 14.07$\pm$0.56& 0.56$\pm$0.07\\

\textbf{H$_3$O$^+$}& $3_{2,1}-2_{2,0}$&
364797.4& 139.7& $2.78\times10^{-4}$&
17.88$\pm$0.35& 8.58$\pm$0.14& 14.41$\pm$0.30& 1.17$\pm$0.09\\

\textbf{SO$_2$}& $25_{9,17}-26_{8,18}$& 
364950.1& 497.1& $8.70\times10^{-5}$&
10.11$\pm$0.26& 12.12$\pm$0.14& 11.08$\pm$0.34& 0.86$\pm$0.08\\

\textbf{CH$_3$OH, $v_t$=1}& $5_3-5_2$ $E$&
364986.8& 452.1& $3.76\times10^{-5}$&
3.59$\pm$0.24& 6.02$\pm$0.36& 10.52$\pm$0.80& 0.32$\pm$0.06\\
 
\textbf{HNC, v$_2$=1}& 4-3& 
365147.5& 709.6& $2.20\times10^{-3}$&
3.79$\pm$0.60& 13.27$\pm$0.21& 7.70$\pm$1.42& 0.46$\pm$0.07\\ 

\textbf{HC$_3$N $v_7$=1}& &
365195.2& 680.2& $3.88\times10^{-3}$&
3.00$\pm$0.28& 13.77$\pm$0.52& 10.33$\pm$1.05& 0.27$\pm$0.09\\

\textbf{p-H$_2$CO$^\star$}& $5_{2,3}-4_{2,2}$&
365363.4& ~~99.7& $1.18\times10^{-3}$&
44.27$\pm$0.74& 7.73$\pm$0.12& 14.23$\pm$0.28& 2.92$\pm$0.16\\

\textbf{H$_2$CN}& $5_{0,5}-4_{0,4}$&
365443.6& ~~52.7& $4.15\times10^{-5}$&
2.98$\pm$0.31& 10.05$\pm$0.49& 9.13$\pm$1.07& 0.31$\pm$0.06\\

\textbf{H$_2$C$^{34}$S}& $11_{1,11}-10_{1,10}$& 
365613.4& 118.5& $7.33\times10^{-4}$&
4.32$\pm$0.37& 9.25$\pm$0.51& 11.83$\pm$1.25& 0.34$\pm$0.07\\

\textbf{CH$_3$COCH$_3$}& $27_{11,16,0}$-$26_{12,15,1}$& 
365696.2& ~~281.7& $1.20\times10^{-3}$&
2.97$\pm$0.41& 9.46$\pm$0.91& 13.63$\pm$2.40& 0.20$\pm$0.06\\

\textbf{$^{34}$SO$_2$}& $24_{1,23}-24_{0,24}$&
365794.6& ~~279.9& $1.75\times10^{-4}$&
4.56$\pm$0.45& 13.44$\pm$0.26& 7.54$\pm$0.68& 0.57$\pm$0.07\\

\textbf{U-line}& & 
365811.5& & &
6.31$\pm$0.84& 10.91$\pm$1.14& 19.03$\pm$3.20& 0.31$\pm$0.07\\

\textbf{SO$_2$, v$_2$=1}& $9_{4,6}-9_{3,7}$&
365904.5& ~~842.5& $3.39\times10^{-4}$&
3.42$\pm$0.53& 11.84$\pm$0.91& 11.60$\pm$2.68& 0.28$\pm$0.06\\

\textbf{SO$_2$, v$_2$=1$^{\rm{blend}/1}$}& $7_{4,4}-7_{3,5}$&
366125.8& ~~826.6& $3.05\times10^{-4}$&
3.04$\pm$0.36& 12.82$\pm$0.71& 10.00& 0.29$\pm$0.07\\

\textbf{SO$_2$, v$_2$=1$^{\rm{blend}/2}$}& $17_{4,14}-17_{3,15}$&
366145.1& ~~943.8& $3.95\times10^{-4}$&
4.28$\pm$0.32& 14.10$\pm$0.50& 10.00& 0.40$\pm$0.07\\

\textbf{SO$_2$, v$_2$=1$^{\rm{blend}/3}$}& $6_{4,2}-6_{3,3}$&
366159.5& ~~820.0& $2.76\times10^{-4}$&
3.68$\pm$0.34& 13.16$\pm$0.67& 10.00& 0.35$\pm$0.07\\

\textbf{SO$_2$~$^\star$}& $15_{2,14}-14_{1,13}$& 
366214.5& ~~119.3& $3.04\times10^{-4}$&
91.77$\pm$1.56& 9.43$\pm$0.13& 15.80$\pm$0.32& 5.46$\pm$0.15\\

\textbf{H$_2^{13}$CO}& $5_{1,4}-4_{1,3}$& 
366270.2& ~~~~64.6& $1.36\times10^{-3}$&
6.96$\pm$0.77& 10.14$\pm$1.06& 14.00& 0.47$\pm$0.08\\

\textbf{$^{33}$SO$_2$}& $6_{3,3}-5_{2,4}$& 
366521.9& ~~~~41.6& $1.31\times10^{-5}$&
7.52$\pm$0.33& 11.61$\pm$0.25& 11.79$\pm$0.64& 0.60$\pm$0.07\\

\textbf{H$_2$CN}& $5_{2,4}-4_{2,3}$&
366637& ~~100.4& $9.42\times10^{-5}$&
2.19$\pm$0.32& 9.06$\pm$1.04& 10.00& 0.21$\pm$0.06\\

\textbf{H$_2$CN}& $5_{2,4}-4_{2,3}$& 
366655.4& ~~100.4& $7.69\times10^{-5}$&
2.96$\pm$0.33& 8.36$\pm$0.80& 10.00& 0.28$\pm$0.06\\
 
\textbf{Atmospheric}& & 
366790.0\\

\textbf{$^{34}$SO$_2$}& $25_{4,22}-25_{3,23}$&
367369.3& ~~337.6& $4.14\times10^{-4}$&
8.97$\pm$0.37& 11.97$\pm$0.26& 12.69$\pm$0.60& 0.66$\pm$0.09\\

\textbf{o-H$_2$S$^{\rm{blend/1}}$~$^\star$}& $3_{2,1}-3_{1,2}$&
369101.4& ~~154.5& $1.90\times10^{-4}$&
145.64$\pm$0.92& 8.54$\pm$0.04& 15.67$\pm$0.12& 8.73$\pm$0.43\\

\textbf{p-H$_2$S$^{\rm{blend/2}}$}& $4_{3,1}-4_{2,2}$& 
369126.9& ~~262.8& $2.20\times10^{-4}$&
15.23$\pm$0.76& 9.69$\pm$0.24& 10.45$\pm$0.56& 1.37$\pm$0.43\\

\textbf{H$_2^{33}$S}& $3_{2,1}-3_{1,2}$&
369170.4& ~~154.4& $2.55\times10^{-5}$&
7.31$\pm$1.32& 11.26$\pm$1.40& 16.12$\pm$3.71& 0.43$\pm$0.13\\

\textbf{H$_2^{34}$S}& $3_{2,1}-3_{1,2}$&
369246.1& ~~154.3& $1.91\times10^{-4}$&
8.92$\pm$0.63& 11.15$\pm$0.42& 11.54$\pm$1.13& 0.73$\pm$0.15\\

\textbf{$^{13}$CS$^\star$}& $8-7$&
369908.6& ~~~~79.9& $1.06\times10^{-3}$&
11.87$\pm$0.45& 10.00$\pm$0.26& 14.02$\pm$0.59& 0.80$\pm$0.10\\

\textbf{SO$_2$~$^\star$}& $9_{6,4}-10_{5,5}$&
370108.6& ~~129.7& $4.19\times10^{-5}$& 
28.15$\pm$0.43& 11.41$\pm$0.09& 12.53$\pm$0.23& 2.11$\pm$0.12\\

\textbf{U-line}& &
370499.6& & &
4.27$\pm$0.58& 10.94$\pm$1.41& 22.64$\pm$4.0&  0.18$\pm$0.06\\

\textbf{$^{34}$SO}& $4_4-3_4$&
370931.7& ~~~~33.4& $1.45\times10^{-5}$& 
2.55$\pm$0.39& 9.63$\pm$1.24& 13.90$\pm$2.46& 0.17$\pm$0.08\\

\textbf{Atmospheric}& &
371036.0& \\

\textbf{SO$_2$~$^\star$}& $6_{3,3}-5_{2,4}$&
371172.5& ~~~~41.4& $3.55\times10^{-4}$&
81.33$\pm$0.57& 9.89$\pm$0.05& 15.66$\pm$0.13& 4.88$\pm$0.13\\

\textbf{SO$_2$, v$_2$=1}& $21_{4,18}-21_{3,19}$&
371264.8& 1017.1& $4.24\times10^{-4}$&
4.11$\pm$0.71& 11.59$\pm$1.25& 13.72$\pm$3.49& 0.28$\pm$0.08\\

\textbf{CH$_3$OH$^\star$}& $17_1-17_0$ A$^\mp$&
371415.7& ~~372.4& $2.52\times10^{-4}$&
5.04$\pm$0.41& 9.57$\pm$0.62& 14.45$\pm$1.14& 0.33$\pm$0.09\\

\textbf{$^{33}$SO$_2$}& $25_{4,22}-25_{3,23}$&
371804.6& ~~343.0& $1.35\times10^{-6}$&
1.89$\pm$0.34& 10.69$\pm$0.85& 8.79$\pm$1.51& 0.20$\pm$0.08\\

\textbf{o-H$_2$CS$^\star$}& $11_{1,11}-10_{1,10}$&
371847.4& ~~120.3& $7.72\times10^{-4}$&
10.58$\pm$0.28& 9.06$\pm$0.18& 13.18$\pm$0.36& 0.75$\pm$0.09\\

\textbf{S$^{17}$O}& $9_9-8_8$&
372113.6& ~~102.5& $6.33\times10^{-4}$&
3.06$\pm$0.48& 9.84$\pm$0.95& 11.42$\pm$1.78& 0.25$\pm$0.09\\

\textbf{HNCO}& $17_{1,17}-16_{1,16}$&
372221.0& ~~	204.1& $7.46\times10^{-4}$&
8.13$\pm$0.64& 11.96$\pm$0.35& 9.23$\pm$0.83& 0.83$\pm$0.08\\

\textbf{$^{34}$SO$_2$}& $13_{7,7}-14_{6,8}$&
372279.8& ~~199.7& $6.00\times10^{-5}$&
1.73$\pm$0.37& 11.74$\pm$0.50& 6.30$\pm$2.23& 0.26$\pm$0.05\\

\textbf{N$_2$H$^+$~$^\star$}& 4-3&
372672.5& ~~~~44.7& $3.33\times10^{-3}$&
54.14$\pm$0.33& 7.44$\pm$0.04& 12.82$\pm$0.09& 3.97$\pm$0.20\\

\textbf{U-line}\footnote{HC$_3$N 41-40 is probable, but the line intensity is not consistent with its other three detected transitions.}& & 
372869.7& & &
7.92$\pm$0.41& 9.33$\pm$0.26& 10.05$\pm$0.64& 0.74$\pm$0.11\\

\textbf{SO$^{18}$O}& $22_{0,22}-21_{1,21}$&
373110.1& ~~221.2& $6.69\times10^{-4}$&
4.18$\pm$0.70& 13.21$\pm$0.90& 10.93$\pm$2.26& 0.36$\pm$0.09\\

\textbf{SO$^\star$}& $4_4-3_4$&
373344.2& ~~~~33.8& $1.51\times10^{-5}$&
43.76$\pm$1.28& 9.78$\pm$0.20& 14.32$\pm$0.51& 2.87$\pm$0.27\\

\textbf{HNCO}& $17_{0,17}-16_{0,16}$&
373600.7& ~~161.4& $7.54\times10^{-4}$&
21.05$\pm$0.55& 10.73$\pm$0.15& 11.84$\pm$0.35& 1.67$\pm$0.13\\

\hline                  

\end{longtable}
\end{small}

\newpage

\begin{figure*}[h!]
\centering
\includegraphics[width=17.6 cm,trim=-0.5cm 0.3cm 0cm 0,clip=true]{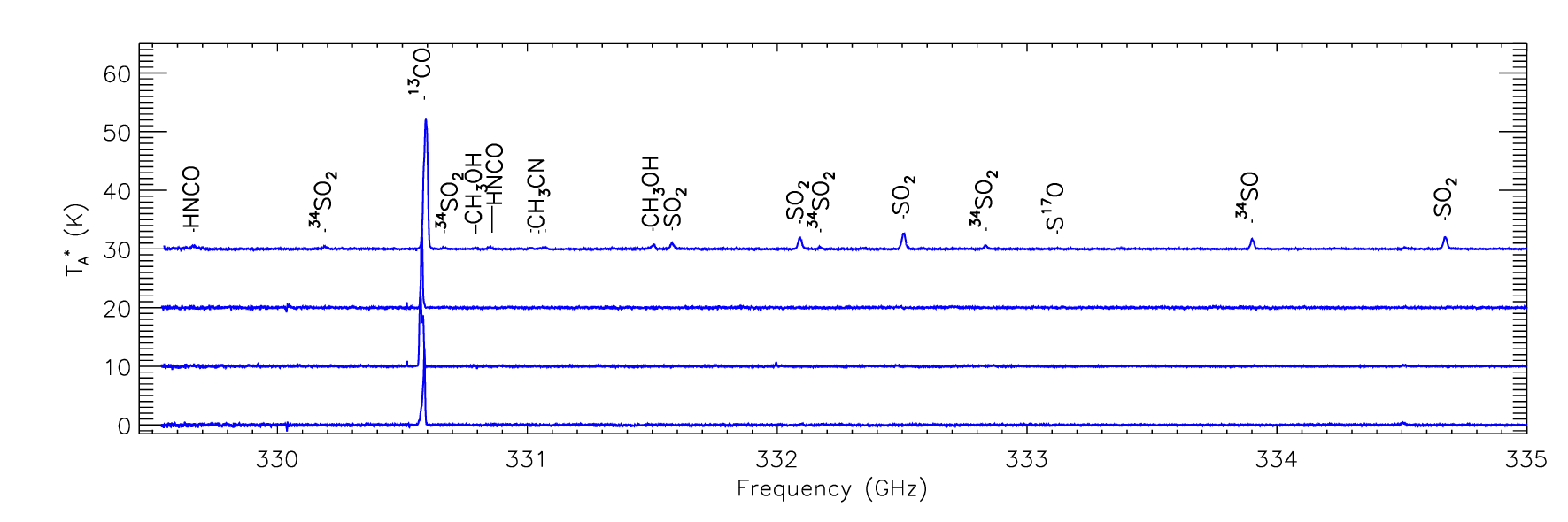}
\vskip-0.5cm
\includegraphics[width=17.6 cm,trim=-0.5cm 0.3cm 0cm 0,clip=true]{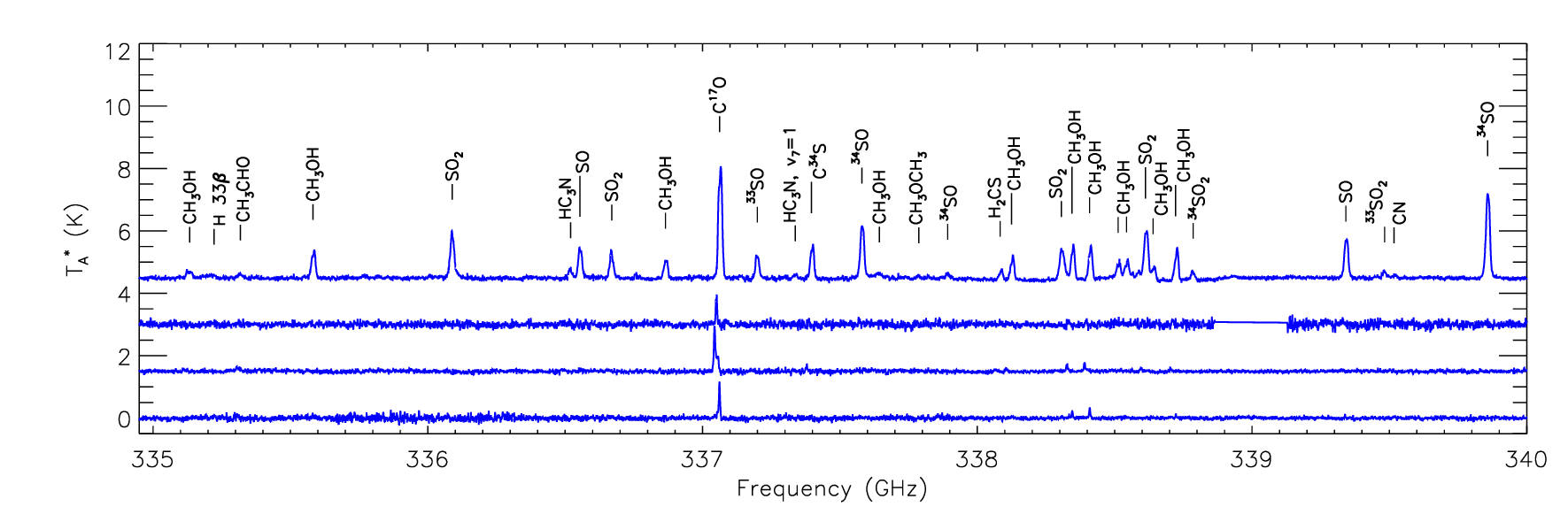}
\caption{The identified lines toward the (from bottom to top) Eastern tail, Northern clump, South-west clump regions, and the center of W49A in the frequency range between 330 and 335 GHz and 335 and 340 GHz.}
\centering
\includegraphics[width=17.6 cm,trim=-0.5cm 0.3cm 0cm 0,clip=true]{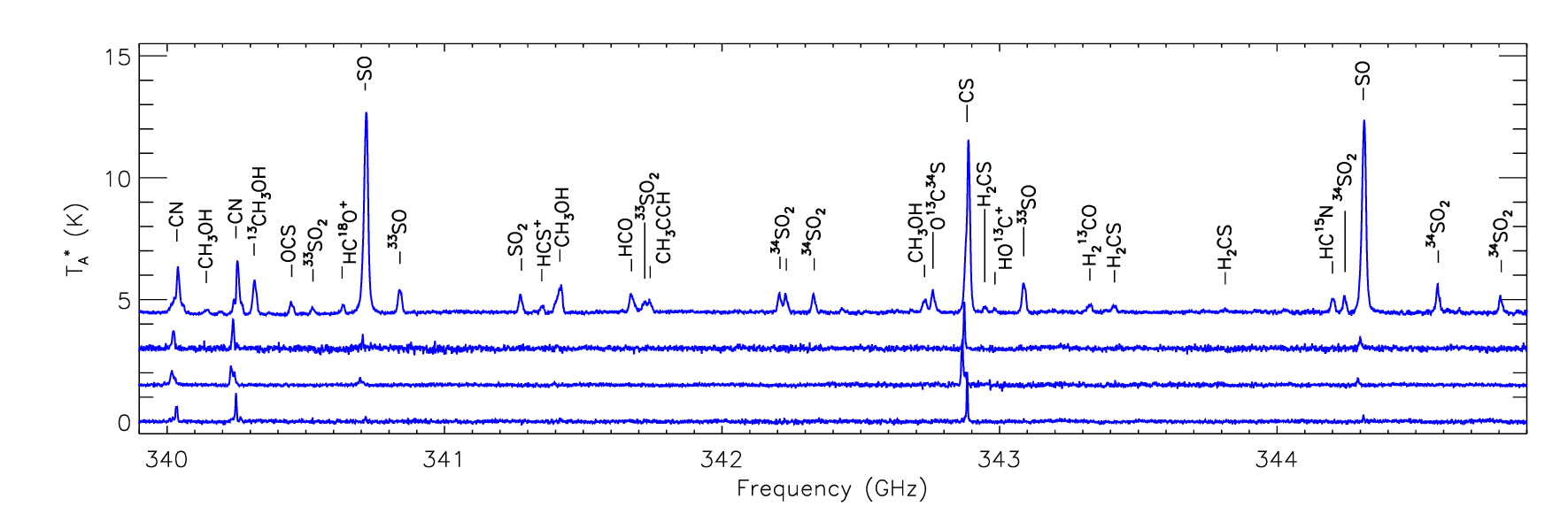}
\vskip-0.5cm
\includegraphics[width=17.6 cm,trim=-0.5cm 0.3cm 0cm 0,clip=true]{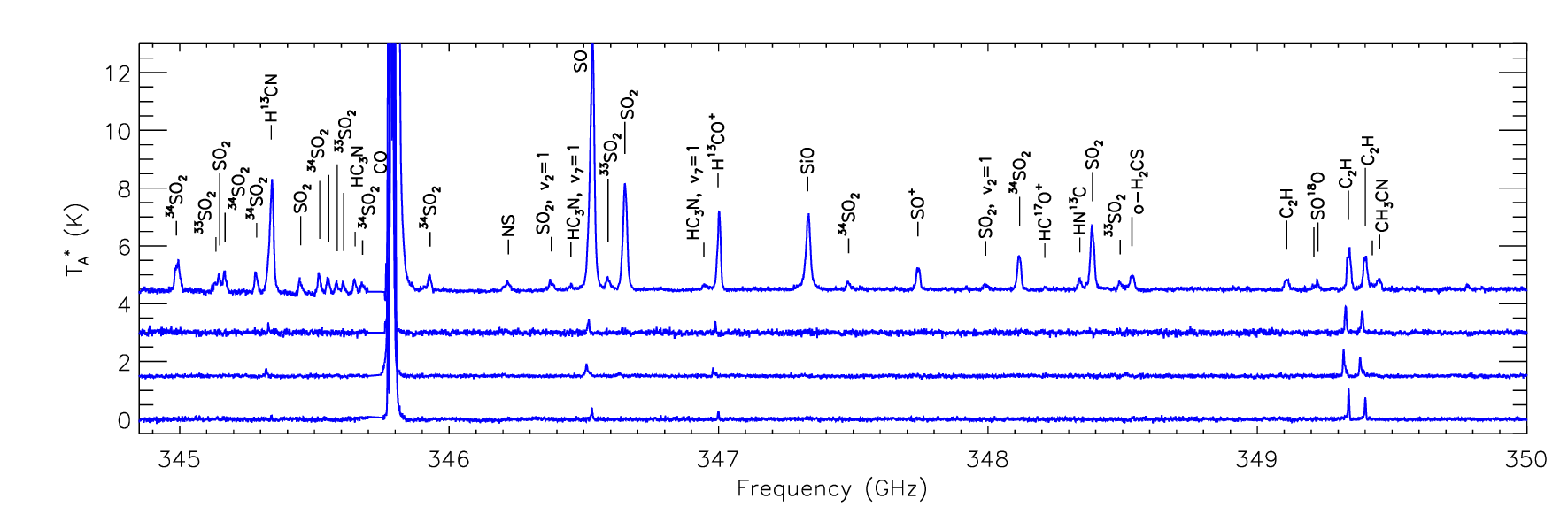}
\caption{The identified lines toward the (from bottom to top) Eastern tail, Northern clump, South-west clump regions, and the center of W49A in the frequency range between 340 and 345 GHz and 345 and 350 GHz.}
\label{spec_330_350}
\end{figure*}

\begin{figure*}[ht]
\centering
\includegraphics[width=17.6 cm,trim=-0.5cm 0.3cm 0cm 0,clip=true]{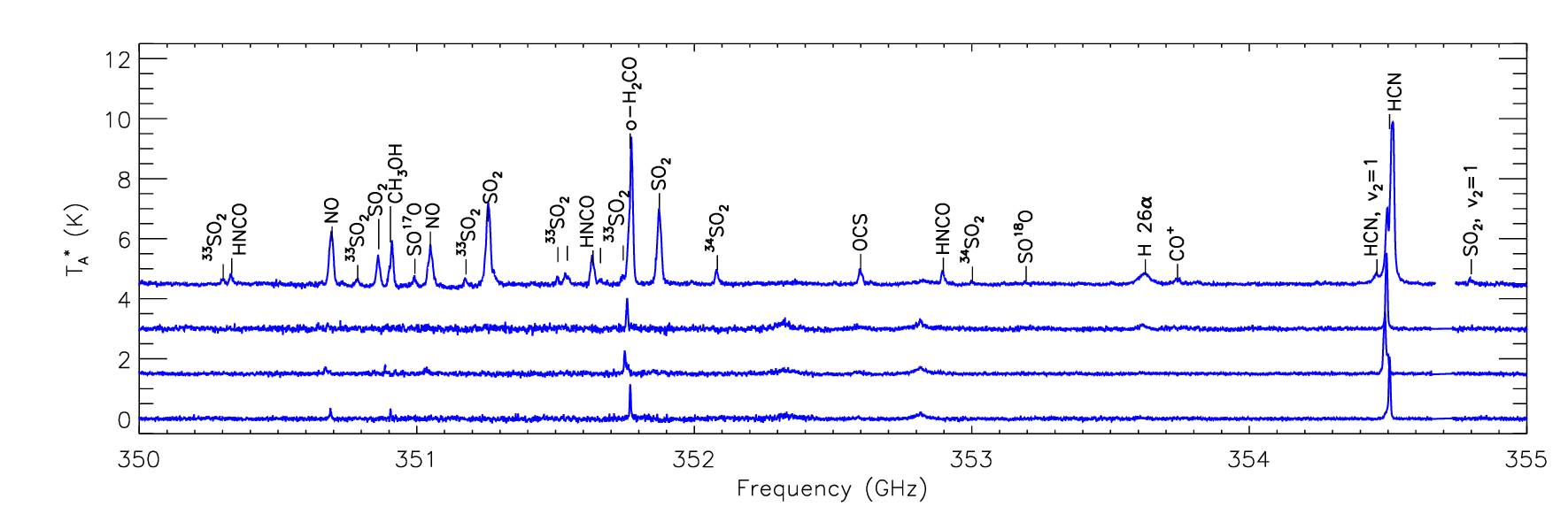}
\vskip-0.5cm
\includegraphics[width=17.6 cm,trim=-0.5cm 0.3cm 0cm 0,clip=true]{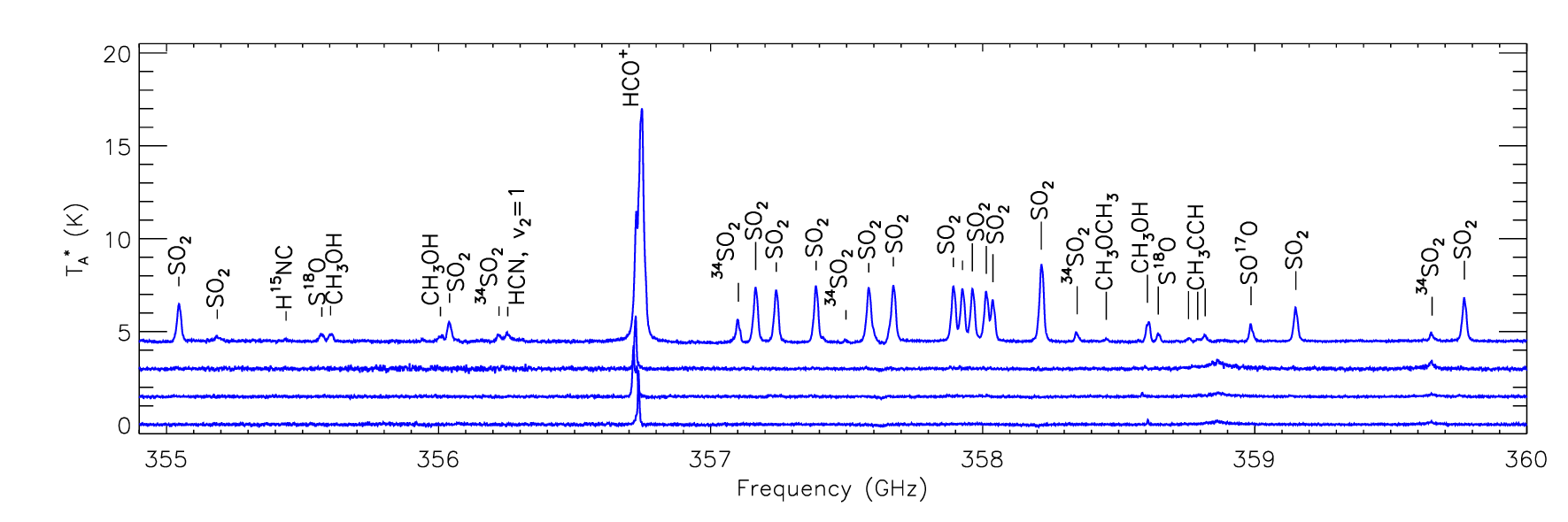}
\caption{The identified lines toward the (from bottom to top) Eastern tail, Northern clump, South-west clump regions, and the center of W49A in the frequency range between 350 and 355 GHz and 355 and 360 GHz.}
\label{spec_350_360}
\end{figure*}

\begin{figure*}[ht]
\centering
\includegraphics[width=17.6 cm,trim=-0.5cm 0.3cm 0 0,clip=true]{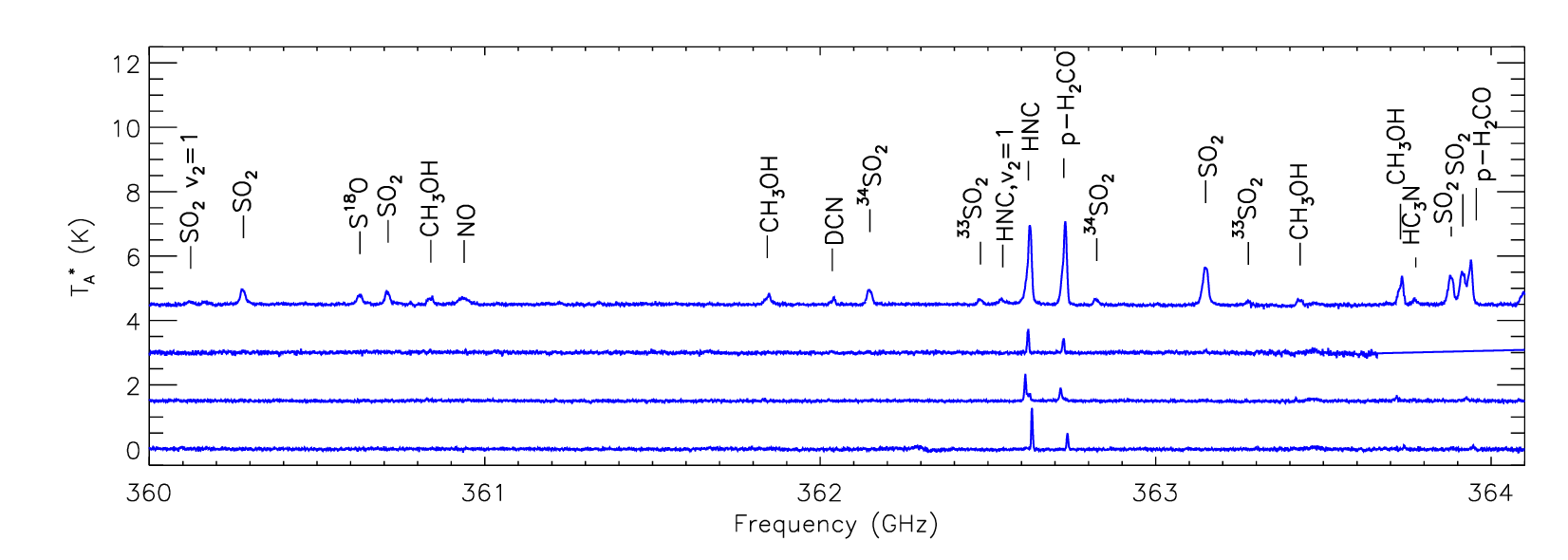} 	    
\vskip-0.5cm
\includegraphics[width=17.6 cm,trim=-0.5cm 0.3cm 0 0,clip=true]{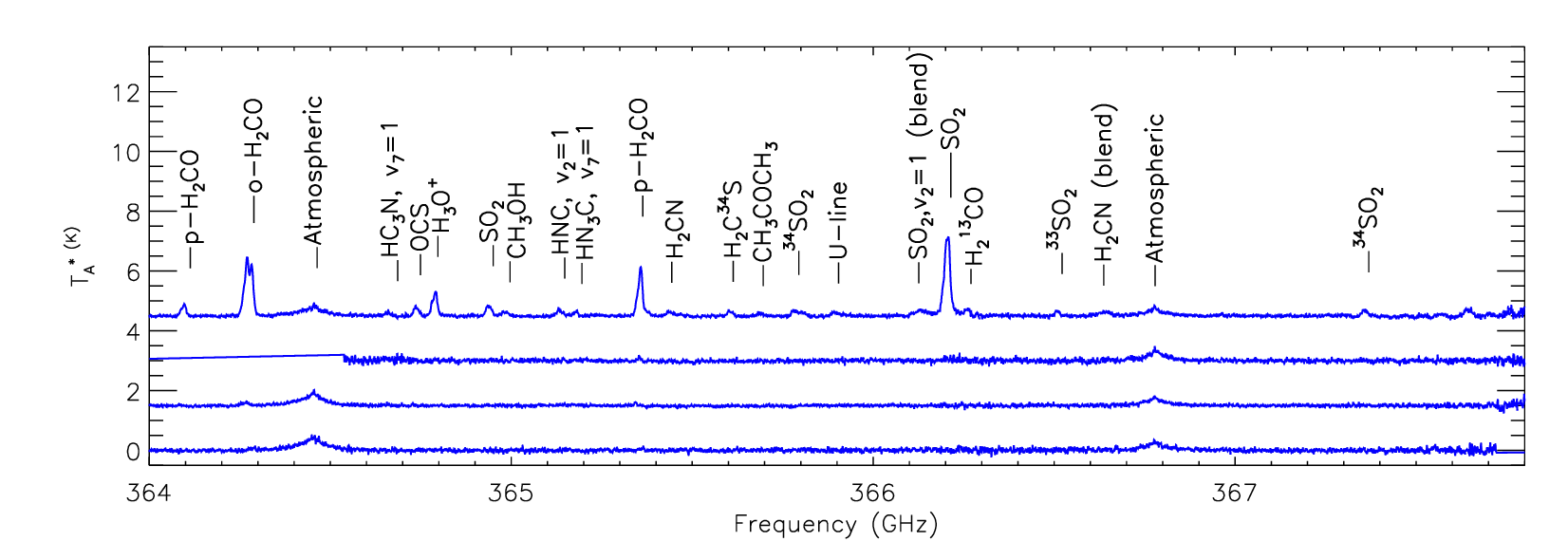} 	    
\caption{The identified lines toward the (from bottom to top) Eastern tail, Northern clump, South-west clump regions, and the center of W49A in the frequency range between 360 and 364 GHz and 364 and 368 GHz.}
\label{spec_subset1}
\end{figure*}

\begin{figure*}[ht]
\centering
\includegraphics[width=17.6 cm,trim=-0.5cm 0.3cm 0 0,clip=true]{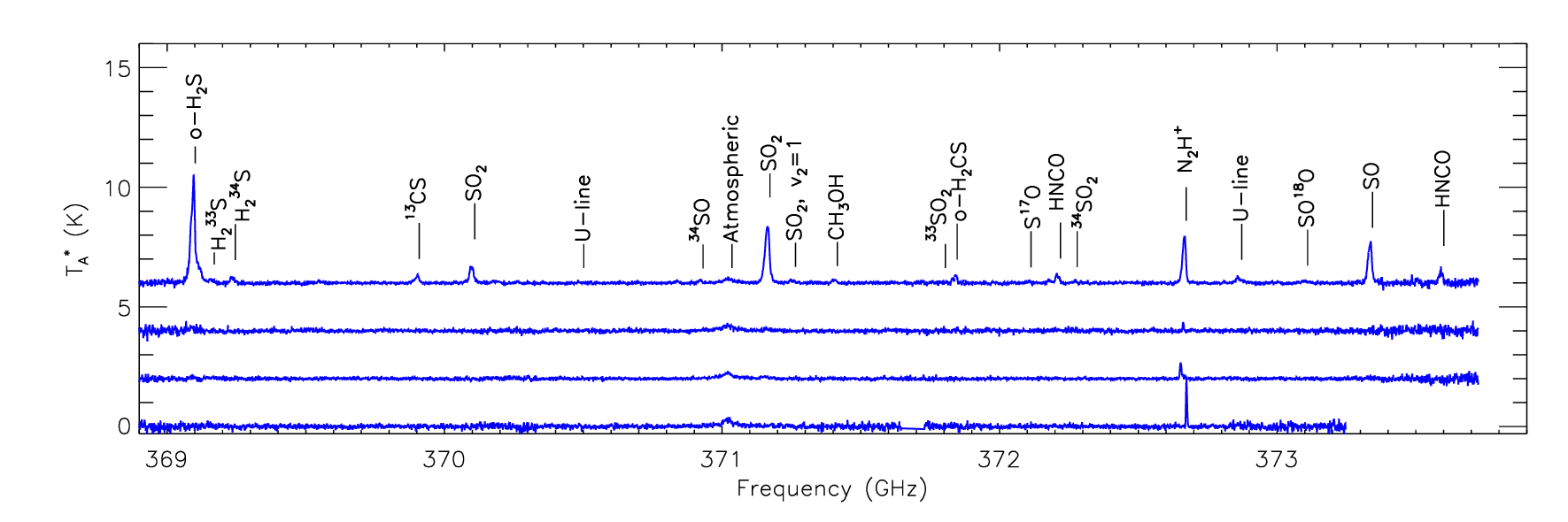} 	    
\caption{The identified lines toward the (from bottom to top) Eastern tail, Northern clump, South-west clump regions, and the center of of W49A in the frequency range between 369 GHz and 374 GHz.}
\label{spec_subset3}
\end{figure*}

\end{appendix}

\end{document}